\documentclass[preprint2]{proto}
\usepackage{comment}
\newcommand{\lt}{\ensuremath <}

\usepackage{times}
\usepackage{xcolor}
\usepackage{graphicx}
\usepackage{float}
\usepackage{amsmath}
\usepackage{float}
\usepackage{dblfloatfix}
\usepackage[outline]{contour}
\usepackage{tikz}
\usepackage{multirow}
\usepackage{widetext}
\usepackage{bigdelim}
\usepackage{enumitem}
\usepackage[makeroom]{cancel} 
\usepackage[normalem]{ulem}
\newcommand{\GG}[1]{}
\newcommand{\arrowIn}{
\tikz \draw[-stealth,line width=1mm] (-1mm,0) -- (1mm,0);
}
\usepackage{multicol,lipsum}
\usetikzlibrary{fit,matrix, shapes, arrows, positioning,calc,scopes,fadings,angles,arrows.meta,arrows.meta,angles,quotes,calc}
\usetikzlibrary{decorations.pathmorphing}
\usetikzlibrary{decorations.pathreplacing,calligraphy}
\usetikzlibrary{decorations.markings}
\usetikzlibrary{decorations.text}

\renewcommand{\comment}[1]{\vskip-\lastskip}
\usepackage{pgfplots}
\usepgfplotslibrary{fillbetween}
\pgfplotsset{
  compat=newest,
  xlabel near ticks,
  ylabel near ticks
}
\let\vec\mathbf

\graphicspath{{Chapter_latex/figures/}}

\begin{document}

\title{\textbf{\LARGE Cometary Ionospheres: An Updated Tutorial}}

\author {\textbf{\large Arnaud Beth}}
\affil{\small\em Umeå Universitet}

\author {\textbf{\large Marina Galand}}
\affil{\small\em Imperial College London}

\author {\textbf{\large Cyril Simon Wedlund}}
\affil{\small\em Österreichische Akademie der Wissenschaften, Institut für Weltraumforschung}

\author {\textbf{\large Anders Eriksson}}
\affil{\small\em Institutet för Rymdfysik}

\begin{abstract}

\begin{list}{ } {\rightmargin 1in}
\baselineskip = 11pt
\parindent=1pc
{\small 
This chapter aims at providing the tools and knowledge to understand and model the plasma environment surrounding comets in the innermost part near the nucleus. In particular, our goal is to give an updated post-Rosetta `view' of this ionised environment: what we knew, what we confirmed, what we overturned, and what we still do not understand.\\~\\~\\~
}
\end{list}
\end{abstract}  

\section{\textbf{INTRODUCTION}}

 \comment{Cyril: Intro reviewed on my part. I think it's really good and interesting! It sets the stage well. Btw, as mentioned below, maybe the metaphor of a stage, with players/actors, scenes, interludes, etc., could be used to tie the whole chapter. Just a suggestion, of course. Also, in the end, we should check where the Rosetta instruments and their acronyms are introduced somewhere, either in intro or after (intro would be the logical place at the end of the section). Finally, \emph{au} for astronomical unit should be defined (although I assume that the editors have an idea about this unit, proper, as we won't be the first ones to use it.}
    
    Comets have aroused the curiosity of Humankind for millennia. 
    Reported in different ways throughout history, for instance through texts and artworks, a few of them have passed Earth close enough to be witnessed with the naked eye. But what makes comets visible and so bright? For those lucky enough to see them in their lifetime, comets in the sky display very singular shapes and colours: a very broad white tail and a thinner blue slightly transparent one, both originating from the same point where the nucleus is located.
    The `white tail' is partly directed anti-sunward (that is, opposed to the Sun's direction), with a component along the comet's trajectory. It originates from the dust reflecting sunlight that is continuously expelled from the comet's surface and pushed away, swept by the solar radiation pressure. This radiation pressure is induced by solar photons which, once they hit the dust, transfer a part of their momentum continuously to it, equivalent to a force. Depending on the size of dust grains, this force, in addition to inertial ones, is the main driver of the dust dynamics, which as a result may overcome the Sun's gravity \citep[Finson-Probstein model, see][]{Finson1968}. 
    
    The `blue' tail itself originates from ions present in the coma. This was first evidenced in the band spectra of Comet C/1907 L2 (Daniel) \citep{Deslandres1907,Evershed1907} although its true origin was unknown at the time \citep{Larsson2012}. Subsequently, \citet{Fowler1907, Fowler1910} investigated these bands by means of laboratory experiments and electric discharges in different gases and concluded that carbon monoxide was the most likely cause. Fowler was \emph{almost} right. The blue colour displayed by this tail comes in fact from the fluorescence of CO$^+$ ions. Although the cometary gas is mainly composed of water (H$_2$O), other species are present such as carbon dioxide (CO$_2$), carbon monoxide (CO), and formaldehyde (methanal, H$_2$CO). As they leave the nucleus, these molecules may break into smaller fragments, neutral atoms/molecules (through photodissociation), be ionised (photoionisation), or undergo both processes at the same time (dissociative photoionisation). The newly formed CO$^+$ may in turn be excited by absorbing solar photons at specific wavelengths and is de-excited by re-emitting a photon at the same wavelength (resonance fluorescence). Amongst cometary species, CO$^+$ is the only one which efficiently emits in the visible, especially in the blue part (so-called `comet tail' A-X bands emitting between 308--720\,nm). 
    
    In almost a century, even though substantial progress has been made in astronomy and instrumentation, only a few other ions have been observed and confirmed at comets through remote sensing from Earth, for example HO$^+$, CO$_2^+$, and H$_2$O$^+$ (see the chapter by Bodewits et al. in this volume). Early on, another mystery was also associated with the `ion' tail: its direction. Indeed, through the comet's course around the Sun, and unlike the dust tail, the blue ion tail is exactly anti-sunward. The discovery of the solar wind was motivated by the observations of cometary ion tails from Earth. Following early works by \cite{Hoffmeister1943}, \citet{Biermann1951} was the first to suggest then that there is a flow of ionised particles from the Sun, `a stream of solar matter' \citep{Biermann1952}, although why this stream could arise was another unanswered question. In 1958, Eugene Parker followed on Biermann's lead and published his seminal paper on the generation and existence of the so-called `solar wind' \citep{Parker1958}, an idea which was initially ill received \citep{Obridko2017}.
    
    Interestingly, contemporarily to these works, the true nature of comets and of their nucleus remained unknown. Only in 1950 did Fred Whipple formulate the idea that comets were made of a conglomerate of ices combined with a conglomerate of meteoric materials \citep{Whipple1950}, known as the `dirty snowball' nucleus model. Our modern view based on recent observations of cometary nuclei shows that comets are rather \emph{dirt balls with some snow}. Based on this model, \citet{Haser1957} \citep[see][for a modern, accessible, and translated version]{Haser2020} later developed the mathematical framework to describe the cometary neutral environment, nicknamed `Haser model' in the literature nowadays. Haser derived the mathematical formulation describing the spatial distribution and number density for neutral species, either those coming from the ice sublimation at the (near-)surface of the nucleus, the `parent' species, or those coming from the dissociation of the parent species, the `daughter' species. For `parent' molecules, such as H$_2$O and CO$_2$, primarily released from the (near-)surface, the flux is conserved; the number density profile of the escaping gas of species $p$ is thus given in m$^{-3}$ by:
    \begin{equation}
        \boxed{n_p(r)= \dfrac{Q_p}{4\pi\,V_p\,r^2}\exp\left(-\dfrac{r}{\mathcal{T}_pV_p}\right)}
    \label{eq:n_p_full}
    \end{equation}
    where $Q_p$ is the total number of `parent' ($p$) molecules expelled from the surface per unit time ([molecules\,s$^{-1}$] or simply [s$^{-1}$] for short), $V_p$ is the radial speed of these molecules assumed constant [m\,s$^{-1}$] estimated around 0.4--0.9~km\,s$^{-1}$ \citep[see the chapter by Biver et al. in this volume and][]{Hansen2016,Biver2019}, $r$ is the distance from the nucleus [m], and $\mathcal{T}_p$ is the lifetime of these molecules against dissociation [s]. This equation remains valid for assymmetric outgassing (i.e. $Q_p$ varies with latitude and longitude at a given $r$ as long as the gas velocity remains radial). Usually, molecules are characterised, in a cometary context, by their so-called characteristic scale length $L_p=V_p\mathcal{T}_p$. This model is adequate for species and molecules that are mainly photodissociated by Extreme Ultraviolet (EUV) solar radiation, for instance H$_2$O into HO+H. It also provides an easy way to retrieve the outgassing rate $Q_p$ of the different cometary neutral parent species based on remote sensing observations. $L_p$ is usually of the order of $10^4$--$10^5$~km corresponding to a typical lifetime of $10^4$--$10^5$~s \citep{Huebner2015}; the exponential correction in Eq.~\ref{eq:n_p_full} should be accounted for when comets are observed remotely from Earth (as the full extent of the coma is probed) or when numerical simulations of the whole cometary environment are developed. Furthermore, not all neutral species may come directly from the nucleus. Some species, such as H$_2$CO and CO, may also originate from dust and photodissociation of larger molecules already present in the coma, referred to as `extended sources', and thus do not follow Eq.~\ref{eq:n_p_full} \citep{Eberhardt1999}. These extended sources must be accounted for as well in large-scale simulations beyond thousands of kilometres (see the chapter by Biver et al. in this volume). However, in the rest of this chapter, the exponential term may be dropped in Eq.~\ref{eq:n_p_full} as we are focusing on {major neutral species (originating primarily from the nucleus)} and on in situ observations at cometocentric distances $r\ll L_p$. 
    In these regions, in situ observations performed at 1P/Halley \citep[hereafter 1P,][]{Krankowsky1986} and at 67P/Churyumov-Gerasimenko \citep[hereafter 67P,][]{Hassig2015} showed that $n_{\text{H}_2\text{O}}\propto 1/r^2$ and Eq.~\ref{eq:n_p_full} reduces to:
    \begin{equation}
        \boxed{n_p(r)\approx \dfrac{Q_p}{4\pi\,V_p\,r^2}\quad\text{for}\quad r\ll L_p}
        \label{eq:n_p}
    \end{equation}
    Although there are neutral species produced in part within the coma and linked to the so-called \emph{extended sources} \citep{Eberhardt1999}, such as CO and O, such a source can often be neglected in plasma models of the inner coma.
    
    Similarly to Eq.~\ref{eq:n_p}, the total neutral number density $n_n$ in the inner coma is given by:
    \begin{equation}
        \boxed{n_n(r) = \sum_p n_p \approx \dfrac{Q}{4\pi\,V_n\,r^2}}
        \label{eq:n_n}
    \end{equation}
    where
    \begin{equation}
    \boxed{V_n = \frac{1}{n_n} \sum_p V_p n_p}
    \label{eq:neutral_speed}
    \end{equation}
    $Q$ is the total outgassing rate (total number of neutral species released by unit time from the (near) surface) and $V_n$, the mean bulk velocity of the neutral species. 
    
    The neutral gas released by sublimation of ices at the surface of the nucleus is primarily made of H$_2$O, CO$_2$, and CO (see the chapter by Biver et al. in this volume). Many other molecules are also present in the coma, such as dioxygen (O$_2$), ammonia (NH$_3$), and glycine (NH$_2$CH$_2$COOH), unambiguously detected in situ at 67P \citep{Leroy2015,Altwegg2016,Gasc2017} thanks to the mass spectrometer ROSINA (Rosetta Orbiter Spectrometer for Ion and Neutral Analysis)/DFMS  \citep[Double Focusing Mass Spectrometer,][]{Balsiger2007}. Along their path in the interplanetary medium, molecules may undergo different fates: They can be excited, dissociated into neutral fragments, or ionised by means of absorption of solar EUV radiation or impact of energetic particles. Although dissociation is more efficient than ionisation and excitation, a fraction of the molecules still undergoes ionisation; this forms a region made of plasma, a mixture of ions and electrons, surrounding the nucleus: the  \emph{ionosphere}.
    
\begin{figure}[ht]
    \resizebox{\linewidth}{!}{
	\begin{tikzpicture}[thick,scale=0.7, every node/.style={scale=0.7}]
    \clip (-9,-4) rectangle (2,4);
	\draw[opacity=0.9,inner color=blue!50!white,outer color=white,draw=white] (0,0) circle (7);

	\fill[yellow] (-30,0) circle (22);
    \fill[black] (0,0) circle (0.5);

    \draw [thick,-{Latex[width=4mm, length=6]},decorate,decoration={snake,amplitude=1mm,segment length=4mm,post length=2mm},yellow,line width=1mm] (-8,0) -- (-4,0);
    \draw [thick,-{Latex[width=4mm, length=6]},decorate,decoration={snake,amplitude=1mm,segment length=4mm,post length=2mm},yellow,line width=1mm] (-7,2) -- (-3,2);
    \draw [thick,-{Latex[width=4mm, length=6]},decorate,decoration={snake,amplitude=1mm,segment length=4mm,post length=2mm},yellow,line width=1mm] (-7,-2) -- (-3,-2);
    \draw [-stealth,thick,color=white!50!black] (-6,-4) -- (-6,3);
    \draw [-stealth,thick,color=white!50!black] (-5,-4) -- (-5,3);
    \draw [-stealth,thick,color=white!50!black] (-7,-4) -- (-7,3);
    \draw [-stealth,thick,color=white!50!black] (-7,-3.5) -- (-7,-2.7) node[pos=0.5,anchor=east]{$\mathbf{B_{SW}}$};
    \draw [-stealth,thick,black] (-7,-3.5) -- (-6.2,-3.5) node[pos=0.5,anchor=north]{$\mathbf{V_{SW}}$};
    \draw[-,thick,black] (-6.5,-3) circle (0.2) node[above=2.5pt]{$\mathbf{E_{SW}}$};
    \draw[-,thick,black] ({-6.5+0.2*cos(45)},{-3+0.2*sin(45)})  -- ({-6.5-0.2*cos(45)},{-3-0.2*sin(45)}) ;
    \draw[-,thick,black] ({-6.5+0.2*cos(45)},{-3-0.2*sin(45)})  -- ({-6.5-0.2*cos(45)},{-3+0.2*sin(45)}) ;
    \def\r{0.4}    
    \pgfmathsetseed{5};
    \foreach \i in {1,2,3,4,5,6}{
        \pgfmathsetmacro{\radius}{1+((rand+1)*0.5)*2}
        \pgfmathsetmacro{\theta}{rand*360}
        \pgfmathsetmacro{\phi}{rand*360}
        \pgfmathsetmacro{\xcoor}{\radius*cos(\phi)}
        \pgfmathsetmacro{\ycoor}{\radius*sin(\phi)}
        \pgfmathsetmacro{\xcoora}{\radius*cos(\phi)+0.1}
        \pgfmathsetmacro{\ycoora}{\radius*sin(\phi)+0.1}
        \pgfmathsetmacro{\xcoorb}{\radius*cos(\phi)-0.1}
        \pgfmathsetmacro{\ycoorb}{\radius*sin(\phi)-0.1}
        \begin{scope}[shift={(\xcoor,\ycoor)},rotate=45]
        \shade [ball color=white,shading angle=45] ({\r*cos(\theta-52.225)},{-\r*sin(\theta-52.225)}) circle (0.2);
        \shade [ball color=white,shading angle=45] ({\r*cos(\theta+52.225)},{-\r*sin(\theta+52.225)}) circle (0.2);
        \shade [ball color=red,shading angle=45] (0,0) circle (0.3);
        \end{scope}
        \draw [thick,black] (\xcoor,\ycoora) -- (\xcoor,\ycoorb);
        \draw [thick,black] (\xcoora,\ycoor) -- (\xcoorb,\ycoor);
    }
    \pgfmathsetseed{1};
    \foreach \i in {1,2,3,4,5,6,7,8}{
        \pgfmathsetmacro{\radius}{1+((rand+1)*0.5)*1.8}
        \pgfmathsetmacro{\phi}{rand*360}
        \pgfmathsetmacro{\xcoor}{\radius*cos(\phi)}
        \pgfmathsetmacro{\ycoor}{\radius*sin(\phi)}
        \pgfmathsetmacro{\xcoora}{\radius*cos(\phi)+0.1}
        \pgfmathsetmacro{\ycoora}{\radius*sin(\phi)+0.1}
        \pgfmathsetmacro{\xcoorb}{\radius*cos(\phi)-0.1}
        \pgfmathsetmacro{\ycoorb}{\radius*sin(\phi)-0.1}
        \begin{scope}[shift={(\xcoor,\ycoor)},rotate=45]
        \shade [ball color=blue,shading angle=45] (0,0) circle (0.15);
        \end{scope}
        \draw [thick,white] (\xcoora,\ycoor) -- (\xcoorb,\ycoor);
    }
    \pgfmathsetseed{8};
    \foreach \i in {1,2,3,4}{
        \pgfmathsetmacro{\xcoor}{-6.5+rand}
        \pgfmathsetmacro{\ycoor}{rand*3}
        \pgfmathsetmacro{\xcoora}{\xcoor+0.1}
        \pgfmathsetmacro{\ycoora}{\ycoor+0.1}
        \pgfmathsetmacro{\xcoorb}{\xcoor-0.1}
        \pgfmathsetmacro{\ycoorb}{\ycoor-0.1}
        \draw [-stealth,thick,black] (\xcoor,\ycoor) -- ({\xcoor+0.8},\ycoor);
        \begin{scope}[shift={(\xcoor,\ycoor)},rotate=45]
        \shade [ball color=white,shading angle=45] (0,0) circle (0.2);
        \end{scope}
        \draw [thick,black] (\xcoora,\ycoor) -- (\xcoorb,\ycoor);
        \draw [thick,black] (\xcoor,\ycoora) -- (\xcoor,\ycoorb);
    }
    \pgfmathsetseed{1};
    \foreach \i in {1,2,3,4}{
        \pgfmathsetmacro{\xcoor}{-6.5+rand}
        \pgfmathsetmacro{\ycoor}{rand*3}
        \pgfmathsetmacro{\xcoora}{\xcoor+0.1}
        \pgfmathsetmacro{\ycoora}{\ycoor+0.1}
        \pgfmathsetmacro{\xcoorb}{\xcoor-0.1}
        \pgfmathsetmacro{\ycoorb}{\ycoor-0.1}
        \draw [-stealth,thick,black] (\xcoor,\ycoor) -- ({\xcoor+0.8},\ycoor);
        \begin{scope}[shift={(\xcoor,\ycoor)},rotate=45]
        \shade [ball color=blue,shading angle=45] (0,0) circle (0.15);
        \end{scope}
        \draw [thick,white] (\xcoora,\ycoor) -- (\xcoorb,\ycoor);
    }
    
    
\draw [decorate, 
        decoration = {calligraphic brace,
        raise=5pt,
        aspect=0.5}] (-2.5,3) --  (2,3)node[pos=0.5,above=5pt]{Cometary ionosphere};
\draw [decorate, 
        decoration = {calligraphic brace,
        raise=5pt,
        aspect=0.5}] (-8,3) --  (-3,3)node[pos=0.5,above=5pt]{Solar wind\vphantom{y}};
    
     \node[anchor=center] at (0, -3.5)   (a) {$\mathbf{E}?$ $\mathbf{B}?$};
     
	\end{tikzpicture}
	}
	\caption{Schematic of the actors of the interaction of the comet (black disk) with its space environment. They include: (i) the cometary ionosphere, here represented by water H$_2$O$^+$ ions (red/white) and electrons (blue around the coma), (ii) the solar extreme ultraviolet (EUV) radiation (yellow), responsible in part for ionising the neutral coma, and (iii) the solar wind plasma, made of H$^+$ (white) and electrons (blue) carrying the interplanetary magnetic field (grey vertical arrows), filling the interplanetary medium and travelling at speeds typically between 400 and 800~km\,s$^{-1}$.
	The transition between the solar wind and cometary ionosphere and the nature of this transition depend on many factors (e.g. outgassing activity, composition, heliocentric distance, etc.), discussed in detail in the chapter by Götz et al. in this volume. \label{schematic}}
\end{figure}
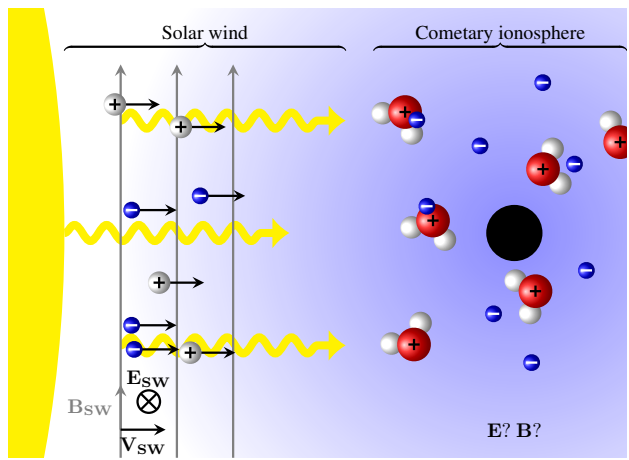

    Ionospheres are commonly present around any solar and extrasolar system body possessing a neutral gas layer, either an atmosphere, like at Earth, Mars, Jupiter and Titan, or an exosphere, like at Mercury or Ganymede. It plays a critical role in the interaction of the Sun (or the star) with the body. Indeed, none of these solar system bodies is isolated: They all bathe in an external plasma, either directly the solar wind for bodies without an intrinsic magnetic field or the magnetospheric plasma (and indirectly the solar wind) for magnetised bodies. Around comets, two broad types of plasma are present (see Fig.\ref{schematic}): 
    \begin{itemize}
        \item The \emph{ionosphere} proper, made of cometary electrons and ions born from outgassed neutral molecules, is dense ($10^2-10^9$~particles\,m$^{-3}$). Cometary ions are heavy (above 12 unified atomic mass units [u] or Daltons [Da]), slow (from 0.5 to a few km\,s$^{-1}$), and cold. 
        \item In contrast, the \emph{solar wind} is rarefied ($1-10$~particles\,m$^{-3}$ at 1~au), made of protons (H$^+$), alpha particles (He$^{2+}$), and electrons ($e^-$); it is fast ($|\mathbf{V}_\text{SW}|=400$--$800$~km\,s$^{-1}$) and hot \citep[ion temperature $T_i \sim 0.8\times 10^5$~K, electron temperature $T_e\sim1.5\times 10^5$~K at 1~au typically scaling to the heliocentric distance $r_h$ as $r_h^{-2/3}$ and $r_h^{-1/3}$, respectively; see][]{Slavin1981}.
    \end{itemize}
    The solar wind transports with itself electromagnetic fields. It carries the solar magnetic field  $\mathbf{B}_\text{SW}$ which is `frozen-in', that is, the magnetic field is dragged into space by the expanding solar wind. Moreover, even though there is no electric field in the solar wind rest frame, as it moves, an observer in a moving frame with respect to the solar wind experiences an electric field from the advected magnetic field by means of relativistic Lorentz transformation. This field is the so-called  convective or motional electric field defined as $\mathbf{E_{SW}}= - \mathbf{V_{SW}}\times \mathbf{B_{SW}}$. A simplistic view of the relationship between a star and comets is shown in Fig.\,\ref{schematic}.
    
    The first spacecraft to visit a comet and its plasma environment was the NASA's \emph{International Cometary Explorer} (abbreviated \emph{ICE}) spacecraft which flew by comet 21P/Giacobini-Zinner {(21P)} on 11 September 1985 at a closest approach of 7800~km through the comet's tail \citep[see][]{Farquhar2001}. This mission was mainly dedicated to plasma measurements and was the first to provide in situ information on the interaction of the cometary ionosphere with the solar wind \citep{VonRosenvinge1986,Cowley1987}. ICE focused on the tail and could not investigate the innermost coma. Although ions of cometary origin were observed up to $4\times 10^6$~km from 21P's nucleus, they were already significantly accelerated, picked-up by the solar wind convective electric field, at speeds around 150~km\,s$^{-1}$ \citep{Ogilvie1986}. At the time, mass spectrometers, which discriminate and separate ions according to their mass per charge (or simply their mass if ions are assumed singly-ionised), had a very limited mass resolution. The resolution in \emph{mass spectrometry} is defined by $R=m/\Delta m$ where $m$ is the mass at which ions are measured and $\Delta m$ is the width of the peak typically at half height \citep[see IUPAC recommendation and nomenclature,][]{IUPAC1978}. With ICE, the mass resolution was relatively poor, with measurements separated by 1--2~u\,q$^{-1}$ (mass unit per charge, {$q$ being the elementary charge in Coulomb}), making it extremely difficult to separate species with relatively close molecular masses, for example HO$^{+}$, H$_2$O$^{+}$, and H$_3$O$^{+}$.
    
    The following year, the \emph{Halley} armada, a fleet of five spacecraft from several space agencies (one from the European Space Agency ESA [\emph{Giotto}], two from the Soviet Union and France [\emph{Vega~1} and \emph{Vega~2}] and two from the Institute of Space and Astronomical Science, now JAXA, in Japan [\emph{Suisei} and \emph{Sakigake}]), flew by the most famous comet of all and one of the brightest. As such, Halley's comet was the first one to be identified as periodic (1P) with its 76-year period determined by Edmond Halley in 1705. In succession, Sakigake, Vega 1, Suisei, Vega 2, and finally Giotto flew by Comet~1P, each probe providing a more precise location of 1P and therefore helping the next probe to get closer to its nucleus. On 13--14 March 1986, the last spacecraft of the fleet, Giotto, achieved a closest approach of 600~km at a heliocentric distance $r_h=0.89$~au from the Sun \citep{Reinhard1986}. The first detected `cometary' ion was O$^+$ (most likely candidate at the peak at 16~u\,q$^{-1}$) at $5.5\times 10^{5}$~km from the nucleus \citep{Krankowsky1986}. Most likely, it resulted from the dissociation of H$_2$O into O followed by ionisation \citep{Balsiger1986}. The mass resolution was high enough to separate HO$^{+}$ from H$_2$O$^{+}$, albeit not O$^{+}$ from NH$_{2}^{+}$ and CH$_{4}^{+}$ \citep[$R\gtrsim 20$,][]{Balsiger1987}.  
    

    As Giotto got closer to 1P's nucleus, the ion composition changed. Between $1.5\times 10^5$~km and $7\times 10^4$~km, the cometary plasma composition was dominated by ion mass 16~u\,q$^{-1}$ (most likely O$^+$), followed by ion masses 17~u\,q$^{-1}$ (most likely HO$^+$) and 18~u\,q$^{-1}$ (most likely H$_2$O$^+$). From $6\times 10^4$~km to $3\times 10^4$~km, the order changed: mass 18~u\,q$^{-1}$ dominated, followed by 17~u\,q$^{-1}$ and 16~u\,q$^{-1}$. From $3\times 10^4$~km to the closest approach, mass 19~u\,q$^{-1}$ (H$_3$O$^+$) exceeded in counts mass 16~u\,q$^{-1}$, and even dominated the overall composition below $2\times 10^4$~km. However, the reader should keep in mind that the main contributor at a given u\,q$^{-1}$ may change during the flyby. For instance, close to the nucleus, the signal at 18~u\,q$^{-1}$ may be associated with NH$_4^+$ while further away, it is more likely  H$_2$O$^+$ due to the ion chemistry. For this reason, ion measurements from Giotto had to be combined with photochemical models to infer the neutral and ion composition of the coma \citep[e.g.][]{Allen1987,Geiss1991}. Furthermore, not only the composition, but also the spatial distribution of the plasma, changed. The ion count associated with cometary ions exhibited a $1/r^2$-dependence for cometocentric distances $r$ beyond $16,000$~km, whereas, in the innermost part of the coma, ion counts varied in $1/r$. 
    
    Giotto was a great mission for exploring large spatial scales around a comet in a very limited time, taking a veritable snapshot of the cometary plasma down to 600~km, the best achievement until the ESA cometary mission \emph{Rosetta}, that escorted comet~67P over a two-year period (2014-2016), from $r_h\sim3.6$\,au to $r_h\sim1.24$\,au (perihelion) and back to 3.8\,au at the end of mission \citep{Glassmeier2007}. Nevertheless, several important aspects should be kept in mind when we compare the Giotto mission to Rosetta. Comet~1P was very active ($Q\approx 6.9\times 10^{29}$~s$^{-1}$), close to the Sun (0.89~au), and, in that respect, its flyby by Giotto and the entire Armada is far from being representative of the whole cometary environment and development.
    
    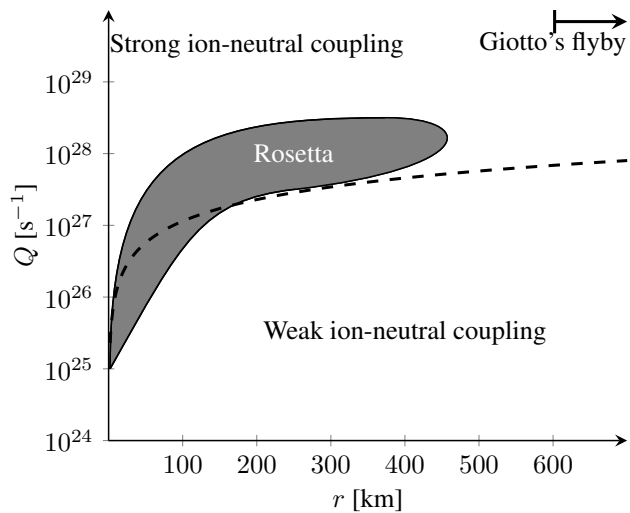
\begin{figure}[!ht]
    \resizebox{\linewidth}{!}{
	\begin{tikzpicture}[thick,scale=0.7, every node/.style={scale=0.7}]

     \begin{axis}[ymin=-3,
                  ymax=3,
                  xmax=7,
                  xmin=0,
                  ytick={-3,-2,-1,0,1,2},
                  xtick={1,2,3,4,5,6},                  xticklabels={$100$,$200$,$300$,$400$,$500$,$600$},                  yticklabels={$10^{24}$,$10^{25}$,$10^{26}$,$10^{27}$,$10^{28}$,$10^{29}$},
                  axis lines = left,
                  xlabel shift= 3pt,
                  ylabel shift= 3pt,
                  xlabel={$r$~[km]},
                  ylabel={$Q$~[s$^{-1}$]}]
                  
    \draw[color=black!50!white] (2.5,0.5) -- (3.75,1.5);
    \draw (0.02,-2) .. controls (0.02,1) and (1,1.5) .. (3.75,1.5) plot[smooth] (3.75,1.5) .. controls (5,1.5) and (5,0.8) .. (2.5,0.5) plot[smooth] (2.5,0.5) .. controls (1.25,0.35) and (1,-0.2) .. (0.02,-2);
    \draw[fill=black!50!white](3.75,1.5) .. controls (5,1.5) and (5,0.8) .. (2.5,0.5);
    \draw[name path=A] (0.02,-2) .. controls (0.02,1) and (1,1.5) .. (3.75,1.5);
    \draw[name path=B] (0.02,-2) .. controls (1,-0.2) and (1.25,0.35) .. (2.5,0.5);
    \addplot[dashed,color=black,domain={0.02:7},samples=1000,thick]{log10(x*10^2*10^28/875)-27};
    \node (none) at (2.5,1) {\textcolor{white}{Rosetta}};
    \addplot[black!50!white] fill between[of=A and B];
    \node (none) at (2,2.5) {\textcolor{black}{Strong ion-neutral coupling}};
    \node (none) at (4,-1.5) {\textcolor{black}{Weak ion-neutral coupling}};
    \draw[|-stealth,thick,color=black] (6,2.84) -- (7,2.84)node[below,pos=0]{Giotto's flyby};
  \end{axis}
    \end{tikzpicture}}
    \caption{Schematic of the cometocentric distance covered by Giotto and Rosetta as a function of the outgassing activity. Excursions have been excluded. The dashed line defines the theoretical limit where ions and neutrals become collisionally decoupled \citep{Gombosi2015} referred to as $R_{i,n}$ in this chapter.\label{fig:schematic}}
\end{figure}

As shown in Fig.~\ref{fig:schematic}, Giotto's closest approach was quite far compared with Rosetta ranges of cometocentric distances, even when scaling these distances to their vastly different outgassing rates. Rosetta not only explored a wide range of distances, from 500~km to the surface, but also more than 3 orders of magnitude in terms of outgassing rate, from $Q\approx 10^{25}$ to $\sim 3\times10^{28}$~s$^{-1}$, as 67P gradually became more active on its journey towards the inner solar system and then faded again on its outward journey. During the mission, the scientific community had thus the unique opportunity to monitor the cometary coma, the ionosphere and its relationship with the solar wind, and their combined evolution over time, providing a wealth of unprecedented scientific data. In the future, we may not have many opportunities to visit other comets with such a dedication; in this way, escorting 67P constitutes a major step forward in our understanding of solar wind-comet interactions.
    
This chapter proposes to look into the lessons learned from the combined past cometary flybys and the recent 2 years of escort by Rosetta, from the formation to the composition and evolution of a cometary ionosphere. It follows the example of the previous books on the matter, {Comets I} and {Comets II}. {Comets I}, published before in situ exploration of cometary environments (prior ICE and the Halley Armada), made use of remote-sensing observations from the Earth and possible theories \citep{IpAxford1982,Huebner1982}, describing the solar wind assimilation into the cometary environment with the formation of plasma boundaries, and the emergence of photochemical models of the inner coma including ion-neutral gas chemistry. {Comets II} focused on the advances brought by the contemporary exploration of a very active comet, 1P/Halley, with the first attempt to compare models against in situ observations \citep{Ip2004}. This chapter is a tentative harmonisation and consensus on our current knowledge of cometary ionospheres, in order to provide some sort of a `continuous' picture from weakly active to very active comets. This chapter must be seen and read as a `tutorial' or `manual' for those who want to become familiar with cometary ionospheres; it sets the stage for future explorations of comets. In Section~\ref{section:2}, we introduce the theory behind the formation of a cometary ionosphere and its chemical loss and discuss its plasma balance, composition, and evolution over time, in the light of both modelling and observations. In Section~\ref{section:3}, we present and focus on the different electron populations and their role observed at comets, while in Section~\ref{section:4} we present and focus on the different ion populations and their behaviour. Section~\ref{section:5} is dedicated to the role of dust on cometary plasmas. Finally, in Section~\ref{section:6}, we summarise the contemporary progress and discoveries made in cometary physics followed by a non-exhaustive list of open questions.
\section{Birth of a cometary ionosphere \label{section:2}}

\subsection{Ion continuity equations\label{section:2:1}}
    
After being ejected from the nucleus' surface via sublimation, desorption, or other mechanisms (see the chapter by Filacchione et al. in this volume), the cometary neutral gas forming the coma is partially ionised by solar radiation ($\lambda\lesssim 100$~nm, see Section \ref{section:2:2:1}), and by energetic electrons (Section \ref{section:2:2:2}) and ions (Section \ref{section:2:2:3}), leading to the formation of an ionosphere. The resulting ionosphere can be described by macroscopic, fundamental plasma quantities, such as the number density, mean velocity, and temperature, all in situ observables from instruments onboard a spacecraft:
\begin{itemize}
    \item The plasma number density is the number of ions (or electrons assuming that ions only carry one positive charge $+q$) per unit volume, often expressed in m$^{-3}$.
    \item The ion (resp. electron)  mean (bulk) velocity is the statistical averaged velocity  over all ions (resp. electrons) at a specific location, expressed in m\,s$^{-1}$.
    \item The ion (resp. electron) temperature represents the variance of the velocity around the mean ion (resp. electron) velocity. It can be expressed in K or in eV (where 1~eV $\approx 1.6\times 10^{-19}\text{~J}\approx k_B \times 11604$~K, $k_B$ being the Boltzmann constant).
\end{itemize}
Note that by virtue of quasi-neutrality of the plasma, ion and electron number densities (noted $n_i$ and $n_e$) must be `almost' equal ($n_i\approx n_e$). However, neither the mean velocities ($\vec{V}_i$ for ions and $\vec{V_e}$ for electrons) nor the temperatures ($T_i$ and $T_e$) are necessarily so.

Let us look at the ion number density. The fate of newly-born ions is manifold  (see Section~\ref{section:2:2} for a detailed description). For instance, ions can be transported, near or away from the cometary nucleus by means of electromagnetic forces. They can also collide along their path with neutral molecules. Through such a collision, the ion may lose its charge to the benefit of the neutral species, the latter itself becoming an ion. Finally, ions may recombine with free electrons and neutralise/dissociate, creating a net loss of ions and electrons. Such processes need to be taken into account when assessing the ion number densities. This can be achieved with the use of a mathematical model based on solving a set of differential equations.

The time evolution of the number density of cometary ion species $s$ is captured by the continuity equation, which describes the evolution of the number of particles over time and space: 
\begin{equation}
    \boxed{\dfrac{\partial n_s(\vec{r},t)}{\partial t}=-\underbrace{\nabla \cdot \left(n_s(\vec{r},t)\vec{V}_s(\vec{r},t)\right)}_{\text{transport}} + \underbrace{S_s(\vec{r},t)}_{\text{source}} - \underbrace{L_s(\vec{r},t)}_{\substack{\text{chemical}\\\text{ loss}}}}
    \label{eq:continuityIon}
\end{equation}
where $\vec{r}$ and $t$, the spatial and temporal coordinates respectively (the spatial reference frame is the comet's inertial frame centred on its nucleus), and where:
\begin{itemize}
    \item $\partial n_s(\vec{r},t)/\partial t$ is the instantaneous temporal variation of the ion number density in the infinitesimal volume $\mathrm{d}\mathcal{V}=\mathrm{d}^3\vec{r}$ at the position vector $\vec{r}$ and time $t$,
    \item $\nabla \cdot \left(n_s(\vec{r},t)\vec{V}_s(\vec{r},t)\right)$ is the transport term and quantifies the rate at which plasma enters and leaves the infinitesimal volume $\mathrm{d}\mathcal{V}$ at $(\vec{r},t)$. It is the net outward flux (here $n_s\vec{V}_s$) through the surface enveloping $\mathrm{d}\mathcal{V}$. If there is no production and no chemical loss of ions $s$ within $\mathrm{d}\mathcal{V}$, the number density $n_s(\vec{r,t})$ depends only on the amount of ions $s$ passing through the surface $\mathrm{d}\mathcal{S}$ at $(\vec{r},t)$, the outer boundary of $\mathrm{d}\mathcal{V}$. Beware that unlike some fluids, plasma is compressible (i.e. $\nabla \cdot \vec{V}\neq 0$),
    \item $S_s(\vec{r},t)$ is the local and instantaneous source (commensurable to a production rate) of the ion species $s$ in the infinitesimal volume $\mathrm{d}\mathcal{V}$ at $(\vec{r},t)$ through ionisation or ion-neutral collisions (see Sections~\ref{section:2:2} and \ref{section:2:3}),
    \item $L_s(\vec{r},t)$ is the local and instantaneous chemical loss rate of the species $s$ in the infinitesimal volume $\mathrm{d}\mathcal{V}$ at $(\vec{r},t)$ through collisions (see Section~\ref{section:2:3}).
\end{itemize}
All four terms in Eq.~\ref{eq:continuityIon}  are expressed in m$^{-3}$\,s$^{-1}$. One should note that there are as many continuity equations as there are ion species. This constitutes a challenging system to solve and therefore assumptions should be made. Physicists tend to reduce the number of independent variables, here $4$: $1$ for time and $3$ for space.

The most practical and well justified assumption is to model the system at/near steady state, that is, in the limit of very slow variations/low frequencies and long timescales: The left-hand side of Eq.~\ref{eq:continuityIon} becomes zero. Even if the system is disturbed, it will naturally converge and recover towards its steady state after a certain typical timescale, ruled here by transport and chemistry.

The second assumption is regarding the symmetry of the system. Although comets have various shapes, it has been shown that the neutral number density follows a $1/r^2$ law, at least for the major species, such as H$_2$O and CO$_2$ \citep[e.g. at 67P,][]{Hassig2015}: This is the same as for a point source or a sphere emitting gas with a constant radial speed, as in Eq.~\ref{eq:n_p}. Close to the nucleus and in the first few hundred kilometres from it, this assumption appears reasonable. Farther away from the nucleus, the loss of neutrals needs to be taken into account (see Eq.~\ref{eq:n_p_full}). 

Source and loss terms of cometary ions, $S_s$ and $L_s$ in Eq.~\ref{eq:continuityIon}, are described in detail in Sections~\ref{section:2:2} (net production of charge in the coma) and \ref{section:2:3} (chemical production and loss of cometary ions through ion collisions with neutrals and with electrons). Solution of Eq.~\ref{eq:continuityIon}, along with the relative importance of the different terms, is discussed in Sections\,\ref{section:2:4:1} when applied to the full plasma and in \ref{section:2:4:2} when applied to individual ion species.

\subsection{Source of the cometary plasma\label{section:2:2}}

We consider three net sources of charge for the cometary ionosphere: photoionisation by solar Extreme Ultraviolet (EUV) radiation (see Section~\ref{section:2:2:1}), ionisation by energetic electrons (see Section~\ref{section:2:2:2}) and ionisation and charge exchange by solar wind ions (see Section~\ref{section:2:2:3}). The relative importance of each of these sources is reviewed in Section~\ref{section:2:2:4}. Complementarily, the reader is directed to other textbooks on the matter \citep{Schunk2009}.

\subsubsection{Photoionisation\label{section:2:2:1}} 
    
Photoionisation is the process of absorption of a sufficiently energetic photon by an atom/molecule $A$ and the subsequent ejection of one or several of its bound electrons. This is a main plasma production process leading to the formation of an ionosphere at any solar system body with a gas envelop:
    \begin{align}
        A + \text{photon (or }h\nu) &\longrightarrow {A}^{m+(*)} + m e^-\nonumber\\
        &\longrightarrow {B}^{m+(*)} + C^{(*)}+m e^-
    \label{eq:photoioni_reaction}
    \end{align}
    where $m$ is the number of electrons released in the process (in most cases $m=1$) and the star $*$ denotes a possible excited state of the ion {${A}^{m+}$, $B^{m+}$, and/or the neutral species $C$}. {Photoionisation can produce new neutral atoms or molecules (here $C$) if $A$ is a molecule: The atoms constituting $B$ and $C$ together constituted $A$.}
    The produced ion {and/or the neutral} may be in an excited state and emissions are produced when the de-excitation is radiative (see the chapter by Bodewits et al. in this volume).
    The efficiency of an photoionisation is linked to the  \emph{cross section} associated with the absorbing neutral species and produced ion species. The cross section depends on the energy (or wavelength $\lambda$) of the photon absorbed, that is $E=h\nu=hc/\lambda$ where $h$ is the Planck constant, and the targetted neutral species. 
    
\begin{figure}[ht]
    \includegraphics[width=\linewidth]{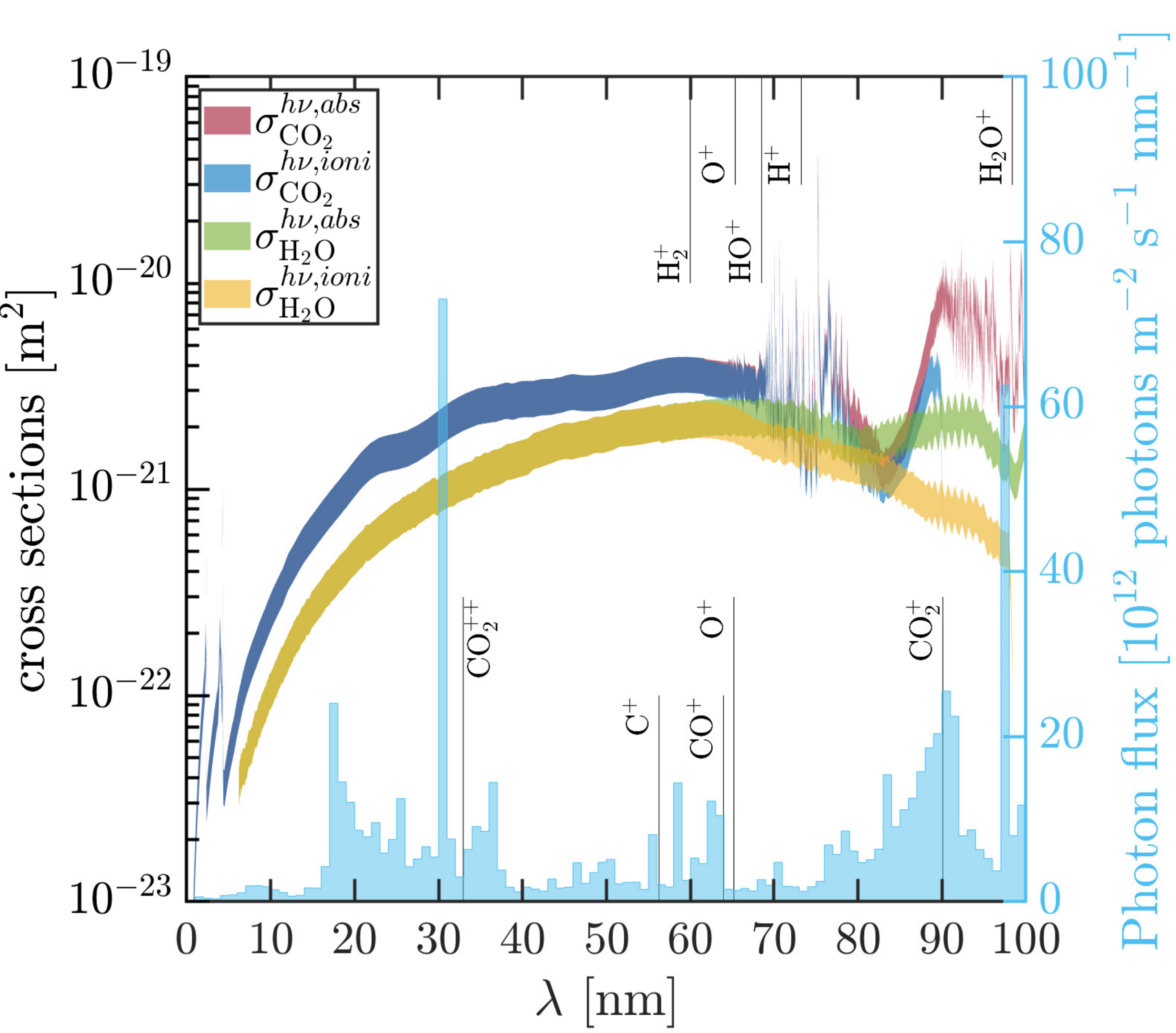}
        \caption{Left axis: Total photoabsorption ($\sigma^{h\nu,\text{abs}}$) and photoionisation ($\sigma^{h\nu,\text{ioni}}$) cross sections for H$_2$O and CO$_2$ as a function of wavelength. Associated wavelength appearances for ionised products are indicated (top for H$_2$O, bottom for CO$_2$). Right axis: EUV solar photon flux on 16 July 2016 between 0 and 100\,nm from TIMED/SEE measured at Earth, normalised at 1~au, integrated over 1\,nm-wide bins. Photoionisation and photoabsorption cross sections are from \citet{Heays2017} and references therein. Wavelength appearances are from \citet{Zavilopulo2005}. \label{fig:solar_spectrum}}
    \end{figure}

    For the most common molecules encountered around comets, photoionisation is triggered by photons of energy $E \gtrsim 12$~eV ($\lambda \lesssim 100$~nm) corresponding to the EUV range. The minimum energy can be lower in the case of alkali metals. Photoionisation depends on the distribution in energy of the incoming photons. First, the EUV spectrum significantly departs from the black body emission (see Fig.~\ref{fig:solar_spectrum}). Second, this spectrum is composed of emission `lines' in the EUV/soft X-ray range originating mainly from the chromosphere, the transition region and the solar corona; They are emitted by excited (often highly ionised) species, such as H, He, C, O, Si, Fe, etc. \citep[][see Fig.~\ref{fig:solar_spectrum}]{Lean1991}. The EUV photon energy distribution strongly depends on the solar activity and solar cycle. 
    
    In order to obtain an expression of the photoionisation source term in Eq.~\ref{eq:continuityIon}, several steps are necessary. These are detailed below. The starting point is to compute the resulting efficiency of the solar EUV flux in ionising neutrals. It is determined by the photoionisation frequency of neutral species $p$ leading to the production of ion species $s$:
    \begin{equation}
        \nu_{p,s}^{h\nu,\text{ioni}}(\vec{r})=\int_{\lambda_\text{min}}^{\lambda^\text{th}_{p,s}}\! \sigma_{p,s}^{h\nu,\text{ioni}}(\lambda)F_{h\nu}(\lambda,\vec{r})\,\mathrm{d}\lambda
        \label{eq:frequency1}
    \end{equation}
    where:
    \begin{itemize}
        \item $F_{h\nu}(\lambda,\vec{r})$~[photons~m$^{-2}$~s$^{-1}$~nm$^{-1}$] is the spectral flux of the solar radiation at the position vector $\vec{r}$ (where the origin is taken at the centre of the nucleus, $|\vec{r}| = r$ corresponds to the cometocentric distance). It is derived from Eq.~\ref{eq:attenuatedflux}, 
        \item $\sigma_{p,s}^{h\nu,\text{ioni}}(\lambda)$ [m$^{2}$] is the photoionisation cross section proportional to the probability to have the neutral species $p$ absorbing a photon of wavelength $\lambda$ and forming the ion species $s$ (see Fig.~\ref{fig:solar_spectrum}),
        \item $\lambda^\text{th}_{p,s}$  is the maximum wavelength triggering the process; it corresponds to a threshold energy typically around $12$~eV, referred to as the \emph{ionisation energy $E_{p,s}^{th}$} (or archaicly \emph{ionisation potential}). For example, water has an ionisation energy of $12.5$\,eV (99.2~nm, see Fig.~\ref{fig:solar_spectrum}); if there is dissociation and/or excitation during the ionisation, the threshold energy has a value higher than for a single, non-dissociative ionisation generating all products in the ground state,
        \item $\lambda_\text{min}$ is the minimum wavelength observable in the solar spectrum, approximately around 0.1~nm. Below this limit, the solar photon flux becomes negligible, even during flares.
    \end{itemize}
    
\begin{figure}[ht]
	    \begin{tikzpicture}[thick,scale=0.7, every node/.style={scale=0.7}]
        \clip (-7,-1) rectangle (5,4);
	    \shadedraw[opacity=0.9,inner color=black!90!white,outer color=white,draw=white] (3,0) circle (7);
	    \draw[stealth-,black!50!white,ultra thick,line width=0.1cm] (-5,0) -- (3,0)node[pos=0, anchor=east]{\LARGE Sun};
	    \draw[-stealth,white, line width=0.1cm](3,0) -- (1,2)node[pos=0.5,sloped,above]{\Large$\vec{r}$};
	    \draw[-, line width=0.1cm,dash pattern=on 4 off 1](1,2) -- (-5,2)node[pos=0,above]{\Large$l=0$} node[pos=1,anchor=south]{\Large$\phantom{_{\infty}}l=l_{\infty}$};
	    \draw[-, thick](1,2.2) -- (1,1.8);
	    \draw[-, thick](-5,2.2) -- (-5,1.8);
        \fill[black] (3,0) circle (0.5);
    
        \draw[stealth-stealth,black](-1,0)--(-1,2)node[pos=0.5,left]{\LARGE$r\sin \chi$};

	    \draw[-latex,white] (1,0) arc (180:135:2)node[pos=0.5,left]{\LARGE$\mathbf{\chi}$} ;
	\end{tikzpicture}
	\caption{Schematic of the geometry of solar photon beams near the cometary nucleus (black disk) illustrating different quantities introduced in Eqs.~\ref{eq:columndens1} and \ref{eq:columndens2}, {see text for further explanation}. 
	\label{fig:opticaldepth}}
\end{figure}
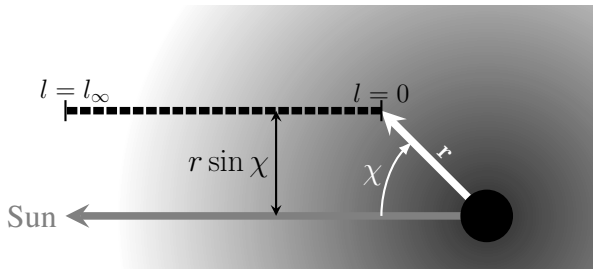

    As photons penetrate the dense coma, the amount of absorbed photons by cometary molecules becomes more and more significant. This results in the solar spectral flux being attenuated as the coma gets thicker and thicker, a process known as \emph{photoabsorption}. Photoabsorption at the position vector $\vec{r}$ depends on the total column density $N_p$ of neutral species $p$ crossed by photons from a position $\vec{r_{\infty}}$ along the line of sight (where the distance $r_{\infty}$ is large enough such that there is no significant absorption of solar photons sunward of $\vec{r_{\infty}}$ along the line of sight) to $\vec{r}$. It is defined as:
    \begin{equation}
        N_p(\vec{r})=\int_{l=0}^{l=l_{\infty}}\!  n_p(\vec{r}^\prime(l))\,\mathrm{d}l 
        \! 
        \label{eq:columndens1}
    \end{equation}
    where $l$ represents the curvilinear abscissa along the Sun-comet line of sight between the point of interest $\vec{r}$ and $\vec{r_{\infty}}$ (see Fig.~\ref{fig:opticaldepth}). The number density $n_p$ decreases fast enough for $N_p$ to be finite. For a comet, the neutral number density decreases as $1/r^2$ (see Eq.~\ref{eq:n_p}) and therefore the neutral column density is given by:
    \begin{equation}
        \boxed{N_p(r,\chi)=\dfrac{Q_p}{4\pi\, V_p}\dfrac{\chi}{r\sin \chi}}
        \label{eq:columndens2}
    \end{equation}
    where $\chi$ is the solar zenith angle between $\vec{r}$ and the direction of the Sun ($\chi = 0^\circ$ in the subsolar direction, see Fig.\ref{fig:opticaldepth}). Eq.~\ref{eq:columndens2} neither accounts for asymmetric outgassing nor adiabatic acceleration of the neutral gas. However, it gives an idea of how strong photoabsorption is. 
    
    To quantify the `thickness' of a coma regarding the penetration of photons of wavelength $\lambda$, the dimensionless `optical depth' $\tau$ is defined as:
    \begin{equation}
        \boxed{\tau(\lambda,\vec{r})= \sum_p \sigma^{h\nu,\text{abs}}_{p}(\lambda) N_p(\vec{r})}
        \label{eq:opticaldepth}
    \end{equation}
where  the subscript $p$ refers to a specific neutral species and $\sigma^{h\nu,\text{abs}}_{p}(\lambda)$ is the total photoabsorption cross section of the cometary neutral species $p$ representing the probability of a photon of wavelength $\lambda$ to be absorbed by the neutral species $p$, regardless of the process (ionisation, dissociation, excitation, see Fig.~\ref{fig:solar_spectrum}). The solar photon spectral flux $F_{h\nu}(\lambda, \vec{r})$ at the location $\vec{r}$ in the coma is then given by a simple Beer-Lambert law:
    \begin{equation}
      \boxed{F_{h\nu}(\lambda,\vec{r})=F_{h\nu, 1\,\text{au}}(\lambda)\exp[-\tau(\lambda,\vec{r})]/r_h^2}
       \label{eq:attenuatedflux}
    \end{equation}
where $F_{h\nu, 1\,\text{au}}$ is the observed spectral flux measured near Earth, for example  by TIMED/SEE  \citep{Woods2005} and normalised to 1~astronomical unit [au], and $r_h$ is the heliocentric distance given in au.
Eq.~\ref{eq:attenuatedflux} does not take into account the temporal variation of the solar EUV flux nor the phase shift correction. The latter stems from the fact that the solar EUV radiation varies with ecliptic solar longitude and that the comet is not aligned with Earth. It is then necessary to take into account the phase shift between Earth and the comet from the Sun in terms of days. For example, if the phase angle (i.e. the angle formed by Earth, the Sun, and the comet in the solar equatorial plane) is $+90^\circ$, the reader interested in the solar EUV radiation at the comet on day $d$ should look for the solar EUV radiation measured at Earth, $\sim6.5$ days before (one fourth of the solar synodic period). Applying Eq.~\ref{eq:attenuatedflux}, Eq.~\ref{eq:frequency1} becomes:
    \begin{equation}
        \boxed{\nu_{p,s}^{h\nu,\text{{ioni}}}(\vec{r})=\int_{\lambda_\text{min}}^{\lambda^\text{th}_{p,s}}\! \sigma_{p,s}^{h\nu,\text{ioni}}(\lambda)\dfrac{F_{h\nu, 1\,\text{au}}(\lambda)}{r_h^2}\exp[-\tau(\lambda,\vec{r})]\,\mathrm{d}\lambda}
    \label{eq:frequency2}
    \end{equation}
    
    For $\tau(\lambda,\vec{r}) \lesssim 0.1$, the coma is optically thin at $\vec{r}$ for photons of wavelength $\lambda$. Photoabsorption is thus negligible and the photoionisation frequency is constant (independent of $\vec{r}$), reducing to:
    \begin{equation}
        \nu_{p,s}^{h\nu,\text{ioni}}=\int_{\lambda_\text{min}}^{\lambda^{th}_{p,s}}\! \sigma_{p,s}^{h\nu,\text{ioni}}(\lambda)\dfrac{F_{h\nu, 1\,\text{au}}(\lambda)}{r_h^2}\,\mathrm{d}\lambda
        \label{eq:frequencycst}
    \end{equation}
    
For $\tau(\lambda,\vec{r}) \gtrsim 0.1-1$, the coma is optically thick at $\vec{r}$ for photons of wavelength $\lambda$. In this case, photoabsorption needs to be taken into account as the solar flux is attenuated at $\vec{r}$. The coma was optically thick in the vicinity of the nucleus but not necessarily at the spacecraft location during the flyby of 1P by Giotto and during the escort of 67P by Rosetta near perihelion, though in both cases the spacecraft was far enough from the nucleus such that the ionospheric density at the location of the spacecraft was not affected by the photoabsorption occurring closer to the nucleus \citep{Heritier2018, Beth2019}. We can now calculate the production rate, through photoionisation, for the ion species $s$ contributing to the source term in  Eq.~\ref{eq:continuityIon}:
    \begin{equation}
        S_{s}^{h\nu}(\vec{r}) = \sum_{p} S^{h\nu}_{p,s}(\vec{r})
        \label{eq:photoionratetotal}
    \end{equation}
    where $S^{h\nu}_{p,s}$ is the production of an ion $s$ from photoionisation of the neutral species $p$ given by:
    \begin{equation}
        \boxed{S_{p,s}^{h\nu}(\vec{r}) = \nu_{p,s}^{h\nu,\text{{ioni}}}(\vec{r}) \ n_p(r)}
        \label{eq:photoionrate}
    \end{equation}
    The total photoionisation rate or photoelectron production rate (as only one electron is usually produced per ionisation process) is then simply given by:
    \begin{equation}
        S_{e^-}^{h\nu}(\vec{r}) = \sum_{p,s} S_{p,s}(\vec{r})
        \label{eq:photoeratetotal}
    \end{equation}
    
What are the typical behaviour and spatial distribution of this photoelectron production rate in the coma? For $\tau(\lambda,\vec{r}) \lt 0.1$, $S_{e^-}^{h\nu}$ is proportional to $n_p$, that is, to ${1}/{r^2}$. For $\tau \gtrsim 0.1-1$, $S_{e^-}^{h\nu}$ has additionally a spatial dependence associated with the optical depth. One might be interested in the location where most of the EUV radiation is absorbed within the coma and deposits its energy. This maximum absorption, which corresponds to a maximum in the photoelectron production rate is taking place as a result of the neutral number density increasing with decreasing $r$, while the solar flux decreases. For a monochromatic radiation, single-species coma, and for a fixed $\chi$, the photoelectron production rate induced by the absorption of photons at wavelength $\lambda$ is maximum where \citep{Beth2019}:
    \begin{equation}
    \boxed{\tau(\lambda,\vec{r})=2}
    \label{eq:tau2}
    \end{equation}
    The value of $2$ at the maximum production  differs from what is usually encountered at planetary objects (including moons with a dense atmosphere such as Titan). At these bodies, the atmosphere can be assumed to be in hydrostatic equilibrium and hence varies exponentially with decreasing altitude: The maximum of absorption and of production rate thus occurs at $\tau=1$ \citep[e.g.][]{Schunk2009} instead of $2$.
    
    From Eq.~\ref{eq:columndens2} and \ref{eq:opticaldepth}, equality~\ref{eq:tau2} is reached at a cometocentric distance corresponding to:
    \begin{equation}
    r^\text{max,abs}_{p,s}(\lambda,\chi)=\dfrac{\sigma_p^{h\nu,\text{abs}}(\lambda)Q_p}{8\pi V_p}\dfrac{\chi}{\sin \chi}
    \end{equation}
    \begin{figure}[!ht]
    \includegraphics[width=\linewidth,trim=0cm 40px 40px 0cm,clip]{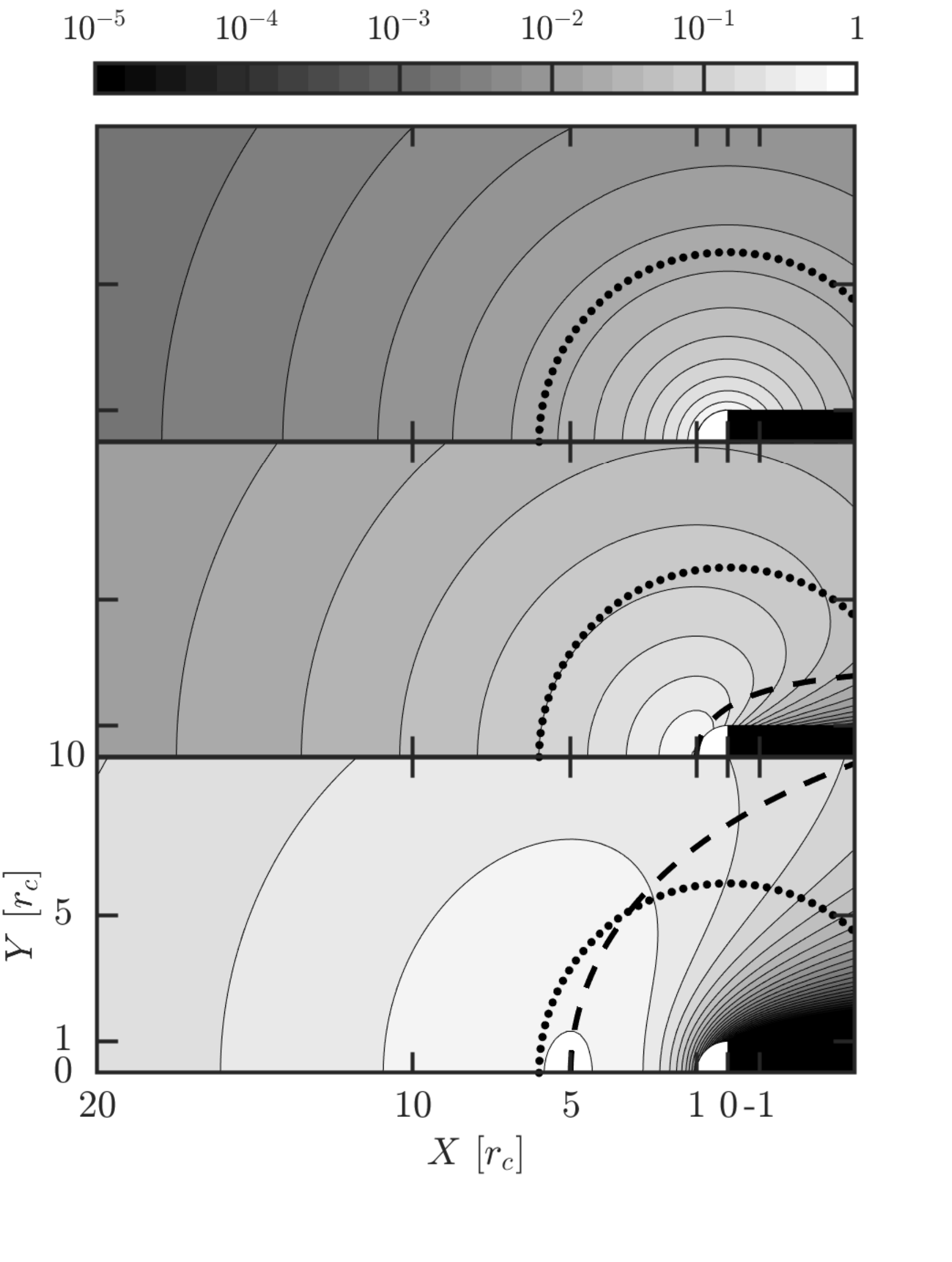}
	\caption{Normalised photoelectron production rate for a monochromatic radiation {in a pure water coma} for different cometary activities using Eq.~\ref{eq:photoeratetotal}. The Sun-comet axis is along $X$, with origin at the nucleus' centre, whereas the terminator plane is along $Y$. Instead of setting the outgassing rate, we set three different optical depths on the surface at the subsolar point ($X = r_c$ at $\chi = 0$): 0.01 (top panel, low activity), 2 (centre panel, intermediate activity), 10 (bottom panel, high activity). The dashed line refers to the optical depth level $\tau=2$, along which the neutral column density is constant. The neutral number density described by Eq.~\ref{eq:n_p} is constant along the dotted lines (spherical symmetry). Cometocentric distances  in the $X-Y$ plane  are expressed in terms of nucleus' radius, $r_c$.     \label{fig:peak_photoelectron}}
    \end{figure}
    As shown in Fig.\,\ref{fig:peak_photoelectron}, the spatial variation in the Sun-comet plane of the photoelectron production rate for a monochromatic solar radiation can be very asymmetric as the comet becomes more active (increasing outgassing rates lead to an increased optical depth). For an optically thin coma ($\tau\ll2$), the photoelectron production only depends on the cometocentric distance: Photoelectrons are produced uniformly, only as a function of the local neutral density. However, for $\tau \geq 0.1-1$, this changes. The neutral column density becomes high enough to significantly attenuate the photon flux, reducing the ionisation efficiency. Photoelectrons are thus produced in most numbers along the Sun-comet axis. As the outgassing rate increases, this maximum is located at a gradually increasing cometocentric distance upstream of the nucleus. 
    
    The photoelectron production rate is not constant along the lines of constant $\tau$ as, for a given value of $\tau$ and $\lambda$, both the neutral number density and the column density of absorption depend on $\vec{r}$. This is of importance as, when considering all wavelengths, the total photoelectron production rate will result from the sum of different production profiles through the coma. Each profile is associated with a different energy distribution of the photoelectrons which varies with cometocentric distance. The energy of the newborn photoelectron is given by: $E_{e^-_{h\nu}} = E_{h\nu} - E^{th}_{p,s}$ where $E_{h\nu}$ is the solar photon energy.
    
    Above 20~eV, $\sigma_{\text{H}_2\text{O}}^{h\nu,\text{abs}}$ decreases with energy, meaning that the maximum of absorption occurs closer to the nucleus for photons at 100~eV than those at 20~eV. Where photons at 100~eV are efficiently absorbed and ionise the neutral species, photons at 20~eV cannot do so as they have been absorbed uptstream due to their larger photoabsorption cross sections. Therefore, the energy distribution of photoelectrons is depleted around 20~eV when the coma starts to be optically thick.  Nevertheless, the 100~eV photons produce highly energetic electrons, which in turn impact neutral molecules to produce secondary electrons. This accounts for the main ionisation source close to the nucleus for highly active comets, above $Q=10^{29}$--$10^{30}$~s$^{-1}$ \citep{Bhardwaj2003}. 

    For more information regarding photoionisation and photoabsorption, the reader is invited to read the historic work of \citet{Chapman1931} initially developed for planetary atmospheres under hydrostatic equilibrium (exponentially decreasing with height) and monochromatic radiation. The theoretical application to comets hosting an expanding coma is presented in full by \citet{Beth2019} .

\subsubsection{Electron-impact ionisation\label{section:2:2:2}}
   
\begin{figure}[!ht]
    \includegraphics[width=\linewidth]{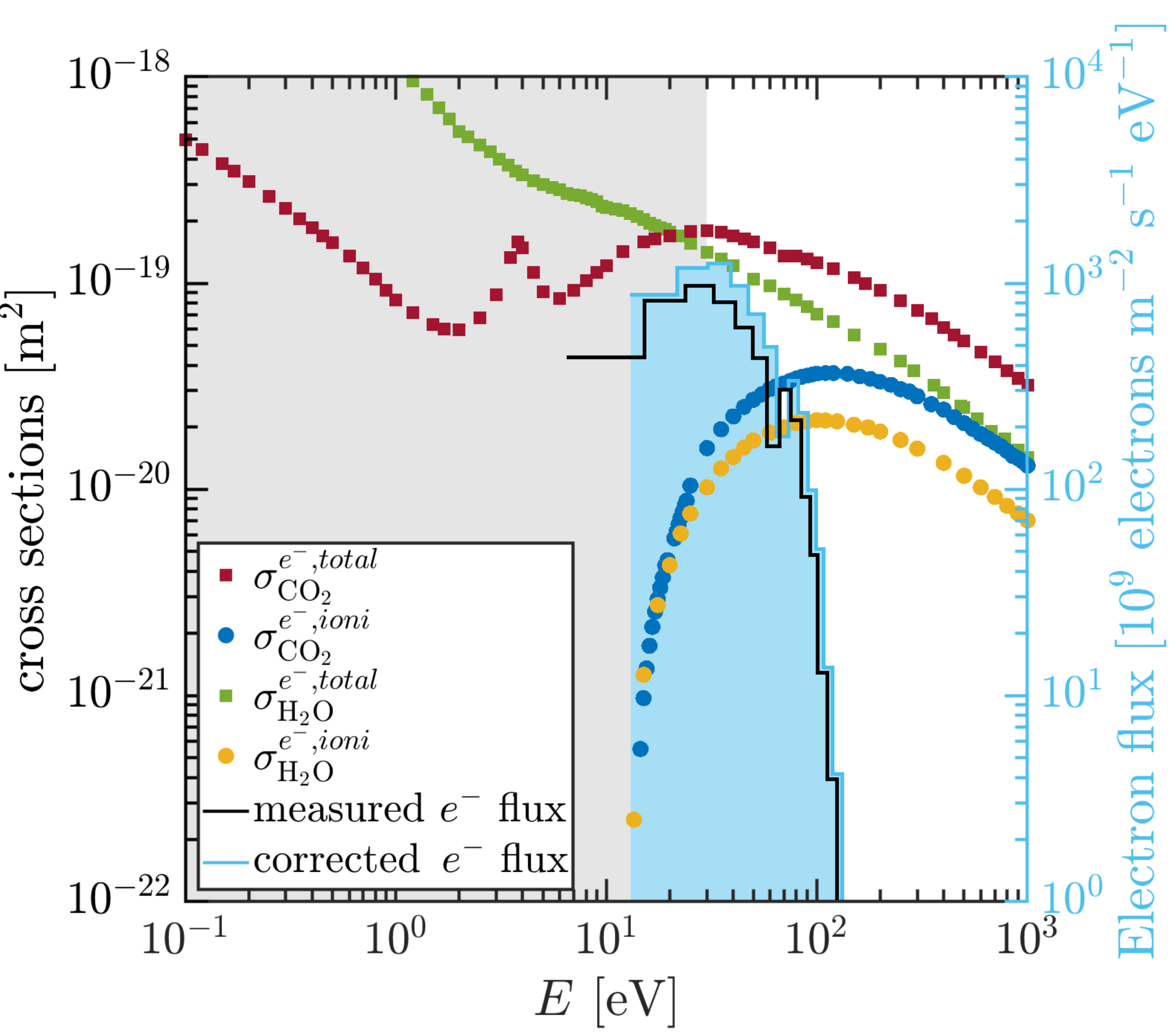}
        \caption{(Left axis) Total and ionisation cross sections resulting from the impact of electrons on H$_2$O and CO$_2$. (Right axis) Raw and corrected (from the spacecraft potential) electron flux by the electron spectrometer onboard Rosetta on 14 January 2015 00:06:59  \citep{Stephenson2021a}. The grey region represents values of $-qV_\text{SC}$ during Rosetta's mission (see Section~\ref{section:3:1:1} for fuller explanation). It represents the minimal energy required by an electron to reach the spacecraft. Cross sections are from \citet{Itakawa2002} for CO$_2$ and \citet{Itakawa2005} for H$_2$O.
        \label{fig:electron_spectrum}}
    \end{figure}

Electron-impact ionisation (sometimes refers as electron ionisation) is another critical process in the birth of a cometary ionosphere \citep[e.g.][]{Heritier2018}. Like photons, free electrons, with an energy above the ionisation energy $E^{th}_{p,s}$ ($\approx12$~eV) efficiently tear off electrons attached to an atom/molecule $A$ following: 
\begin{align}
    A + e^-_\text{prim} &\longrightarrow {A}^{m+(*)} + e^-_\text{prim} + (m-1) e^-_\text{sec}\nonumber\\
    &\longrightarrow {B}^{m+(*)} + C^{(*)}+ e^-_\text{prim} + (m-1) e^-_\text{sec}
    \label{eq:electronIonisationReaction}
\end{align}
where  subscripts `$\text{prim}$' (resp. `$\text{sec}$') refer to the primary impacting electron (resp. the secondary electron,  freed during the ionisation). 
The impacting electron can be a photoelectron, a solar wind electron, or a secondary electron (see \ref{section:3:1:3}). It loses energy through the ionisation: It loses the ionisation energy to the neutral target{, and eventually more whether one or more products are left in excited states and/or if $A$ is dissociated,} plus a fraction of it that is passed on to the secondary electron as kinetic energy. The  corresponding electron-impact ionisation frequency is given by:
\begin{equation}
    \boxed{\nu^{e^-,\text{ioni}}_{p,s}(\vec{r})=\int_{E^{\text{th}}_{p,s}}^{+\infty}\! \sigma^{e^-,\text{ioni}}_{p,s}(E)F_{e^-}(\vec{r},E)\, \mathrm{d}E}
\end{equation}
where $F_{e^-}(E)$ is the electron flux (also referred sometimes as differential flux) given in (electrons)\,m$^{-2}$\,s$^{-1}$\,eV$^{-1}$ and $\sigma^{e^-,\text{ioni}}_{p,s}$ is the  electron-impact cross section of the neutral species $p$ leading to the ion species $s$ (see Fig.~\ref{fig:electron_spectrum}). This formula is very similar to Eq.~\ref{eq:frequency1} except that ($i$) solar spectral flux is replaced by electron flux and ($ii$) the problem is defined in terms of energy instead of wavelength. 

We can now calculate the electron-impact production rate for the ion species $s$ contributing to the source term in  Eq.~\ref{eq:continuityIon}:
\begin{equation}
        \boxed{S_{s}^{e^-}(\vec{r}) = \sum_{p} S^{e^-}_{p,s}(\vec{r})}
        \label{eq:elimpratetotal}
    \end{equation}
    where $S^{e^-}_{p,s}$ is the electron-impact ionisation rate 
of cometary neutral $p$ leading to the production of a new cometary ion $s$ and given by:
\begin{equation}
    \boxed{S_{p,s}^{e^-}(\vec{r}) = \nu_{p,s}^{e^-,\text{ioni}}(\vec{r}) \, n_p(r)}
        \label{eq:elecrate}
\end{equation}

Although it is  an endothermic reaction (the ionisation energy is lost), the repartition of the remaining energy between the primary and secondary electrons varies. In extreme cases, when the incident electron is very energetic, above 200--300\,eV, two electrons can be kicked out through the ionisation process. This can also occur with energetic photons of the same energy. Such high-energy primaries may lead to a process known as the \emph{Auger effect} \citep{Auger1923}, caused by the ejection of an electron from the innermost atomic or molecular shells (K-shell)  and the subsequent filling of that gap with a higher-shell electron, itself leading to photoemission and/or emission of a so-called \emph{Auger electron} \citep{Fox2008}.

It is important to note that the produced ion $A^+$ in Reaction~\ref{eq:electronIonisationReaction} can be in an excited state $A^{+*}$, in a similar way to photoionisation: its radiative de-excitation towards a lower excited state (ultimately ground state) leads to the emission of a photon in the FUV-Visible range, a process at play most famously in cometary ion tails, but also in aurora-like structures throughout the solar system and  recently discovered at comets \citep[see][and the chapter of Bodewits et al. in this volume for more detail]{Galand2020}.
\subsubsection{Charge-exchange and ionisation by solar wind ions\label{section:2:2:3}}

In addition to solar photons and energetic electrons, energetic ions emitted by the Sun, such as protons (H$^+$) and alpha particles (He$^{2+}$), may ionise the neutral species surrounding the comet. This generally occurs either through \citep{SimonWedlund2019c,SimonWedlund2020}:
\begin{itemize}
    \item direct impact ionisation, as for energetic electrons; we refer to it as \emph{Solar Wind Ionisation} (SWI),
     \item \emph{charge-exchange processes} (with no \emph{net} production of charge), referred to as SWCX in the following.
\end{itemize}

As solar wind energetic ions pass through the coma, they may strip electrons from the neutral atoms and molecules they encounter, thereby lowering their charge state. The first evidence of such a phenomenon was reported by \citet{Lisse1996}, who observed at Comet Hyakutake C/1996 B2 a strong sunward emission of the coma in the X-Ray and EUV ranges. Quickly after, \citet{Cravens1997} identified the mechanism behind these emissions.  Solar wind plasma is not only formed of protons and alphas, but also of heavier ions amounting to a small proportion ($<0.1\%$) of the total solar wind density. These ions, highly charged because of their origin in the deep solar corona, are for example O$^{m+}$, C$^{m+}$, and N$^{m+}$ (noted $X^{m+}$ in the following), where $m\geq4$ is their charge number \citep{Steiger2000}. Such ion species have large and very different first ionisation energies $E_{i}$, that is, the minimum energy required to eject an electron from an isolated atom/molecule in the gaseous phase \citep{Muller1994}. As they stream through the coma, multiply-charged heavy ions undergo collisions with cometary molecules, such as water:
\begin{eqnarray*}
    X^{m+} + \text{H$_2$O} &\longrightarrow& X^{(m-1)+*} + \text{HO} + \text{H}^+
    \end{eqnarray*}
a dissociative ionisation process followed by radiative de-excitation:
    \begin{eqnarray*}
    X^{(m-1)+*} &\longrightarrow& X^{(m-1)+} + h\nu
\end{eqnarray*}
The emitted photons have typical wavelengths below about $12$\,nm ($\sim100$\,eV), placing them in the soft X-ray (roughly 0.1--10\,nm) and EUV ranges \citep[][]{Cravens2002,Krasnopolsky2004}. SWCX is responsible for global X-ray and EUV emissions not only at comets, but also throughout the Universe \citep{Dennerl2010}. Starting with comets, it has only recently become a rich subject of investigation and neutral gas diagnostic in astrophysics and space physics, from supernova remnants \citep{Lallement2012} to planetary magnetospheres and exospheres \citep{Bhardwaj2007}, with applications in fusion plasmas \citep{Dennerl2012,Sembay2012}. Moreover, laboratory measurements of charge-transfer processes have also experienced an unprecedented `boom' over the last two decades \citep[see][and the chapter by \emph{Bodewits et al.} in this volume]{Wargelin2008,SimonWedlund2019a}. 

SWCX does not only concern heavy solar wind ions: Protons and alphas constituting the bulk of the solar wind are also subject to it \citep{SimonWedlund2019b}, although they do not produce such energetic photons. In this case, the main effect of SWCX is to neutralise the solar wind (creating H and He energetic neutral atoms or ENAs, from incoming protons alphas, respectively) and implant heavier cometary ions into the solar wind flow so that, for example 
for alpha particles:
\begin{equation}
\begin{array}{lllllll}
    \text{He}^{2+}_\text{fast}\ \xrightarrow{\text{H$_2$O}}\text{He}^+_\text{fast}\ (+ \text{H$_2$O}^+) \xrightarrow{\text{H$_2$O}} \text{He}_\text{fast}\ (+ \text{H$_2$O}^+)
\end{array}
\end{equation}
This in turn partakes of the so-called \emph{mass/momentum-loading} of the solar wind as the newly-born $\text{H$_2$O}^+$ molecules are accelerated by the convective electric field (see Section~\ref{section:4:2} and the chapter by Götz et al. in this volume).

For a solar wind ion species $sw$ (e.g. H$^+$) charge-exchanging with a heavy neutral species $p$ (e.g. H$_2$O) thereby creating heavy ion species $s$ (e.g. H$_2$O$^+$), the charge-exchange frequency $\nu_{p,s}^{sw,\text{CX}}$ can be expressed as:
\begin{equation}
    \boxed{\nu_{p,s}^{sw,\text{CX}}(\vec{r}) = \int_{E_{p,s}^{\text{th,CX},sw}}^{+\infty} \sigma_{p,s}^{sw,\text{CX}}(E)\,F_{sw}(\vec{r},E)\,\mathrm{d}E}
\end{equation}
where $F_{sw}$ [m$^{-2}$\,s$^{-1}$\,eV$^{-1}$] is the attenuated flux of solar wind ions $sw$ at the position $\vec{r}$, $\sigma_{p,s}^{sw,\text{CX}}$ the corresponding charge-exchange cross section, and $E^{\text{th,CX},sw}_{p,s}$ the threshold energy of the reaction involving solar wind ion species $sw$ reacting with the neutral species $p$ to form ion species $s$. SWCX is a cumulative process, hence the flux of solar wind ions is progressively attenuated the deeper they penetrate in the coma, whereas keeping most of its initial energy (at least to a first approximation). Assuming straight-line trajectories with no energy change of the solar wind along the Sun-comet axis even after charge-exchange, this translates to \citep{SimonWedlund2016,SimonWedlund2019b}:
\begin{equation*}
    F_{sw}(\vec{r},E) = F_{sw}^{\infty}(E) \exp\left(-\sum_{p,s}\sigma_{p,s}^{sw,\text{CX}}(E)\, N_p(r,\chi)\right)
\end{equation*}
where $F_{sw}^{\infty}$ is the upstream `undisturbed' solar wind flux, $N_p(r,\chi)$ the neutral column density at distance $r$, assuming cylindrical symmetry along the Sun-comet axis of the neutral coma, and solar zenith angle $\chi$, defined in Eq.~\ref{eq:columndens2}. 

Similarly to photoionisation and electron ionisation processes, and summing through all solar wind ion species $sw$, the source term for charge-exchange reactions leading to the production of a new cometary ion $s$:
\begin{equation}
    \boxed{S_{p,s}^\text{CX}(\vec{r}) = \sum_{sw}{\nu_{p,s}^{sw,\text{CX}}(\vec{r}) \, n_p(r)}}
        \label{eq:chargeexchangerate}
\end{equation}
Compared to photoionisation and electron ionisation, SWCX is expected to be a minor source of new plasma in the inner coma. That said, large-scale 3D numerical simulations of the cometary plasma environment have shown that it is one of the most important processes responsible for the formation of boundaries upstream of the inner ionosphere \citep[][and the chapter of Götz et al. in this volume]{SimonWedlund2017}.

Because charge-exchange processes have large interaction cross sections (and thus efficiencies), SWCX production rates usually dominate {those of SWI}. This is especially true outside transient solar wind phenomena such as Coronal Mass Ejections (CME) or Co-rotating Interaction Regions (CIR), which carry away particles at much larger bulk speeds (up to $\sim1-3\times 10^{3}$~km\,s$^{-1}$) than the nominal `slow' solar wind ($\sim400$~km\,s$^{-1}$ or $\sim833$~eV\,u$^{-1}$). The explanation is straightforward: SWCX cross sections peak below $1$~keV\,u$^{-1}$ whereas proton and helium ionisation cross sections in H$_2$O and CO$_2$ have their maximum above that limit \citep{SimonWedlund2019a}.

\subsubsection{Which ionisation sources matter at comets?\label{section:2:2:4}}

{The relative importance of each separate source of cometary ions, namely photoionisation (see Section~\ref{section:2:2:1}), electron-impact ionisation (see Section~\ref{section:2:2:2}}), and solar wind charge-exchange (see Section~\ref{section:2:2:3}),  depends on outgassing activity and cometocentric distance \citep[][]{Heritier2018,SimonWedlund2019c,SimonWedlund2020}. {Whereas solar wind charge exchange plays an important role at large cometocentric distances, typically far upstream of the comet, the production of ions in the inner ionosphere is heavily driven by electron-impact ionisation and photoionisation. For example, at comet~67P, Rosetta made it possible to monitor the ionisation frequencies of each of these sources individually from low to high outgassing rates. Outside of the so-called \emph{solar wind ion cavity} (see Section~\ref{section:4:3}, the chapter by Götz et al. in this volume, and Fig.~\ref{fig:ionfluxes_sw_cometary_ions}), the electron-impact ionisation frequency was on average $5$--$10$ times larger than that of photoionisation, the latter a factor $10$ (respectively, $100$) or so larger than solar wind charge-exchange (ionisation) frequencies, except for exceptional solar transient events when charge exchange could briefly rival the other two main ionisation sources. Inside the \emph{solar wind ion cavity}, photoionisation frequencies at the cometocentric distances probed by the spacecraft steadily increased with decreasing heliocentric distance, to progressively dominate over electron-impact ionisation frequencies by the time the comet reached perihelion \citep{Heritier2018}.}

\subsection{Centre of chemical reactions\label{section:2:3}} 
Beside the net production of charge in the coma presented in Section~\ref{section:2:2}, there are additional chemical processes which contribute to the source and loss terms, $S_s$ and $L_s$, in Eq.~\ref{eq:continuityIon} (see Section~\ref{section:2:1}). Firstly, there are ion-neutral collisions, which heavily influence the cometary ion composition. Through such processes, there is no net change in charge in the coma, but there is a change in cometary ion species: it is a loss for the reacting ion and a production for the new ion species formed. Secondly, there are electron-ion dissociative recombination reactions, through which there is a net loss of charge. These two types of processes are presented hereafter.

At low outgassing activity (at large heliocentric distances), a coma can be described as a quasi-collisionless medium. In the case of comet~67P, farther than 3~au, the outgassing rate was lower than $10^{26}$~s$^{-1}$, which corresponds to two  1.5-l bottles of water released to space each second. Hence, the probability of collisions between ions and neutrals was low. As the comet gets closer to the Sun and the outgassing activity grows, ion-neutral collisions occur more often and chemistry becomes increasingly effective within the inner coma, generating a rich zoo of ion species \citep{Beth2020}.

What are the different key chemical processes contributing to the sources $S_s$ and losses $L_s$ in Eq.~\ref{eq:continuityIon}? These are:
\begin{itemize}[leftmargin=0mm]
    \item[] {$\bullet$  \bf charge exchange}: {through the collision of a cometary ion $A^+$ with an atom/molecule $B$, $A^+$ captures an electron from $B$}, such that 
    \begin{equation*}
        A^+ +B\longrightarrow A + B^+
    \end{equation*}
    For example
        \begin{equation*}
        \text{O}^+ +\text{H}_2\text{O}\longrightarrow \text{O} +\text{H}_2\text{O}^+
    \end{equation*}
    This is similar to an oxidation-reduction reaction but here in the context of gases: $B$ is oxidised and $A$ is reduced. Mathematically, this translates into the ion source/loss rates, as follows:
    \begin{equation}
        S_{B^+}=L_{A^+}=k_{A^+,B}(T)\,n_{A^+}\,n_{B}
        \label{eq:27}
    \end{equation}
    where $k(T)$ [m$^{3}$\,s$^{-1}$] is the reaction rate coefficient, which depends on the temperature of the gas $T$, $n_{B}$ [m$^{-3}$] is the local number density of neutral species $B$, and $n_{A}^+$ [m$^{-3}$] is that of ion species $A^+$.
    This process is a chemical source for $B^+$ ($S_{B^+}$), while it is a chemical loss for $A^+$ ($L_{A^+}$). For single electron capture, there is no net production of ions since we start with one ion and end up with another. 
    \item[] {$\bullet$ \bf proton transfer}: {through the collision of a cometary ion $AH^+$ with an atom/molecule $B$, a proton is transferred from $AH^+$ to $B$} such that:
    \begin{equation*}
        AH^+ +B\longrightarrow A + BH^+
    \end{equation*}
    For example
        \begin{equation*}
        \text{H}_3\text{O}^+ +\text{NH}_3\longrightarrow \text{H}_2\text{O} +\text{NH}_4^+
    \end{equation*}
    Mathematically, this translates into the ion source/loss rates, as follows:
    \begin{equation}
        S_{BH^+}=L_{AH^+}=k_{AH^+,B}(T)\,n_{AH^+}\,n_{B}
        \label{eq:28}
    \end{equation}
    This process is a chemical source for $BH^+$ ($S_{BH^+}$), while it is a chemical loss for $AH^+$ ($L_{AH^+}$). Compared with charge transfer, this reaction produces `unique' ions. For example, whereas H$_2$O$^+$ can be produced both through ionisation of H$_2$O and charge transfer, NH$_4^+$ can only be produced through protonation of NH$_3$. Thus, the existence of BH$^+$ (e.g. NH$_4^+$) does not imply that of BH (e.g. NH$_4$). This process is governed by the proton affinity (PA) of both neutral species $A$ and $B$. This reaction occurs only if $B$ has a higher PA than $A$. In the context of cometary ionospheres, this is of importance and pointed out initially by \citet{Aikin1974}, the first to apply ion-neutral gas chemistry at comets. Through ionisation, the main ion produced is H$_2$O$^+$ (which may be loosely seen as protonated hydroxyl radical HO-H$^{+}$). Any neutral with a higher proton affinity than HO (PA(HO)$=6.15$\,eV) may steal the proton from H$_2$O$^+$ to be in turn protonated. Amongst cometary neutrals, candidates include: H$_2$O (7.16~eV),  H$_2$S (7.30~eV), H$_2$CO (7.39~eV), HCN/HNC (7.39/8.00~eV), and NH$_3$ (8.84\,eV) \citep{Hunter1998}. Ammonia, the latter, is thus at the top of the `proton food' chain. In the coma, once protonated ammonia NH$_4^+$ (ammonium ion) is formed, it is only destroyed through recombination with electrons or transport, and therefore it should persist for a long time {(as such, it is referred to as terminal ion)}. The mass resolution of the ROSINA/DFMS mass spectrometer onboard Rosetta was high enough to unambiguously identify it for the first time at a comet, differentiating it from H$_2$O$^{+}$ \citep[][see also Fig.~\ref{fig:ion_zoo}]{Fuselier2016, Beth2016}.
\end{itemize}
Ion-neutral chemistry affects ion composition but, as there is no net production of charge, the total ion and electron number densities remain unchanged. Only one known process affects the electron density, reduces the net amount of charges, and efficiently removes ions and electrons from a plasma. It is the so-called:
\begin{itemize}[leftmargin=0mm]
    \item[] {$\bullet$ \bf electron-ion dissociative recombination (DR)}: a molecular ion recombines with an electron, which yields fragmented neutral species.
    For example
        \begin{equation*}
        \text{H}_3\text{O}^+ + e^- \longrightarrow \text{OH} +\text{H}_2
    \end{equation*}
    ``Dissociative'' refers to the breaking of the molecular ion into neutral fragments in order to dissipate the exceeding energy. The temperature-dependent loss rate for the DR reaction is given by:
    \begin{equation}
        L_{\text{DR},s}(T_e) = \alpha_s(T_e)\,n_s\,n_e
        \label{eq:29}
    \end{equation}
    where $\alpha_s$ stands for the DR reaction rate coefficient (a function of the electron temperature $T_e$), $n_s$ for the number density of ion species $s$, and $n_e$ for the electron number density ($n_e\approx\sum_s n_s$). 
\end{itemize}
Some words of caution regarding the DR reaction. Firstly, it depends on (i) the electron temperature (the colder the electrons are, the more efficient the DR is), and (ii) the molecular ion species \citep{Heritier2017b}. In fact, $\alpha_s(T_e)$ exhibits very different $T_e$-dependency. $\alpha_s(T_e)$ is often assumed to behave like $\alpha_s(T_e)\propto 1/\sqrt{T_e}$ as theoretically derived \citep{Wigner1948}, an approximation that works well with many ion species \citep{FlorescuMitchell2006}. In reality, this dependency might only hold for a specific electron temperature range whether the molecular ions are diatomic or polyatomic \citep{McGowan1984}. For instance, $\alpha_{\text{H}_2\text{O}^+}=4.3\times(300/T_e)^{0.74}\times10^{-13}$~[m$^{3}$\,s$^{-1}$] \citep{Rosen2000} for $T_e<1000$~K and
$\alpha_{\text{CO}_2^+}=4.2\times(300/T_e)^{0.75}\times10^{-13}$ [m$^{3}$\,s$^{-1}$] \citep{Viggiano2005} without constrains on $T_e$, where $T_e$ is expressed in K. Nevertheless, these recombination rates are of the same order of magnitude for similar $T_e\lesssim10^3$~K so that assuming initially a common value is reasonable. {When the range of $T_e$ encountered at comets is considered, $\alpha_s$ may span over several orders of magnitude (see Section~\ref{section:2:4:1}).} 

Secondly, DR introduces non-linearity into the continuity equation. For ion-neutral chemistry in the inner coma, one may safely assume that the neutral density profile $n_p$ is decreasing as $1/r^2$ (see Eq.~\ref{eq:n_p}). DR introduces non-linear terms, such as $n_s^2$ and $n_sn_e$. Last but not least, $\alpha_s$ is often greater than the coefficient rate $k$ associated with ion-neutral chemistry. Despite ions being much less abundant than neutrals in the inner part of the coma by a factor $\sim \nu^{\text{ioni}}\,r/V_n$ (see Eqs.~\ref{eq:n_p} and \ref{eq:multi_ni}), $\nu^{h\nu, \text{ioni}}$ (a main contributor of ionisation for $r_h<2$~au) increases as the comet gets closer to the Sun and the gap reduces. In addition, as the activity rises, collisions occur more frequently, cooling the electrons: The electron temperature decreases, the loss through DR increases, and potentially overcoming the loss through transport (see Section \ref{section:2:4:1}). This appears to be the case for outgassing rates $\gtrsim10^{28}$ s$^{-1}$ \citep{Heritier2018,Beth2019}.

\subsection{The Ionosphere: Putting it all together\label{section:2:4}}

\subsubsection{Total ionospheric density\label{section:2:4:1}} 

As shown in Section~\ref{section:2:1}, one has to solve a complex system made of $s'$ continuity equations, one for each ion species $s$, in order to model the cometary ion composition under steady-state conditions. This can be summarised by combining Eqs.~\ref{eq:27}, \ref{eq:28}, and \ref{eq:29} for the chemical reactions, as follows:

\begin{widetext}
\begin{equation}
\boxed{%
\def\arraystretch{1.5}
\begin{array}{lclcclc}
\hspace{-0.5em}\ldelim\{{3}{*}[\textit{s'} eqs.]&&\overbrace{\nabla \cdot (n_{A^+} \vec{V}_{A^+}\vphantom{\vec{V}^i})}^{\text{transport}}&=&\overbrace{\nu_{A^+}^\text{ioni} n_n\vphantom{\vec{V}^i}}^{\text{ionisation}}&\overbrace{-\xcancel{k_{A^+,B} n_{A^+}n_{B}}\pm\ldots\vphantom{\vec{V}^i}}^{\text{ion-neutral chemistry}}&\overbrace{-\alpha_{A^+}n_in_{A^+}\vphantom{\vec{V}^i}}^{\text{electron-ion recombination}}\\
&+&\nabla \cdot (n_{B^+}\vec{V}_{B^+})&=&\nu_{B^+}^\text{ioni} n_n&+\xcancel{k_{A^+,B} n_{A^+}n_{B}}\pm\ldots&-\alpha_{B^+}n_in_{B^+}\\
&&&\vdots&&&\\
\hline
\text{sum}&&\nabla \cdot (n_{i} \vec{V}_{i})&=&\nu_{i}^\text{ioni} n_n&+0&-\left(\sum_{s=1}^{s'} \alpha_{s}n_{s}\right)n_i
\end{array}
}
\label{eq:continuityIonosphere}
\end{equation}
\end{widetext}
where 
\begin{align}
\vec{V}_i&=\sum_s n_s\vec{V}_s/n_i\\
\nu^\text{ioni}_s &=  \sum_p{n_p\left(\nu^{h\nu,\text{ioni}}_{p,s} + \nu^{e-,\text{ioni}}_{p,s}\right)/n_n} \label{eq:ionFreq}\\
\nu_i^\text{ioni} &=  \sum_s \nu^\text{ioni}_s \label{eq:totalIonFreq}
\end{align}
$n_n$ stands for the total neutral density, $n_i$ for the total ion number density (equal to the electron density, assuming no doubly-charged ions), 
$\vec{V}_i$ for the mean ion velocity, $\nu_i^\text{ioni}$ for a mean/effective ionisation rate (via photoionisation, see Section\,\ref{section:2:2:1}, and electron impact, see Section\,\ref{section:2:2:2}), and $\alpha_s$ for the electron-ion dissociative recombination rate coefficient associated with ion species $s$. 
Indices $p$ and $s$ refer to a specific neutral and a specific ion species, respectively. Note that in Eq.~\ref{eq:ionFreq}, only photoionisation and electron-impact ionisation terms are included, as solar wind ionisation and charge-exchange terms are usually negligible in the inner coma (see Sections~\ref{section:2:2}, \ref{section:2:2:3}, and \ref{section:2:2:4}). 

Once all ion continuity equations are summed, the ion-neutral chemistry terms (second set of terms on the right-hand side of Eq.~\ref{eq:continuityIonosphere}) cancel out between lost and produced ion species. From Eq.~\ref{eq:continuityIonosphere}, the continuity equation ruling the total ion number density is weakly depending on the ion composition through the last term on the right-hand side (electron-ion recombination) as the kinetic rates are different between ion species (see Section~\ref{section:2:3}).

Let us look more closely at the transport term (left-hand side of  Eq.~\ref{eq:continuityIonosphere}). Its contribution in non-cometary environments, such as a planetary ionosphere in the plane-parallel approximation \citep{Schunk2009}, is limited: Vertical transport between atmospheric layers remains small and is dominated by eddy or molecular diffusion. 
In contrast, at comets, the neutrals, and therefore the newly-born ions, are moving at speeds $V_n$ ranging from 300~m\,s$^{-1}$ close to the surface, up to 1~km\,s$^{-1}$ farther away, assuming adiabatic expansion. The radial velocity thus plays a significant role at comets. In spherical symmetry and under steady-state conditions, physical quantities depend only on the cometocentric distance $r$ and the bulk cometary ion velocity $V_i(r)$ being mainly radial, although these assumptions are only valid for the inner cometary ionosphere and may significantly break at larger cometocentric distances (see the chapter by Götz et al. in this volume). Under such conditions, the transport term is reduced to:
\begin{equation}
    \nabla \cdot (n_{i} \vec{V}_{i}) = \frac{1}{r^2}\frac{\mathrm{d} (n_i V_i r^2 )}{\mathrm{d} r}= {V}_{i}\dfrac{\mathrm{d} n_{i}}{\mathrm{d} r}+n_{i}\dfrac{\mathrm{d} {V}_{i}}{\mathrm{d} r}+\dfrac{2 n_{i} {V}_{i}}{r}\label{eq:transportIonosphere1}
\end{equation}
The spherical shell enclosed in $[r, r+\mathrm{d}r]$ has a volume $4\pi r^2\,\mathrm{d}r$ such that ions move into shells of increasingly larger volumes the more outwards they move. The last term $2 n_{i} {V}_{i}/r$ in Eq.~\ref{eq:transportIonosphere1} accounts for this geometrical effect.

As a whole, Eq.~\ref{eq:continuityIonosphere} represents the balance between the source of the ionospheric plasma through ionisation of the cometary neutrals and ionospheric loss terms, namely radial transport and electron-ion dissociative recombination (net chemical loss). Unfortunately, it is not possible to solve Eq~\ref{eq:continuityIonosphere} analytically or without making assumptions/approximations as one needs the radial ion velocity and electron temperature profiles.

Before going further, the first question should be: Which of these terms are the most relevant to consider and are there any that can be neglected? One approach is to look at a typical physical quantity characterising each term. If Eq.~\ref{eq:continuityIonosphere}, combined with Eq.~\ref{eq:transportIonosphere1}, is  now divided by $n_i$:
\begin{equation}
-\underbrace{\vphantom{\sum_i}\dfrac{1}{n_ir^2}\dfrac{\mathrm{d}(n_i\, {V}_i\,r^2)}{dr}}_{1/\mathcal{T}_\text{transport}}\underbrace{+\vphantom{\sum_i}\dfrac{\nu_i^{\text{ioni}}\,n_n}{n_i}}_{1/\mathcal{T}_\text{ionisation}}\underbrace{-\sum_s \alpha_{s}\,n_{s}}_{1/\mathcal{T}_\text{recombination}}\approx 0
\label{eq:1_timescale}
\end{equation}
an equation whose dimension is the inverse of time. Large (resp. small) terms in Eq.~\ref{eq:1_timescale}  are associated with fast (resp. slow) processes, that is, short (resp. long) timescales.

\begin{figure*}[!ht]
\begin{tikzpicture}[thick,scale=1, every node/.style={scale=1}]
        \node[inner sep=0pt](B) at (0,2.8)
        {\includegraphics[width=.33\textwidth]{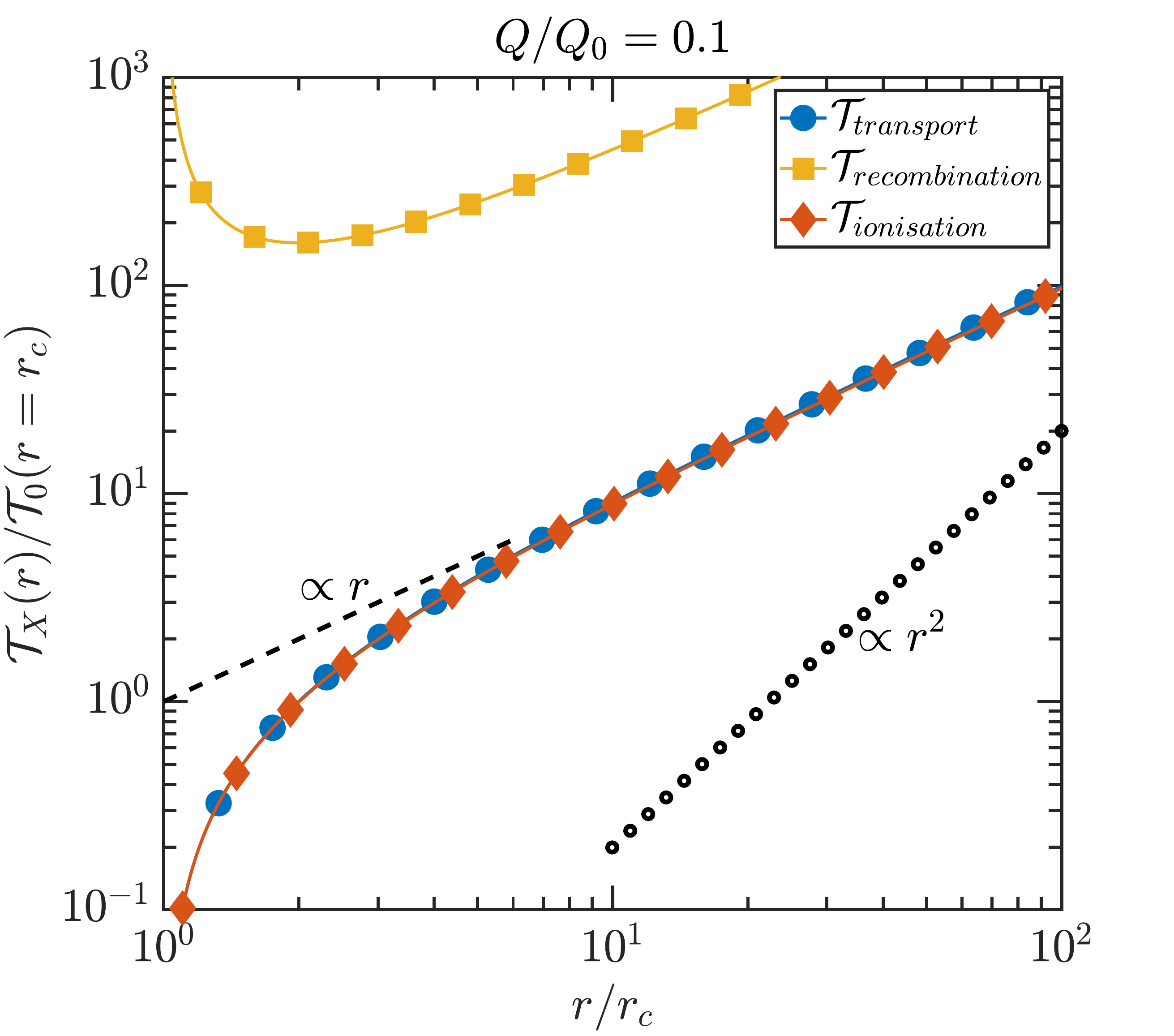}\includegraphics[width=.33\textwidth]{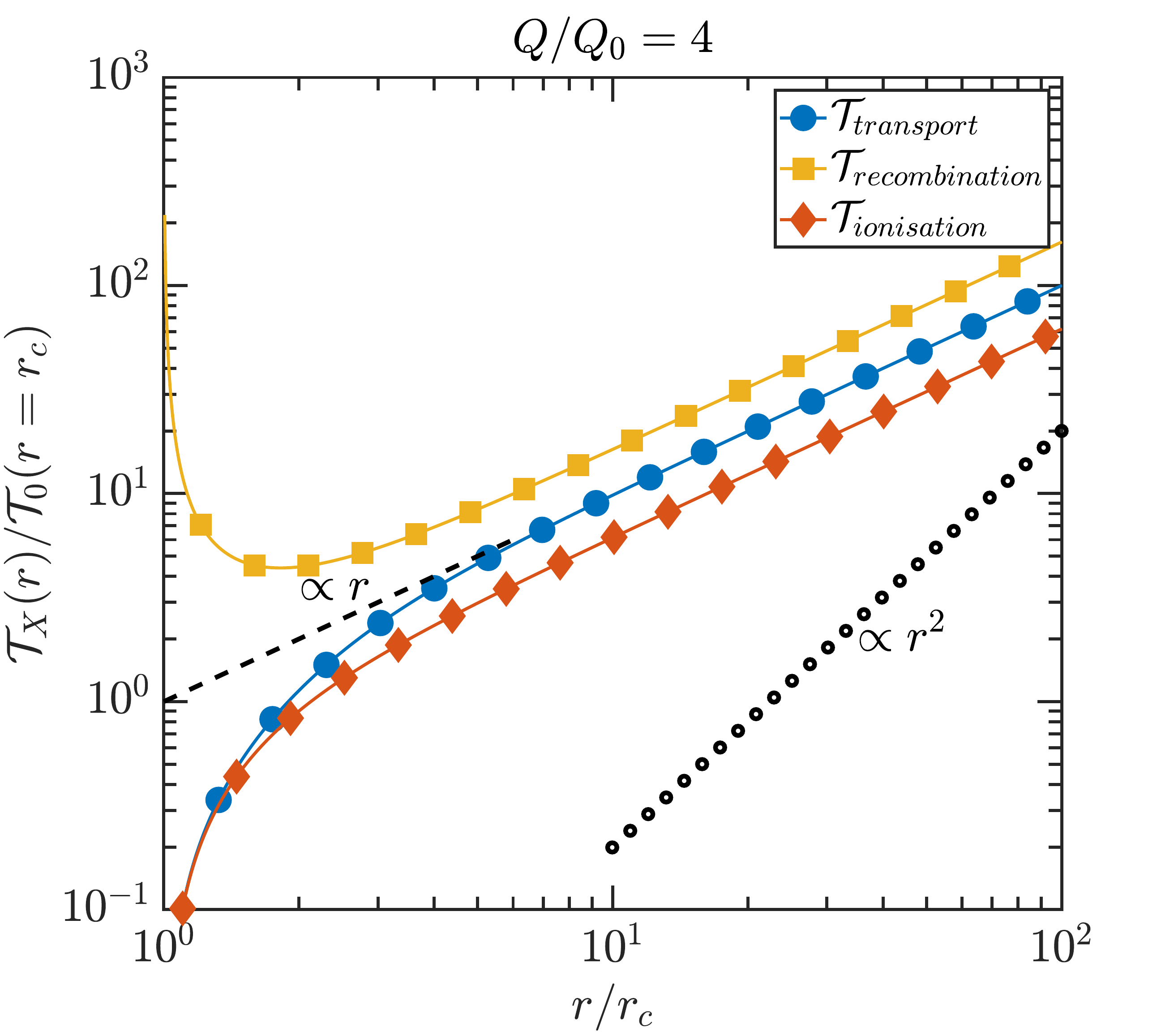}
        \includegraphics[width=.33\textwidth]{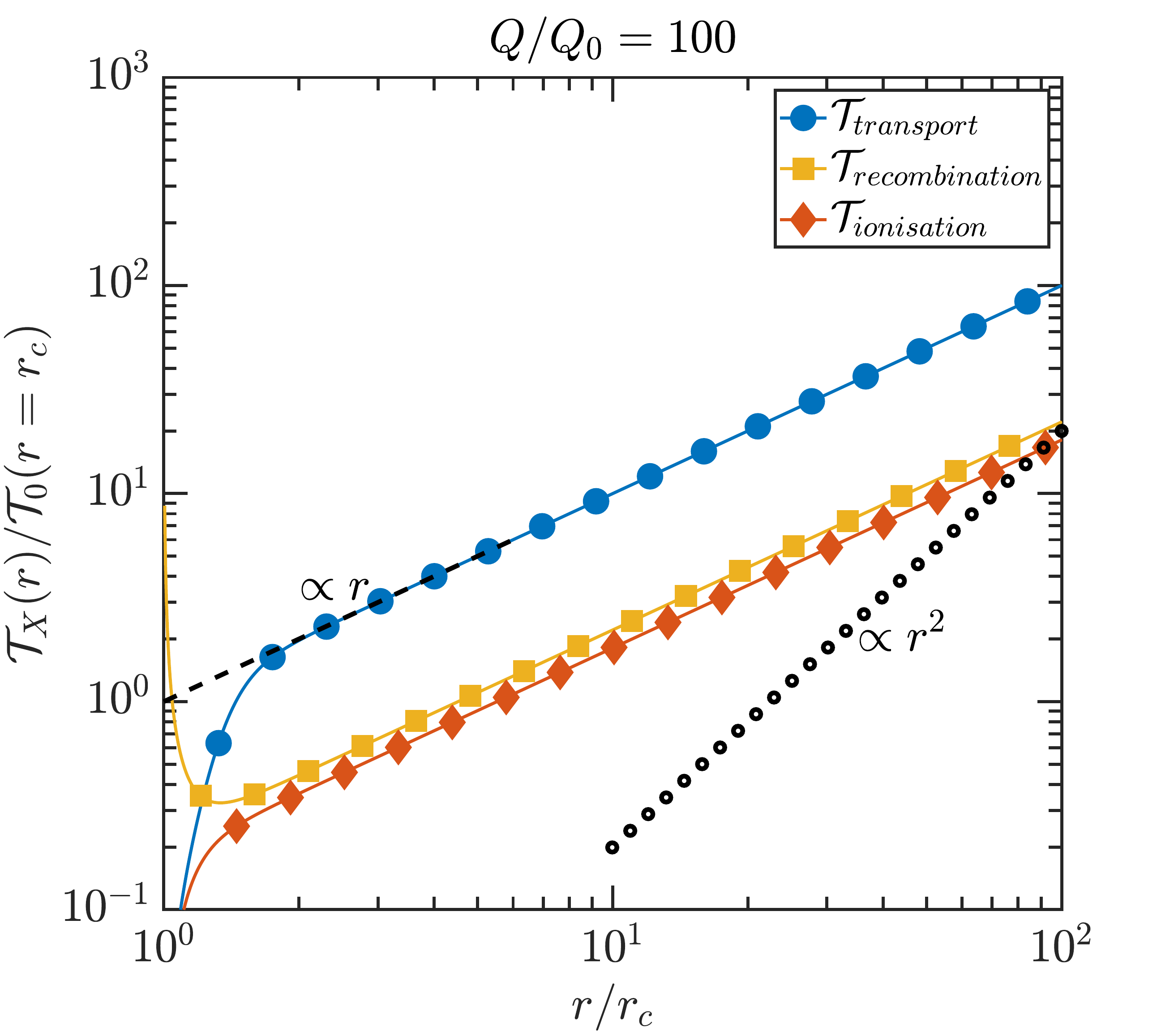}};
        \draw[-stealth,thick](-8.5,5.5) -- (8.5,5.5)node[above,pos=0.5]{\large $Q/Q_0$ increases: $\nearrow$ outgassing rate/ionisation rate/DR rates or $\searrow$ ion/neutral speeds};
\end{tikzpicture}

    \caption{Timescales for the ion number density profile from Eq.~\ref{eq:iondensityradial} for different activities normalised to $Q_0$: $Q/Q_0=0.1$ (left, loss dominated by transport), $Q/Q_0=4$ (middle, losses through transport and recombination are similar), and $Q/Q_0=100$ (right, loss dominated by recombination) as a function of cometocentric distances normalised to $r_c$ and expressed in terms of the approximated transport timescale $\mathcal{T}_0$ at the nucleus' surface (Eq.~\ref{eq:t_transport}). Timescale dependencies in $r$ and $r^2$ (losses in $1/r$ and $1/r^2$) are indicated for reference. Photoabsorption is ignored for simplicity.\label{fig:timescales}
    }
\end{figure*}

Let us review each loss term first (terms with negative signs in Eq.~\ref{eq:1_timescale} which represent a physical sink), that is, transport and dissociative recombination, and assess their respective timescale. The transport timescale is given by: 
\begin{equation}
\dfrac{1}{\mathcal{T}_\text{transport}}=
V_i\dfrac{\mathrm{d} \log n_i}{\mathrm{d} r}+\dfrac{\mathrm{d} V_i}{\mathrm{d} r}+\dfrac{2 V_i} {r} 
\label{eq:t_transport_true}
\end{equation}
independent of the cometary activity. From observations, the plasma density in the inner cometary ionosphere of 1P and 67P alike  varies in $1/r$  \citep{Balsiger1986,Edberg2015, Heritier2017}
such that $\mathrm{d} \log n_i/\mathrm{d} r\sim -1/r$. Regarding the variation of the radial speed, there is no evidence of a strong velocity gradient (acceleration or deceleration) except near the contact surface at 1P \citep{Goldstein1989}. However, within the diamagnetic cavity of 1P, the ion velocity was relatively constant \citep{Schwenn1987}. At 67P, there is no evidence of strong velocity gradients either. This term can thus be neglected and we end up with: 
\begin{equation}
\dfrac{1}{\mathcal{T}_\text{transport}}\sim \dfrac{V_i}{r}\text{ or }\dfrac{V_n}{r}
\label{eq:t_transport}
\end{equation}
In fact, Eq.~\ref{eq:t_transport} is an approximation which only holds beyond a few cometary radii, beyond the location where the plasma density peaks, where $n_i\propto 1/r$. Below and up to the location of this maximum, this is an underestimate and Eq.~\ref{eq:t_transport_true} must be used (see Fig.~\ref{fig:timescales}).

By comparison, the electron-ion recombination timescale for the whole plasma population (electron sink term) is given by: 
\begin{equation}
\dfrac{1}{\mathcal{T}_\text{recombination}}
 \approx \alpha_i(T_e)\,n_i(r)
\sim \dfrac{\alpha_i(T_e)\,n_{i,\text{obs}}\,r_\text{obs}}{r}
\label{eq:recombtimescale}
\end{equation}
where:
\begin{equation*}
    \alpha_i = \frac{\sum_s \alpha_s n_s}{n_i}
\end{equation*}
and the subscript `obs' corresponds to a location of reference where the cometocentric distance and number plasma density are known, {for instance, probed and observed at a spacecraft}. The numerator in Eq.~\ref{eq:recombtimescale} has the dimension of a speed and should be compared with the ion speed. The recombination timescale may span over several orders of magnitude because it depends both on electron temperature (through the recombination coefficient $\alpha_i$, see Section~\ref{section:2:3}) and plasma density. 

Experimentally, $\alpha_i$ ranges between $10^{-12}$ and $10^{-15}$ m$^{3}$\,s$^{-1}$ for $10^2<T_e<10^4\text{ K}$ \citep{Gombosi1996,Rosen2000}. The electron temperature may be very different throughout the coma or even between comets. The lowest range for $T_e$ typically corresponds to the inner cometary ionosphere at 1P under active outgassing conditions. As further discussed in Section \ref{section:3:1:2}, frequent electron-neutral collisions are expected to bring the electron temperature close to the nucleus down to that of the neutral gas, which is consistent with indirect observations at comet 1P \citep[see][]{Eberhardt1995}. The highest range has been observed at 67P under low outgassing conditions when the bulk of the thermal population is not thermalised \citep{Wattieaux2019,Wattieaux2020}, see Section~\ref{section:3:1:1}). During Giotto's flyby, ion number densities reached almost $10^{10}$~m$^{-3}$ at $r_\text{obs}\sim10^4$~km \citep{Altwegg1993} such that $n_{i,\text{obs}}\,r_{\text{obs}}\approx 10^{17}$~m$^{-2}$. Because $\alpha_i\sim 10^{-14}$~m$^{3}$\,s$^{-1}$ in those conditions, the dissociative recombination timescale was $\mathcal{T}_\text{recombination}\sim~10^4$~s \citep[or $\sim 2\text{ h }45$~min,][]{Gombosi1996}. Compared with the transport timescale of about $r/V_i\sim 10^4$~s at the same location, transport and dissociative recombination were contributing equally to the ion sink at that location at 1P \citep{Beth2019}. However, for $r\lesssim10^4$~km, $T_e$ fell drastically by 2--3 orders of magnitude:  The recombination timescale was then shorter than the transport timescale and hence dissociative recombination was the main ion sink in the innermost of 1P's ionosphere.

Unlike the transport timescale, the recombination timescale depends on cometary activity: The higher the outgassing activity, the larger the ion number density, the more efficient the recombination. We can then try to find out for which activity we should consider/neglect the recombination. Let us consider a very simplistic case in the coma, with a uniform electron temperature and therefore a constant $\alpha_i$. Under steady-state conditions and spherical symmetry, the ionospheric continuity equation (Eq.~\ref{eq:continuityIonosphere}), combined with Eqs.~\ref{eq:n_p} and \ref{eq:transportIonosphere1}, reduces to:
\begin{equation}
    \dfrac{1}{r^2}\dfrac{\mathrm{d} (n_i V_i  r^2)}{\mathrm{d}r}=\dfrac{\nu_i^\text{ioni} Q}{4\pi V_n  r^2}-{\alpha_i} n^2_i
    \label{eq:continuityIonosphere_2}
\end{equation}
where $Q$ is the total outgassing rate for the comet (all major neutral species included). Assuming constant and radial ion speed $V_i$, ionisation frequency $\nu_i^\text{ioni}$, and recombination rate coefficient $\alpha_i$ to be independent of $r$, the solution of Eq.~\ref{eq:continuityIonosphere_2} is given by \citep{Beth2019}:
\begin{equation}
    n_i(r)=\dfrac{V_i}{2\alpha_ir}\left(\delta-1\right) \dfrac{r^\delta-r_c^\delta}{r^\delta+\dfrac{\delta-1}{\delta+1}\,r_c^\delta}\label{eq:iondensityradial}
\end{equation}

{where}:

\begin{equation*}
\delta=\sqrt{1+\dfrac{Q}{Q_0}}\text{\hfill\quad , \hfill\quad } Q_0=\dfrac{\pi V_n V^2_i}{\nu_i^\text{ioni}\,\alpha_i},
\end{equation*}
and $r_c$ is the nucleus' radius. $Q_0$ is homogeneous to an outgassing rate [s$^{-1}$].  The different timescales (namely transport, recombination, and ionisation) associated with this profile are given for illustration in Fig.~\ref{fig:timescales}. Even though ion-neutral chemical reactions do not affect the total ion density, their typical timescale (varying in $\propto r^2$) is indicated. 

Regarding the assessment of $Q_0$, and unlike $\alpha_i$, the order of magnitude for $\nu_i^\text{ioni}$, $V_n$, and $V^2_i$ can be reasonably estimated (resp. see Section~\ref{section:2}, the chapter by Biver et al. in this volume, and  Section~\ref{section:4}). In the case of active comets, $\nu_i^\text{ioni}$ is not uniform (even when assuming spherical symmetry for the neutral density) and can be significantly reduced closer to the nucleus in the inner coma due to photoabsorption (see Section~\ref{section:2:2:1}). This has also the consequence of limiting the effect of  recombination near the nucleus \citep{Beth2019}: As the ion number density is  near zero at the nucleus' surface, the recombination is negligible there. 

Interestingly, Eq.~\ref{eq:continuityIonosphere_2} gives a constraint on the flux. By integrating Eq.~\ref{eq:continuityIonosphere_2} over radial distance $r$, one gets:
\begin{equation}
    n_i(r) V_i(r) \leq\dfrac{1}{r^2} \int_{r_c}^r\!\dfrac{\nu_i^\text{ioni}Q}{4\pi V_n}\,\mathrm{d}r= \dfrac{\nu_i^\text{ioni}Q}{4\pi V_n} \dfrac{r-r_c}{r^2}<\dfrac{\nu_i^\text{ioni}Q}{4\pi V_n r}
    \label{eq:ion_flux}
\end{equation}
Equation~\ref{eq:ion_flux} may be used in two ways:
\begin{itemize}
    \item[] {$\bullet$} If the ion number density is known based on in situ observations, it provides an upper limit for the ion radial velocity.
    \item[] {$\bullet$} When recombination is negligible compared to transport, Eq.\,\ref{eq:ion_flux} reduces to the equality (see Eq.\,\ref{eq:continuityIonosphere_2}):
    \begin{equation}
        n_i(r)\,V_i(r) = \dfrac{\nu_i^\text{ioni}\,Q}{4\pi V_n}\ \dfrac{r-r_c}{r^2}
        \label{eq:ion_flux_2}
    \end{equation}
    This corresponds to a solution to the ion continuity equation reduced to the simple balance between transport (loss) and ionisation (source). In contrast with Eq.\,\ref{eq:iondensityradial}, Eq.\,\ref{eq:ion_flux_2} is valid for $V_i(r)$ varying with $r$ but still purely radial.
\end{itemize}

For weakly active comets, such as 67P, the recombination  timescale is much longer than the transport timescale \citep{Galand2016}. This corresponds to the `equality' case given by Eq.\,\ref{eq:ion_flux_2}. Applying Eq.\,\ref{eq:n_p}, Eq.\,\ref{eq:ion_flux_2} can be re-written as:
    \begin{equation}
        n_i(r) = \dfrac{\nu_i^\text{ioni}\, n_n\, (r-r_c)}{V_i(r)}
        \label{eq:multi_ni}
    \end{equation}
This equation was successfully used at 67P at low activity to retrieve the electron number density observed by the Mutual Impedance Probe and Langmuir Probe (see Section~\ref{section:3:1:1}) onboard Rosetta  \citep{Galand2016,Heritier2018}. A multi-instrument approach was applied to cometocentric distances from 10~km to 80~km, for heliocentric distances larger than 2~au.  In those conditions, photoabsorption and dissociative recombination were negligible  \citep{Heritier2018}. The ions were assumed not to have undergone any significant acceleration at Rosetta's position and to have a bulk velocity similar to that of the neutrals ($V_i(r)\approx V_n$ at $r=r_\text{Rosetta}$). Electron-impact ionisation frequency was derived from the electron spectrometer onboard Rosetta (see Section~\ref{section:3:1:3}), whereas photoionisation frequency was extrapolated from solar flux measurements from Earth. Neutral density and neutral composition were derived from Rosetta pressure gauge and double-focusing mass spectrometer \citep[][see the chapter by Biver et al. in this volume]{Balsiger2007}. Equation~\ref{eq:multi_ni} captured very well the observed electron density from the balance between transport of non-accelerated ions and the major sources of ionisation, photoionisation and electron-impact ionisation. The latter source was found to dominate at times, especially post-perihelion for large heliocentric distances \citep[$>2$~au, see][]{Heritier2018}. Eq.~\ref{eq:multi_ni} held even under solar-wind disturbed conditions, such as CMEs and CIRs \citep{Galand2016, Heritier2018}. 

\begin{figure*}
\includegraphics[width=\textwidth]{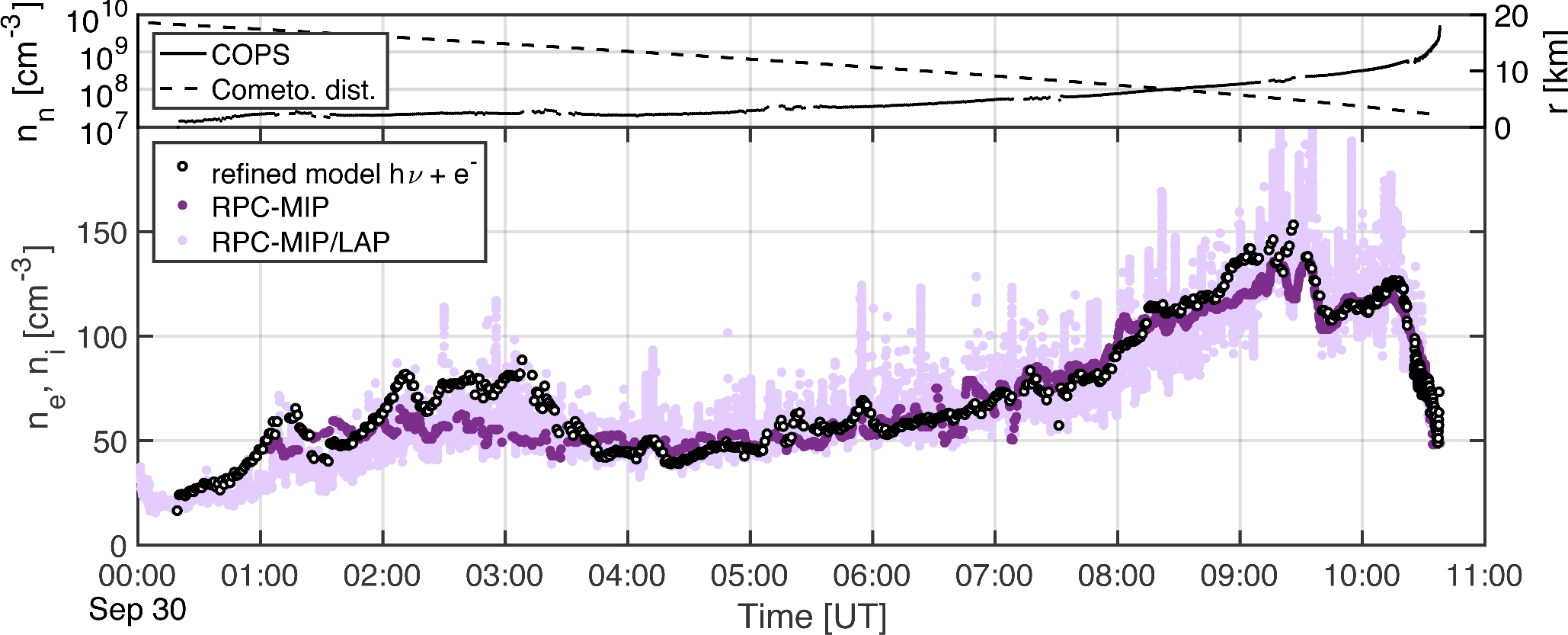}
\caption{Top panel: Time series of neutral number density from ROSINA pressure gauge (solid line) and of cometocentric distance (dashed line). Bottom panel: Time series of the modelled (black dots) and observed (magenta) ionospheric number densities at the end of Rosetta's mission, from 20\,km down to the surface on 30 September 2016. The outgassing rate was around $4\times 10^{25}$\,s$^{-1}$. The dark magenta dots correspond to RPC Mutual Impedance Probe measured electron density, while the light magenta dots represent the high time resolution RPC Langmuir Probe data cross-calibrated with RPC Mutual Impedance Probe. \citep[Fig. 9 from ][]{Heritier2017}
\label{fig:ne_eom}}
\end{figure*}

This multi-instrument approach was also applied all the way down to the surface at the end of the Rosetta mission \citep{Heritier2017}. In that case, the model was refined by taking into account adiabatic expansion and acceleration of the cometary neutrals following sublimation near the surface \citep{Huebner2000}. Excellent agreement was found between observed and modelled ionospheric densities, as shown in Fig.~\ref{fig:ne_eom}. The only exception was around 3~UTC near the neck of the comet which had a concave shape and which is a region where the local outgassing has a complex structure and the gas may be collimated (see the chapter by Marschall et al. in this volume).

\subsubsection{A rich diversity of ions: a real ion zoo\label{section:2:4:2}} 

After the focus on the total ionospheric density (Section~\ref{section:2:4:1}), let us now have a look at the ion composition in the coma. We consider here individual ion species, whose number density is governed by the continuity equation given in Eq.\,\ref{eq:continuityIon}. In terms of loss, three processes are in competition depending on the outgassing activity: transport, ion-neutral collisions, and electron-ion dissociative recombination (Sections~\ref{section:2:3} and \ref{section:2:4:1}). Note that the recombination timescale for  ion species $s$ is simply given  by $\mathcal{T}_{s,\text{recombination}}\approx (\alpha_s n_i(r))^{-1}$ (Eqs.~\ref{eq:1_timescale} and \ref{eq:recombtimescale}). Because it involves the electron density (or total ion density, $n_i$), it is almost the same for the different ion species. A given ion species also undergoes loss through ion-neutral collisions (see Section~\ref{section:2:3}  and Eq.~\ref{eq:continuityIon}). The loss timescale for an ion species $s$ through ion-neutral collisions is given by:
\begin{equation}
    \dfrac{1}{\mathcal{T}_{s,\text{ion-neut loss}}}=\sum_p k_{s,p}(T)\, n_p(r) = k_{s,n}\, n_n(r) \propto \dfrac{1}{r^2}
    \label{eq:l_ionneutr}
\end{equation}
where 
\begin{equation}
    k_{s,n} = \frac{1}{n_n} \sum_p k_{s,p}\, n_p(r)
\end{equation}
is the average reaction rate coefficient between ion species $s$ and the set of neutral species $n$. The subscript $p$ refers to the neutral species $p$ and is incremented over all neutral species chemically reacting with ion species $s$. Excluding extended sources, the neutral density of a neutral species follow a spatial dependence in $1/r^2$ (Eq.~\ref{eq:n_p}),  thus the timescale $\mathcal{T}_{s,\text{ion-neut loss}}$ varies in $r^2$, whereas dissociative recombination and transport timescales vary in $r$ (see Fig.~\ref{fig:timescales}). Therefore, the loss timescale through ion-neutral chemistry increases much faster with cometocentric distance $r$ than that through transport or recombination. The ion-neutral chemistry and coupling are relevant in the region where $\mathcal{T}_{s,\text{ion-neut loss}}\leq \mathcal{T}_\text{transport}$ \citep{Gombosi2015}, that is, within a region around the nucleus bounded by:
\begin{equation}
R_{s,n}=\dfrac{k_{s,n}\,Q}{4\pi\, V_s\, V_n}\lesssim\dfrac{k_{s,n}Q}{4\pi\, V^2_n}
\label{eq:Rsn}
\end{equation}
where the distance $R_{s,n}$ corresponds to $\mathcal{T}_\text{transport}$ (combining Eqs. \ref{eq:n_p}, \ref{eq:t_transport}, and \ref{eq:l_ionneutr}). Ions and neutrals are assumed to flow radially outward at the same speed (see Sections~\ref{section:2:4:1} and \ref{section:4:1}). However, recombination may shrink this region of coupling (i.e. reduce $R_{s,n}$) if $\mathcal{T}_\text{recombination}\leq \mathcal{T}_{\text{transport}}$.

\begin{figure*}[ht]
	    \begin{tikzpicture}[thick,scale=0.95, every node/.style={scale=0.95}]
	    \node[inner sep=0pt](A) at (0.5,7.9)
        {\includegraphics[width=.32\textwidth]{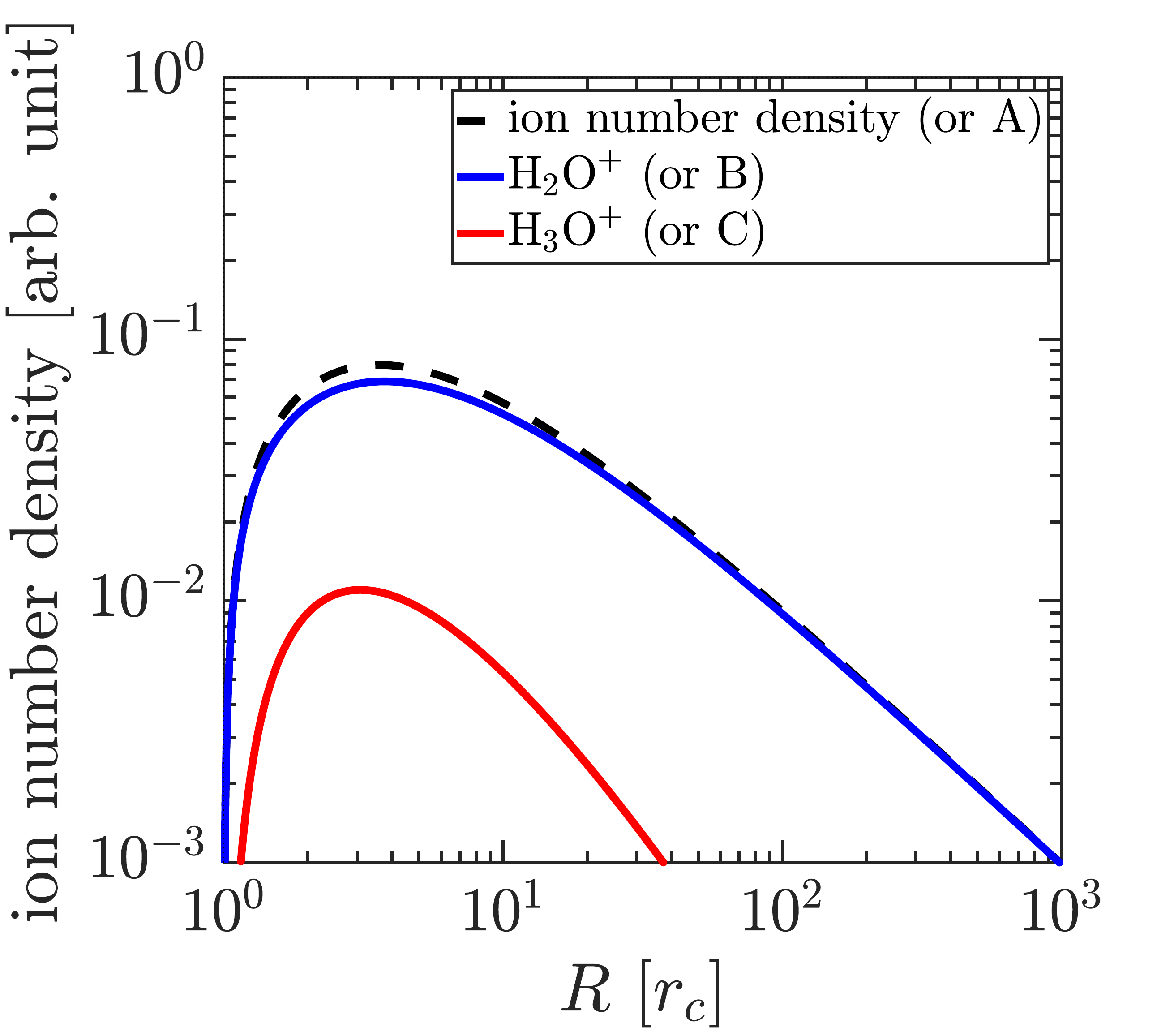}
        \includegraphics[width=.32\textwidth]{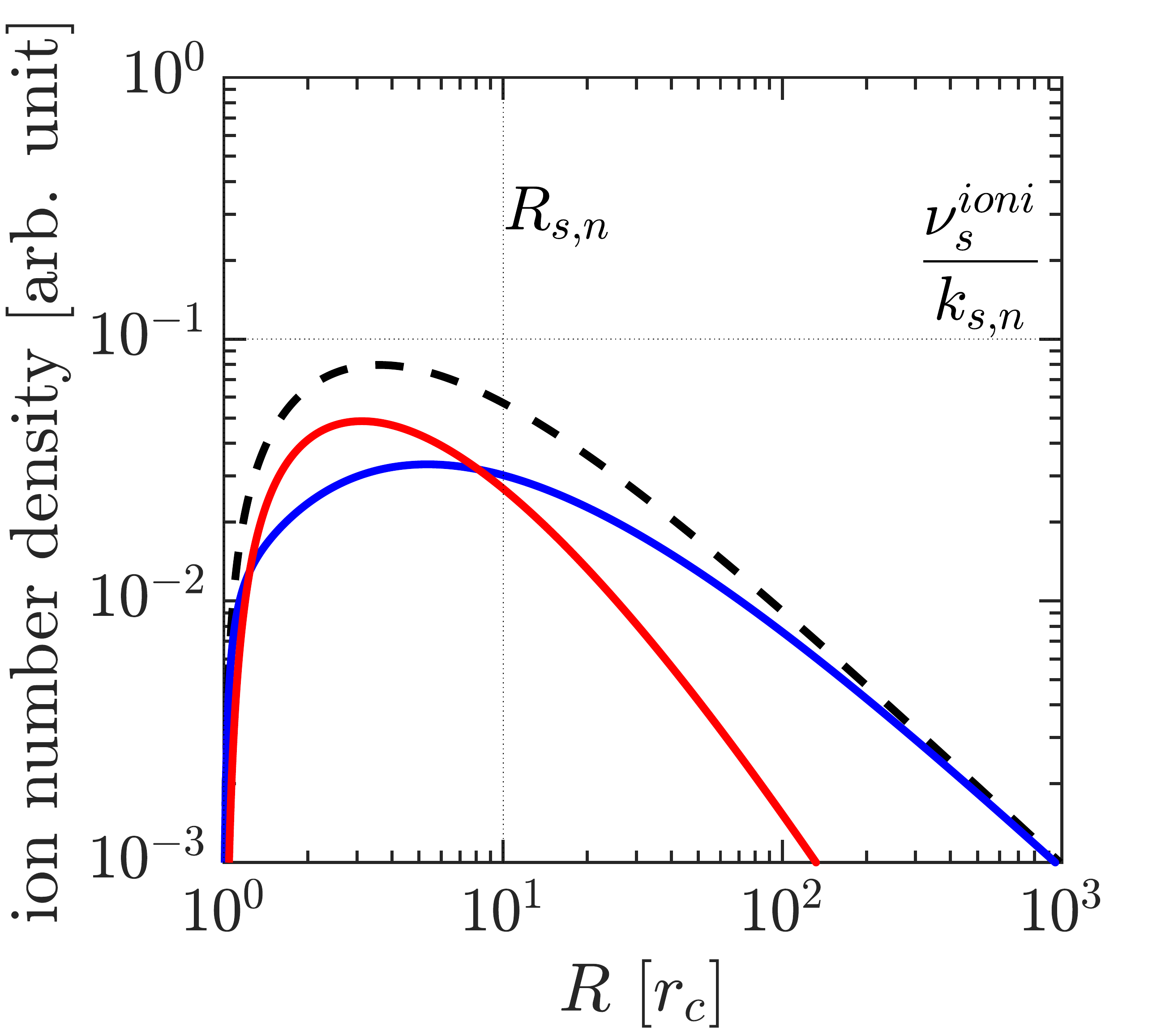}
        \includegraphics[width=.32\textwidth]{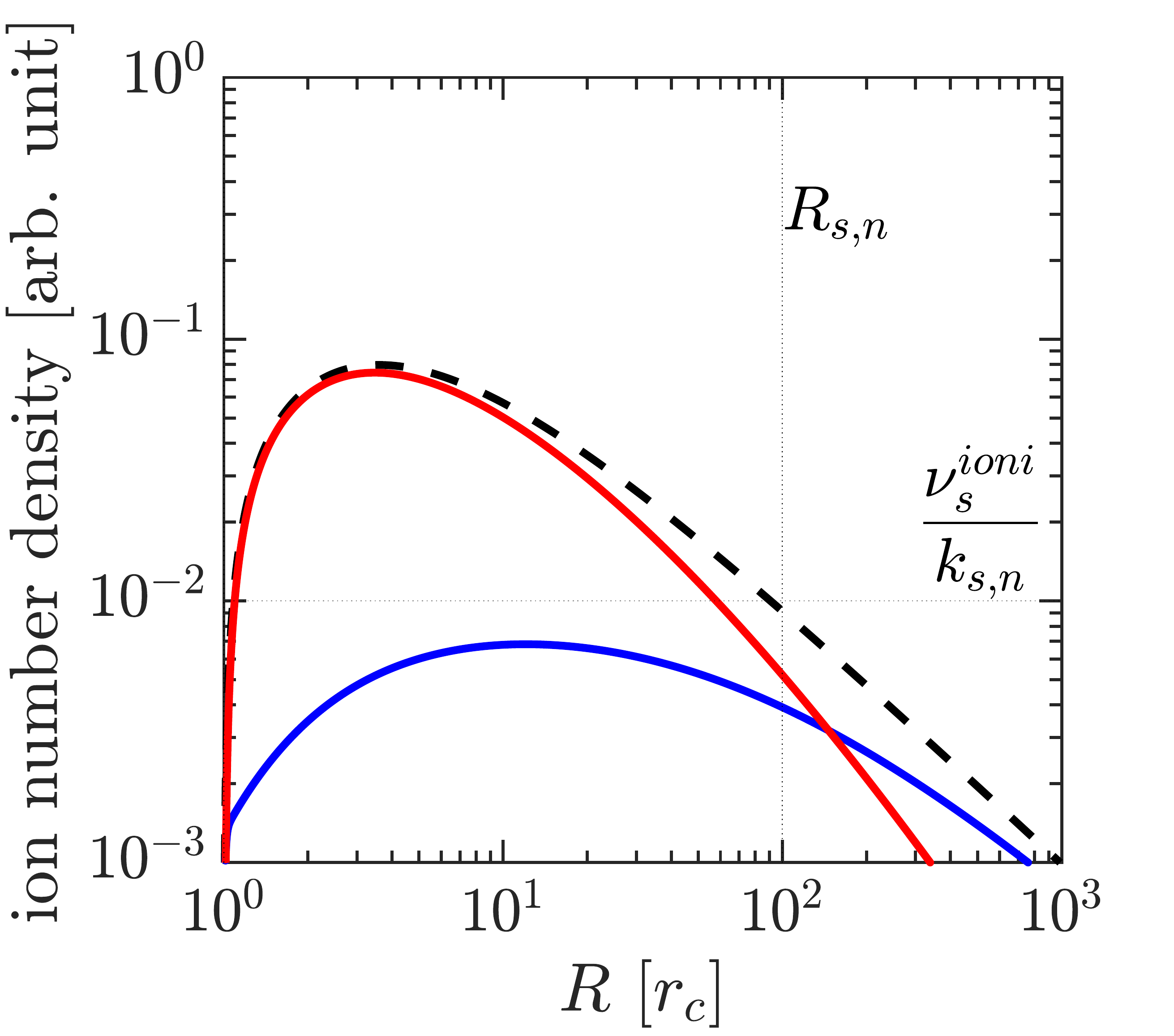}};
        \node[inner sep=0pt](B) at (0.5,3.1)
        {\includegraphics[width=.32\textwidth]{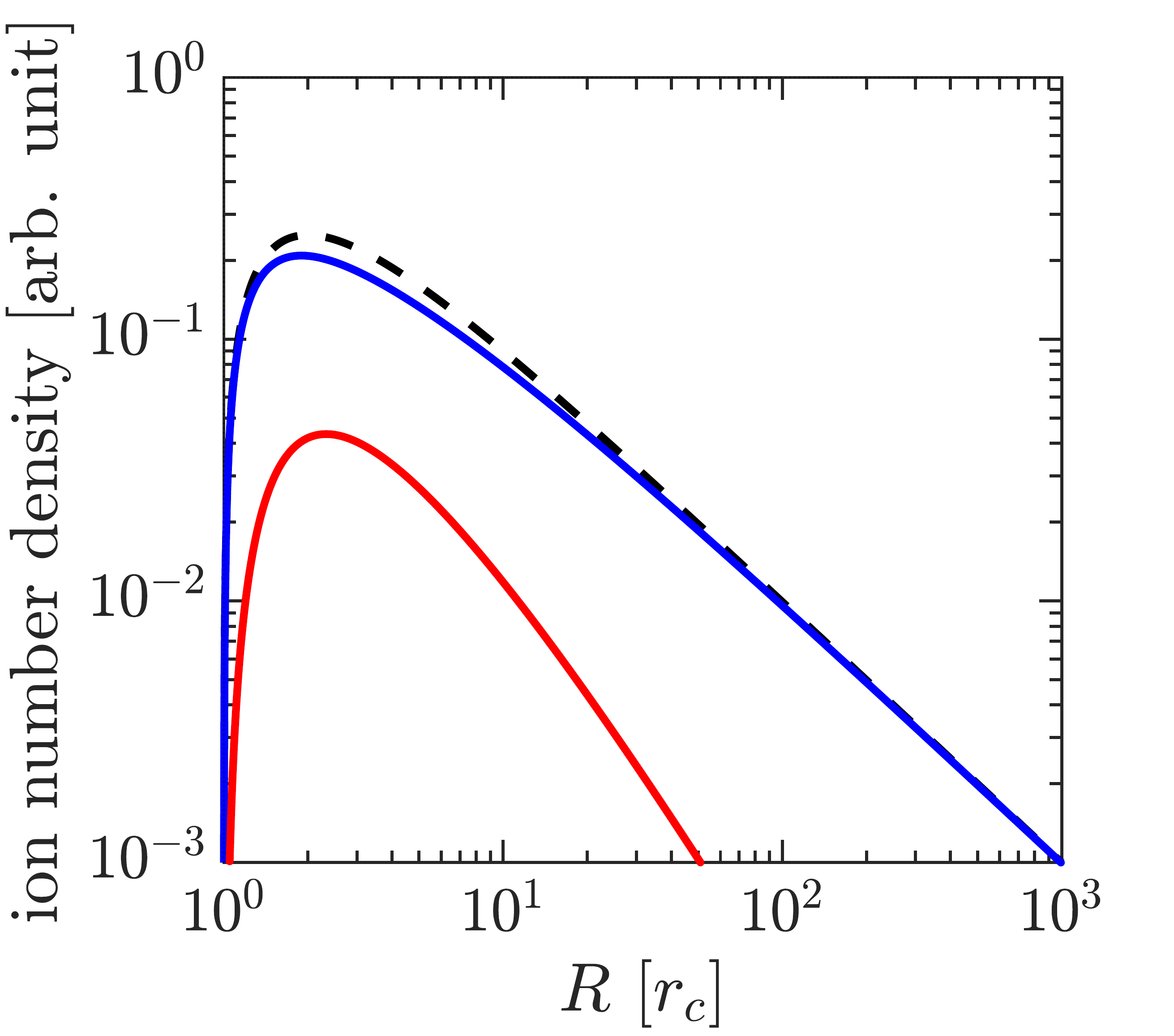}
        \includegraphics[width=.32\textwidth]{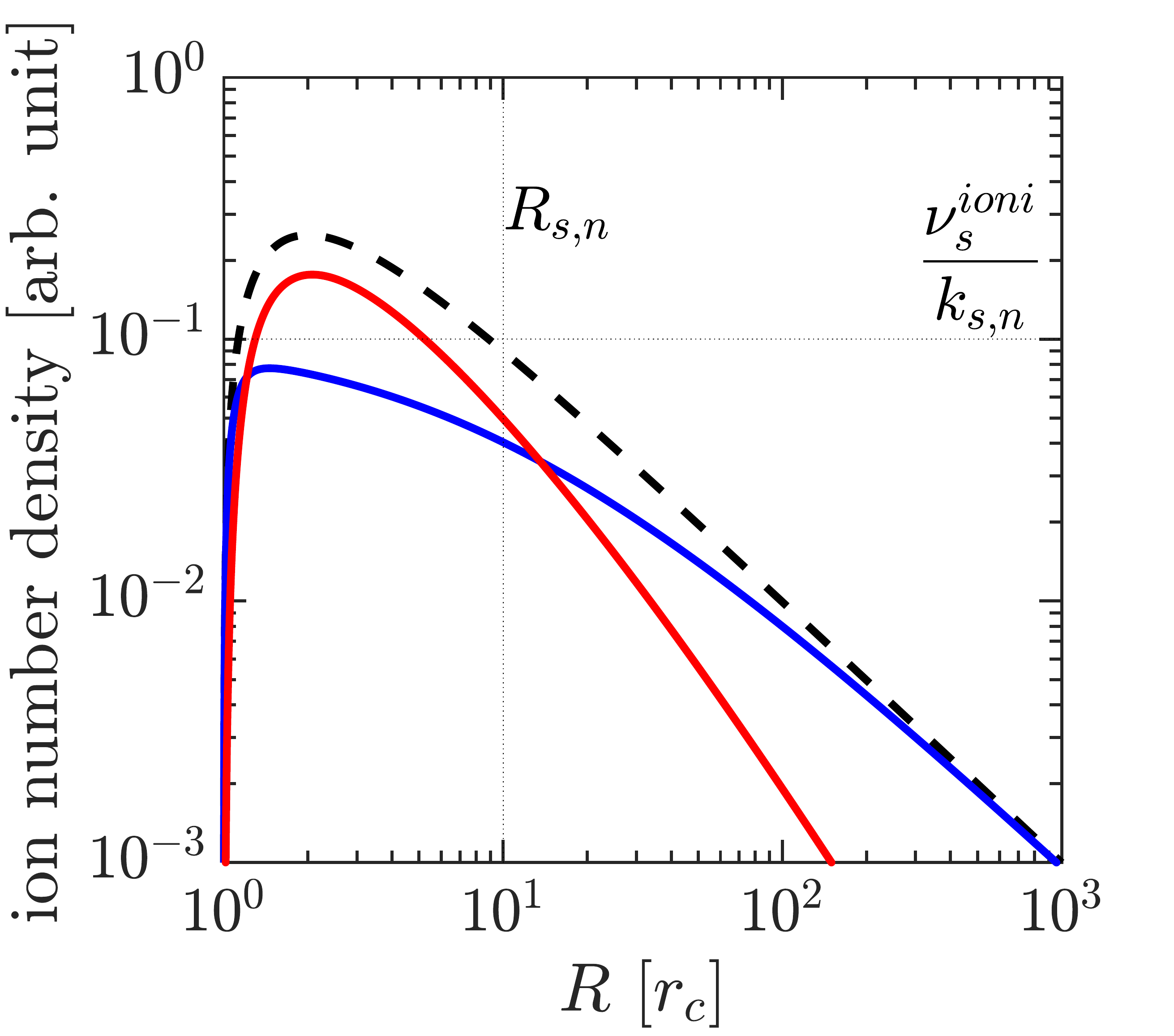}
        \includegraphics[width=.32\textwidth]{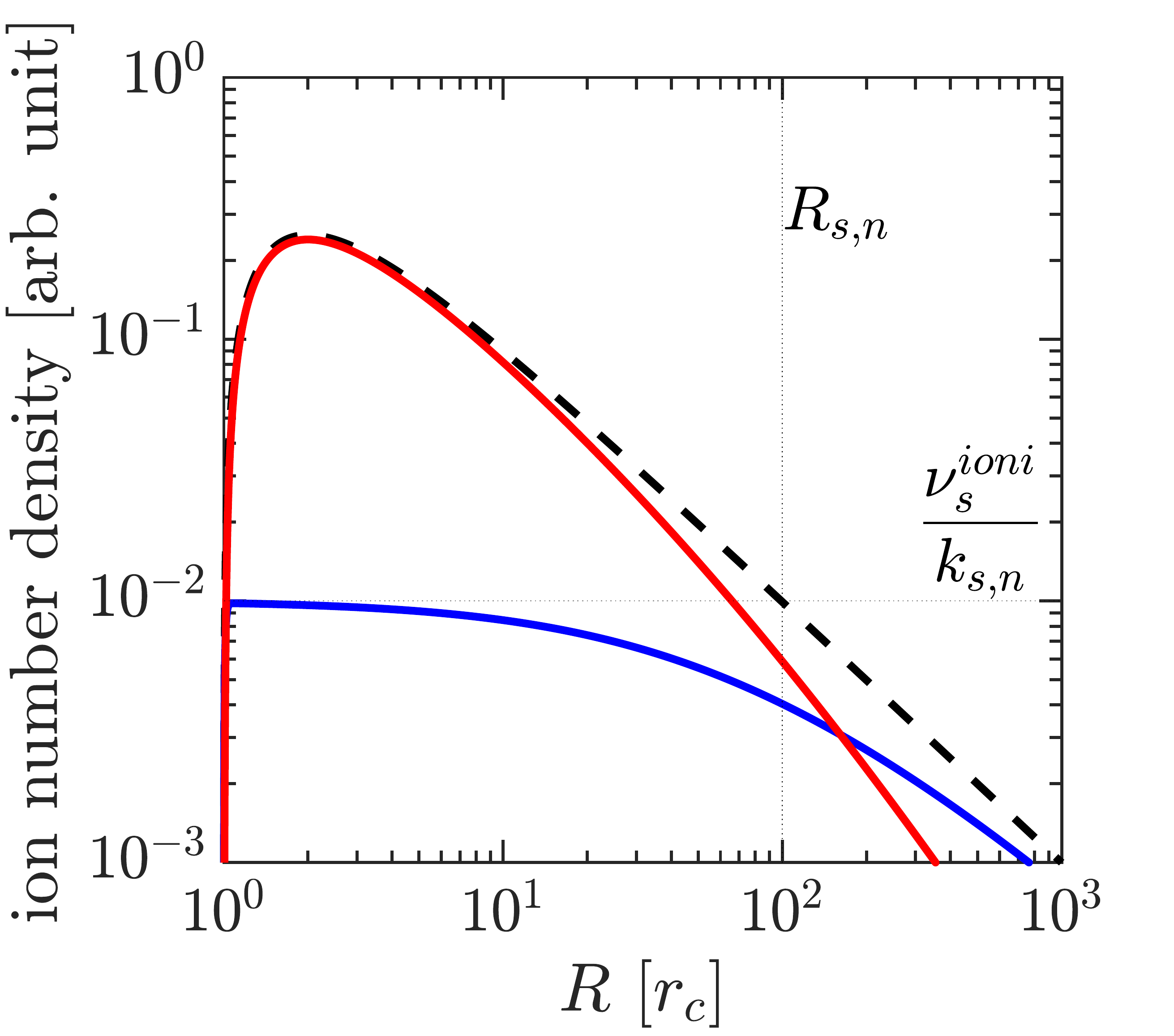}};
        \draw[-stealth,thick](-8.5,0.4) -- (8.5,0.4)node[below,pos=0.5]{ ion-neutral chemistry};
        \draw[-stealth,thick](-8.5,0.4) -- (-8.5,10.3)node[above,pos=0.5,rotate=90]{ photoabsorption};
	    \end{tikzpicture}
	    \caption{Total ion number density profile (dashed, representative of the category as well) and those of H$_2$O$^{+}$ (blue, representative of category B) and H$_3$O$^+$ (red, representative of category C) for a coma made of H$_2$O only, in arbitrary unit using the analytical model from \citet{Beth2020}. Electron-ion dissociative recombination is ignored. Although photoabsorption and ion-neutral chemistry are both linked to the outgassing rate, different conditions are achieved by increasing either photoabsorption (from bottom to top) or ion-neutral chemistry (from left to right) separately. The H$_2$O$^+$ number density at photochemical equilibrium (Eq.~\ref{eq:n_s_photochem}, horizontal dotted line) and the ion-neutral coupling limit (Eq.~\ref{eq:Rsn} vertical dotted line) are indicated. \label{fig:typicalionprofile} }
\end{figure*}

Ion species can be separated into three broad categories whose $r$-dependency of their number density is summarised in Table~\ref{tab:r_dependency} and exemplified in Fig.~\ref{fig:typicalionprofile}:
\begin{itemize}
    \item (A) Ions produced from ionisation but not (or barely) reacting  with neutrals, that is, $R_{s,n}< r_c$ even for high outgassing activity (see Eq.~\ref{eq:n_s_trans}), 
    \item (B) Ions produced from ionisation and which are able to react efficiently with the main neutral cometary species for high enough outgassing activity ($r_c<R_{s,n}$),
    \item (C) Ions only produced through ion-neutral chemistry, such as protonated molecules (except H$_2$O$^+$), and lost through mainly transport or chemistry (Section~\ref{section:2:3}). They are born from collisions between neutrals and ions from category (B).
\end{itemize}

When $R_{s,n}\lesssim r_c\leq r$, ions from category A and ions from category B for weak outgassing activity are mainly produced through ionisation and are primarily lost through transport, which implies:
\begin{equation}
\dfrac{1}{r^2}\dfrac{\mathrm{d}(n_s V_s r^2)}{\mathrm{d}r} \approx \nu_{s}^\text{ioni} n_n(r) 
\end{equation}
Applying Eq.~\ref{eq:n_p} leads to:
\begin{equation}
n_s(r)\approx \dfrac{\nu^\text{ioni}_s Q}{4 \pi V_n V_s(r)}\dfrac{r-r_c}{r^2}\approx \dfrac{\nu^\text{ioni}_s (r-r_c)}{V_s(r)}\,n_n(r)
\label{eq:n_s_trans}
\end{equation}

When $R_{s,n}> r_c$ (moderate to high activity), for $r_c\lesssim r \lesssim R_{s,n}$, ion species $s$ from category B, produced through ionisation, is lost primarily through ion-neutral chemistry (note that usually such an ion species is not terminal and hence not lost through electron-ion dissociative recombination); it is said to be in \emph{photochemical equilibrium}, meaning that the ion number density is only governed by chemical reactions:
\begin{equation}
0\approx \nu^\text{ioni}_s n_n(r) - k_{s,n} n_n(r) n_s(r)
\end{equation}
This implies that:
\begin{equation}
n_s(r)\approx \dfrac{\nu^\text{ioni}_s}{k_{s,n}}\approx \text{constant}
\label{eq:n_s_photochem}
\end{equation}
This value is a plateau and an upper limit for the ion species $s$ which cannot be surpassed. In addition, photoabsorption  may reduce this value in an optically thick coma (see Section~\ref{section:2:2:1}).

Ions from category C, so-called \emph{protonated ions} are produced through proton-transfer reactions, either from H$_2$O$^{+}$ or through the production of other protonated ions, such as H$_3$O$^+$. Neglecting dissociative recombination in Eq.\,\ref{eq:continuityIonosphere}, the total ion density of protonated molecules $AH^+$ (i.e. the sum of the number densities of the protonated molecules such that $n_{AH^+}=n_{\text{H}_3\text{O}^+}+n_{\text{NH}_4^+}+\text{etc.}$) obeys approximately the equation:
\begin{equation}
\dfrac{1}{r^2}\dfrac{\mathrm{d}\left(n_{AH^+}V_{AH^{+}}r^2\right)}{\mathrm{d}r}\approx k_{\text{H}_2\text{O}^+, A}\,n_{A}\,n_{\text{H}_2\text{O}^+}
\label{eq:AHP_low}
\end{equation}

The loss through dissociative recombination did not play a key role to assess ion densities at the location of Rosetta during its escort phase \citep{Heritier2018}, hence this process was disregarded in Eq.~\ref{eq:AHP_low}. However, in specific conditions, such as high enough outgassing rates at a certain range of cometocentric distances coupled with a high sensitivity of the recombination rate coefficient $\alpha_i$ with $T_e$, dissociative recombination may represent a key loss for protonated ions: This was for example the case during Giotto’s flyby of 1P \citep{Rubin2009}.

To solve Eq.~(\ref{eq:AHP_low}), one needs to know the dependence of $n_{\text{H}_2\text{O}^+}$ with $r$ as it plays the role of precursor for the formation of protonated molecules  \citep[see e.g.][]{Aikin1974}. Indeed, the protonated molecules dominate the ion composition near the nucleus only when H$_2$O$^+$ is close to photochemical equilibrium and triggers the formation of H$_3$O$^+$ (see e.g.  Fig.\,\ref{fig:typicalionprofile}). On the one hand, in the region $r_c \lesssim r \lesssim R_{\text{H}_2\text{O}^+,n}$ occurring only for high enough outagssing rates, H$_2$O$^{+}$ is in photochemical equilibrium and thus the left-hand side of Eq.~\ref{eq:AHP_low} is $\propto1/r^2$. Therefore, we have:
\begin{equation}
    n_{AH^+}(r)\propto \dfrac{r-r_c}{r^2} \text{ for $ r \lesssim R_{\text{H}_2\text{O}^+,n}$}
\label{eq:AHP_low1}
\end{equation}
On the other hand, for $r>R_{\text{H}_2\text{O}^+,n}$, H$_2$O$^+$ dominates again the ion composition and behaves $\propto1/r$. The right hand side of Eq.~\ref{eq:AHP_low} is $\propto1/r^3$ and thus:
\begin{equation}
    n_{AH^+}(r)\propto \dfrac{\log r}{r^2} \text{ for $r \gtrsim R_{\text{H}_2\text{O}^+,n}$}
    \label{eq:AHP_low2}
\end{equation}

\begin{table}[ht]
\centering
\caption{Different radial number density dependency for each category below and above the ion-neutral coupling limit $R_{s,n}$ (Eq.~\ref{eq:Rsn}) when transport dominates the ion loss ($Q/Q_0\lesssim1$). The $r$-dependencies given are for an optically thin coma. $^{*}$For category A, $R_{s,n} < r_c$, so third column refers to $r_c<r$. $^{\dagger}$For category C, $s$ in $R_{s,n}$ is not associated with the ion itself but to its parent ion, for example, $s=\text{H}_2\text{O}^+$ and $n=\text{H}_2\text{O}$ for H$_3$O$^+$ and $s=\text{H}_2\text{O}^+/\text{H}_3\text{O}^+$ and $n=\text{NH}_3$ for NH$_4^+$. For category C, transport is here assumed to be the ion loss process. \label{tab:r_dependency}}
\renewcommand{\arraystretch}{2}
\begin{tabular}{lccr}
      \multicolumn{1}{l}{Cat.}&  \multicolumn{2}{c}{{Ion density $n_s$ when:}} &  \multicolumn{1}{r}{Ref. Eqs.} \\
    \cline{2-3}
    & $r_c<r<R_{s,n}$ & $r_c<R_{s,n}<r$ &\\
    \cline{1-4}
     A& N/A ({$R_{s,n}\ll r_c$})& $\propto \dfrac{r-r_c}{r^2}^{*}$ & \ref{eq:n_s_trans}\\
     B&$\approx \text{constant} \leq \dfrac{\nu}{k}$&  $\propto \dfrac{1}{r}$ & \ref{eq:n_s_trans}, \ref{eq:n_s_photochem}\\
     C$^{\dagger}$ &$\propto \dfrac{r-r_c}{r^2}$ &$\propto\dfrac{\log r}{r^2}$ & \ref{eq:AHP_low1}, \ref{eq:AHP_low2}\\
    \cline{1-4}
     A+B+C&$\propto \dfrac{r-r_c}{r^2}$&$\propto \dfrac{r-r_c}{r^2}$ & $-$
\end{tabular}
\end{table}

 At low activity ($Q\leq 10^{25}$\,s$^{-1}$), ion species from category B exhibit an ion density profile (blue, solid line) similar to that of ions from category A (black, dashed line) and given by Eq.~\ref{eq:n_s_trans}, as shown in the left panels in Fig.~\ref{fig:typicalionprofile}. Ions from B, although reacting with H$_2$O and CO$_2$ like H$_2$O$^+$, have their associated ion-neutral coupling limit $R_{B,n}$ smaller than $r_c$ due to the low outgassing rate. So effectively, they behave as ions from category A. Ions from category C are hence barely produced as they require at least one ion-neutral collision and their density decreases very quickly with increasing $r$ (see red profile, left panels, Fig.~\ref{fig:typicalionprofile}). A and B-type ion species are dominating the ion composition. 

As the outgassing activity increases ($Q\gtrsim 10^{25}$~s$^{-1}$), so does the number density of the neutral species (e.g. H$_2$O and CO$_2$), and $R_{B,n}$ passes above the surface ($R_{B,n}>r_c$) meaning there exists a region where ion-neutral chemistry dominates over transport. Within this region, ions from category B are efficiently reacting with neutrals and their density profile is near photochemical equilibrium (Eq.~\ref{eq:n_s_photochem}, see blue profile, middle and right panels in Fig.~\ref{fig:typicalionprofile}). In the same region, ions from category C are efficiently produced, such as H$_3$O$^+$ (and NH$_4^+$), and may dominate the composition region with a profile following the red, solid line, middle and right panels, in Fig.~\ref{fig:typicalionprofile}. Beyond $R_{B,n}$, transport or electron-ion recombination govern the loss. Ions from B (resp C) are barely lost (resp. produced) through ion-neutral collisions. A-type and C-type ions are dominating the ion composition at low $r$ distances, whereas A-type and B-type ions are dominating at high $r$.

\begin{figure}[!ht]
    \centering
    \includegraphics[width=\linewidth]{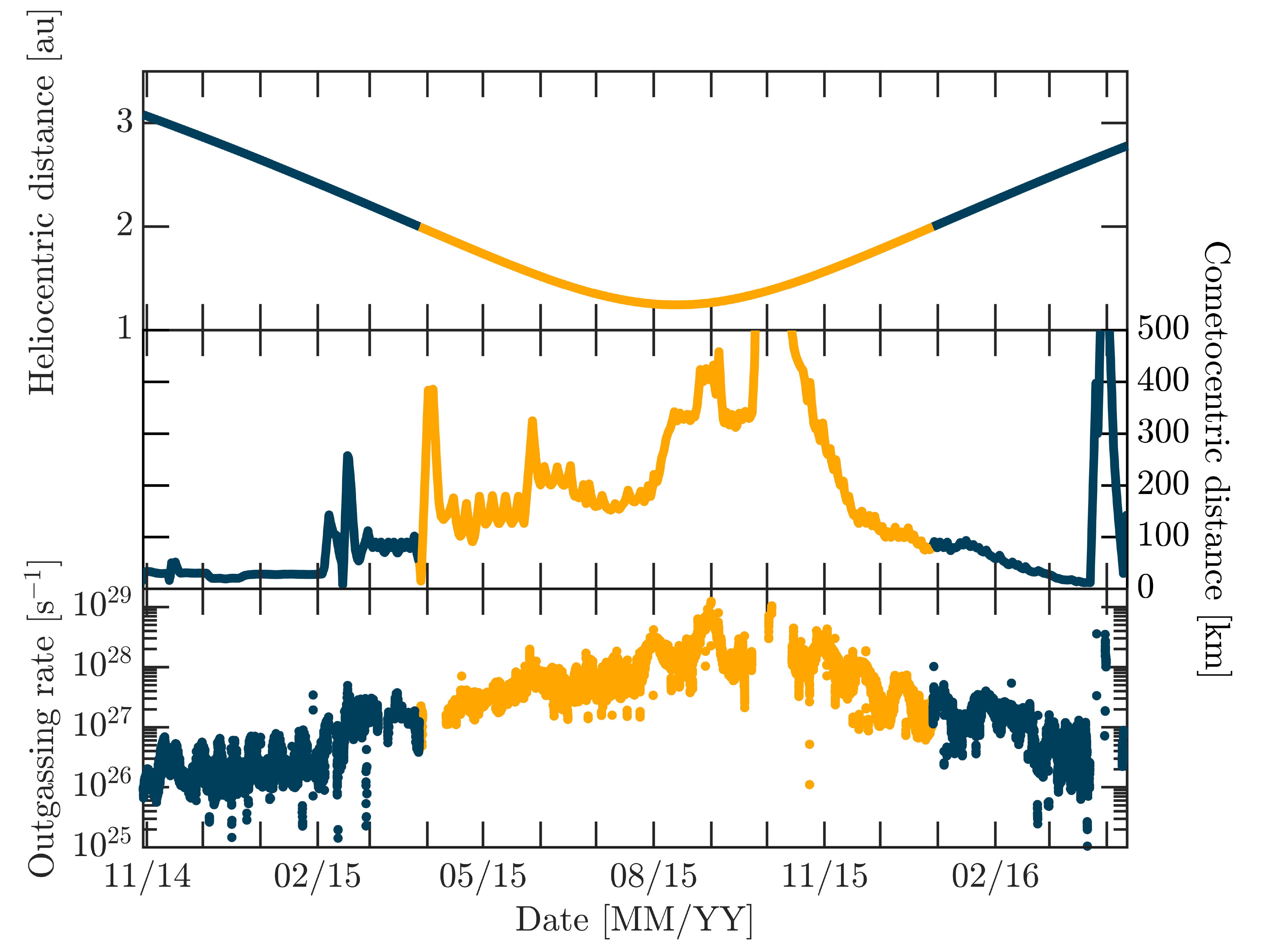}\\
    \includegraphics[width=\linewidth]{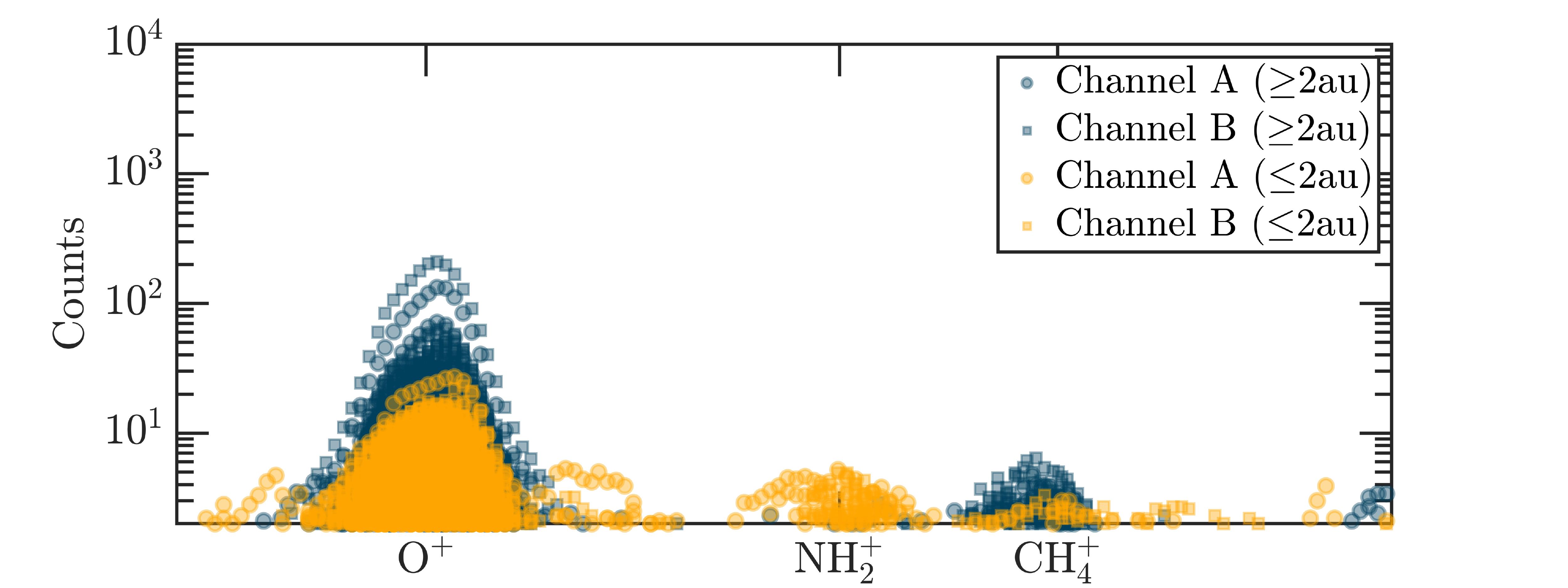}\\
    \includegraphics[width=\linewidth]{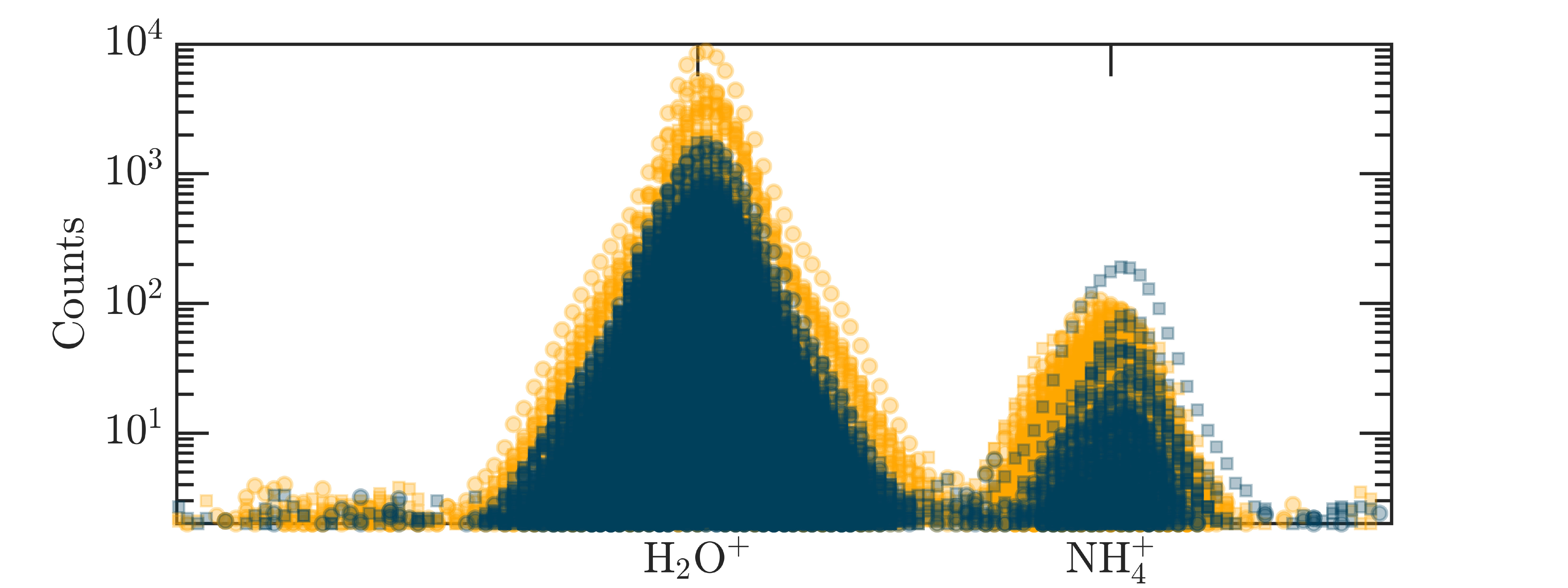}
    \caption{Top: Overview of the heliocentric and cometocentric distances of Rosetta, and of the outgassing rate during the operational phase of ROSINA-DFMS mass spectrometer in ion mode. Middle and bottom: Spectra at 16~u and 18~u from ROSINA/DFMS mass spectrometer during Rosetta's mission. Both colours are associated with different heliocentric distances: above (blue) or below (yellow) 2 au. Adapted from \citet{Beth2020}. \label{fig:ion_zoo}}
\end{figure}

Regarding the detection of ion species in a cometary environment, Rosetta has revolutionised the field \citep{Beth2020}. If the mass resolution of the ion mass spectrometer onboard a cometary probe is high enough, the mass, or mass-per-charge, may work as a unique identifier of ion species. For example, the mass of H$_2$O$^+$ is 18.015~u while that of NH$_4^+$ is 18.039~u. We thus need an instrument with a minimum resolving power of $\Delta m\approx0.025$~u to separate both species. During the Giotto flyby of comet 1P, a snapshot of the ion density profiles was obtained at different integer masses \citep{Altwegg1993}; ion species of the same integer mass could however not be distinguished due to the too low mass resolution of the ion mass spectrometer. In contrast, at comet 67P, ion species were identified individually over a two-year period and a wide range of outgassing activities thanks to the ROSINA double-focusing mass spectrometer (DFMS) aboard Rosetta (see Fig.~\ref{fig:ion_zoo}-top). This spectrometer was one of the most powerful spectrometers ever flown on a planetary probe in terms of mass resolution, reaching up to $R\approx3000$. Thanks to its high accuracy, more than 20 ion species were unambiguously identified in comet 67P's ionosphere \citep[][see Fig.\,\ref{fig:ion_zoo}-centre and Fig.~\ref{fig:ion_zoo}-bottom]{Beth2020}, with the serendipitous detection of the CO$_{2}^{++}$ dication.

\section{Electron population and its impact\label{section:3}}

\subsection{Different electron populations\label{section:3:1}}

At any given time, the electrons in a cometary environment can be classified into three broad populations: \emph{cold}, \emph{warm}, and \emph{hot}.
Although the energy range of each population may slightly vary between published studies, this classification allows to distinguish different electron sources and processes affecting electrons. The cold and warm electron populations, which are primarily of cometary origin with some contribution from the solar wind to the warm electrons, as discussed below, represent a dense population, compared with the rarefied hot electron population. Note that, while cold and warm populations cohabit, one of them may be greatly dominating over the other, depending on the outgassing activity, neutral composition, and cometocentric distance. These three electron populations were first reported from in situ observations at around 12000\,km of comet 21P during its flyby by ICE \citep{Zwickl1986}. During the course of the Rosetta escort mission, these three populations were also identified.  They were differentiated by their signatures as detected by different  Rosetta Plasma Consortium \citep[RPC,][]{Glassmeier2007} sensors onboard Rosetta. The characteristics, prime origin (photoelectrons, Section~\ref{section:2:2:1}, solar wind electrons, or their secondaries, Section~\ref{section:2:2:2}), and detection of each of the three electron populations are discussed in more details hereafter: 
\begin{itemize}
    \item the newly-born, warm, cometary electrons  (5-15~eV), which dominate the plasma density at large heliocentric distances, but are always present (Section~\ref{section:3:1:1}). They include photoelectrons (for coma optically thin to EUV radiation) and electrons from electron-impact ionisation by hot electrons.
    \item the cold, cometary electrons ($<1$~eV), which are energy-degraded warm electrons and whose relative importance increases with outgassing activity (Section~\ref{section:3:1:2}),
    \item the more rarefied, hot electrons ($>20$~eV). When the coma is optically thin to EUV at large heliocentric distances, they are mainly accelerated, solar wind electrons, and are responsible for the bulk  ionisation of the neutral coma and for the auroral FUV emissions (Section~\ref{section:3:1:3}). When the coma is optically thick (e.g. during the flyby of 1P by Giotto), they are mainly photoelectrons in the inner coma \citep{Bhardwaj2003,Beth2019}.
\end{itemize}

\subsubsection{Newly-born warm, cometary electrons\label{section:3:1:1}}

The warm electron population refers to electrons produced through ($i$) photoionisation in an optically thin coma (see Section~\ref{section:2:2:1}) and ($ii$) electron-impact ionisation (see Section~\ref{section:2:2:2}). Photoionisation of a neutral species $p$ by solar radiation  of incident energy $E_{h\nu}$ above ionisation energy $E^{th}_{p,s}$ yields the production of a photoelectron  and an ion (see Eq.~\ref{eq:photoioni_reaction}). The rate at which neutrals $p$ are ionised per unit time is given by the photoionisation frequency, $\nu^{h\nu,\text{ioni}}_p$ (see Eq.~\ref{eq:frequency2} and Fig.~\ref{fig:opticaldepth}). 
The total photoelectron production rate, $S_{e^-}^{h\nu}$, depends on the neutral composition and total neutral density (Eq.~\ref{eq:photoeratetotal}). During photoionisation of a neutral $p$ yielding the production of a photoelectron and an ion $s$, the excess energy, $\Delta E= E_{h\nu}- E^{th}_{p,s}$, goes mainly to the released photoelectron, in the form of kinetic energy. 
However, as this process is largely isotropic, the electrons are uniformly scattered in every direction.  The energy distribution of the produced photoelectrons is as structured as the incident solar radiation flux, some strong solar lines producing a peak of photoelectrons at a given energy, like at 27--28~eV  \citep[photoionisation of H$_2$O or CO$_2$ by the HeII 30.4~nm (40.8~eV) solar line,][]{Korosmezey1987,Bhardwaj2003,Vigren2013}. Nevertheless, in an optically thin coma, the average photoelectron energy is around 10--15~eV \citep{Huebner1992}. These electrons are referred to as \emph{warm}, prior to efficient cooling in the coma through collisions (see Section~\ref{section:3:1:2}). In an optically thick coma, the mean photoelectron energy increases in regions of large neutral column densities \citep{Bhardwaj2003}. Indeed, as illustrated in Fig.~\ref{fig:solar_spectrum}, photoabsorption cross sections have the largest values above 20~nm. Hence it is this part of the solar spectrum which is going to be absorbed first, at an optical depth $\tau = 2$ (Eq.~\ref{eq:tau2}). Inward of the boundary corresponding to $\tau = 2$ at 20~nm, photoelectrons are produced by the remaining solar photons of wavelength less than 20~nm (energy larger than $\sim60$~eV). Subtracting the ionisation energy  (typically 10--15\,eV), this means that all photoelectrons in this region are born with an energy larger than 45~eV. In such a case, photoelectrons are \emph{hot}, no longer warm.

Electrons themselves can be energetic enough to ionise the neutral species. This is the case if their energy is above the ionisation energy (see Section~\ref{section:2:2:2}). As energetic electrons at 67P were found to have typical energies of 30--40~eV \citep{Myllys2019}, the energy of the secondary electron produced through ionisation is less than 15~eV, because the primary electron retains at least half of the remaining energy (where the remaining energy corresponds to the primary electron original energy minus the ionisation energy). Thus, electrons produced by electron-impact ionisation also contribute to the warm population.

The warm electron population was detected throughout the Rosetta escort phase by several instruments. Firstly, the Ion and Electron Sensor \citep[IES,][]{Burch2007}, part of the RPC consortium, measured energy distributions of negatively-charged particles (primarily electrons) above the noise level over an energy range from 4.3~eV\,q$^{-1}$ to a few 100~eV\,q$^{-1}$ with an energy resolution $\Delta E/E\sim 8\%$. The electron velocity/energy distribution, fitted by a suprathermal double-kappa function, exhibited signatures from two populations:  one warm and one hot \citep{Broiles2016b,Myllys2019}. Whereas the hot population was rarefied (see Section~\ref{section:3:1:3}),  the warm population was  dense ($\sim10$--$200$\,cm$^{-3}$) with temperatures typically below 10~eV \citep{Myllys2019}. 
At the beginning of Rosetta's mission, on 14 November 2014, \citet{Broiles2016} identified two dense populations in the RPC-IES dataset: One with temperatures above 8.6~eV was observed when the local neutral number density at Rosetta was high ($n_n>8\times 10^{6}$~cm$^{-3}$) and one with temperature below $8.6$~eV when $n_n$ was generally low ($n_n<8\times 10^{6}$~cm$^{-3}$). 
\citet{Myllys2019} did not find that the electron temperature of the warm population $T_w$ decreased with heliocentric distance but the spread in electron temperature estimates was found to range from 1~eV to 15~eV. 
It should be pointed out that the warm population is \emph{not} thermalised, its distribution is \emph{not} following a Maxwellian and the double-kappa fit is often \emph{not} suitable, especially at large heliocentric distances \citep{Myllys2019}. Furthermore, part of the warm population is unseen by RPC-IES \citep{Madanian2016}, as a result of the combination of the lowest energy of detection (4.3~eV for electrons) and of the precarious effect of the spacecraft potential. Prior to launch, it was not anticipated that the spacecraft would `charge' so negatively. In space, there is no ground/earth and the spacecraft may charge positively or negatively, depending on the material it is made of and the interaction with the surrounding plasma environment. At 1P, the spacecraft potential of Vega-1 and Vega-2 was only a few volts, negative ($V_\text{SC}<0$) or positive ($V_\text{SC}>0$)  \citep{Pedersen1986}.  In contrast, at 67P, the warm and dense cometary electron population was responsible for the negative spacecraft potential, often between $-15$~V and $-5$~V, rarely below $-20$~V \citep{Odelstad2017}. Biased elements on solar panels attracting cold electrons also contributed to the negative spacecraft potential \citep{Johansson2020}. The presence of a negative spacecraft potential prevented low-energy electrons ($0<E_{k,e^-}<-qV_\text{SC}$, where $E_{k,e^-}$ is the electron kinetic energy and $q>0$)  to reach the spacecraft itself. In contrast, those with energies  $E_{k,e^-}>-qV_\text{SC}$  could reach the spacecraft with a reduced kinetic energy, $E_{k,e^-}+qV_\text{SC}$ (see Fig.\,\ref{fig:electron_spectrum}). This should be accounted and corrected for when analysing the RPC-IES electron spectrometer dataset \citep{Broiles2016, Galand2016}. 

The warm electron population was detected at 67P by two other instruments less sensitive to the spacecraft potential: the spherical LAngmuir Probe \citep[LAP]{Eriksson2007} and the Mutual Impedance Probe \citep[MIP]{Trotignon2007}, both part of the RPC consortium. The RPC-LAP, based on a century-old technique, provides insight on both the ion and electron populations. The principle is relatively simple. A metallic sphere is set at different voltages, scanning from $-30$\,V to $+30$\,V in a short period of time, and the total current from charged particles in the plasma collected by the probe is measured. It results in a characteristic $I-V$ curve (collected current $I$ vs applied voltage $V$) which, upon strong assumptions on the velocity distribution function of the ions and electrons, and on the ion mass, may give access to macroscopic properties of the surrounding plasma, including number density, electron temperature, and ion speed \citep{MottSmith1926}. The second instrument, RPC-MIP, is able to probe the dielectric characteristics of the plasma to derive electron number density and temperature. In active mode, it consists in two emitting/transmitting electrodes excited by an intensity source of very high impedance. The excited electrodes inject an alternating current $I(\omega)$ (where $\omega$ is the angular frequency) into the medium, here the cometary plasma environment. This induces a voltage difference $V(\omega)$ between the two other receiving electrodes, itself a function of the frequency, $\omega$. Expanding the concept of resistance to alternating currents, the ratio $Z(\omega)=V(\omega)/I(\omega)$ is therefore an impedance containing information on both amplitude and phase (complex number).  Between emitting and receiving electrodes, the surrounding plasma plays the role of a `filter' as its dielectric constant $\varepsilon$ depends on the frequency: $Z(\omega)$ is maximum near the electron plasma frequency $\omega_{p,e}$, that is, $\max Z(\omega)\approx Z(\omega_{p,e})$ where:
\begin{equation}
    \omega_{p,e}=\sqrt{\dfrac{q^2n_e}{m_e\varepsilon_0}},
    \label{eq:omega_pe}
\end{equation}
(with $m_e$ the electron mass and $\varepsilon_0$ the vacuum permittivity) as the plasma is in resonance. Indeed, $\omega_{p,e}$ corresponds to the natural frequency at which electrons oscillate in response to a small charge separation. Beside this simple approach to derive the electron density in a non-magnetised  plasma characterised by only one electron temperature, more elaborated processing taking advantage of the full RPC-MIP mutual impedance spectra as a function of frequency can be applied to the dataset in order: (1) to derive electron density and temperature for two different electron populations, here warm and cold (when the densities are high enough and Debye length suitable for the operation mode). Indeed, the Debye length is defined by
\begin{equation}
    \lambda_{\text{Debye}}=\sqrt{\dfrac{\varepsilon_0 k_B T_e}{q^2 n_e}}=\dfrac{v_{\text{th},e}}{\omega_{p,e}}
    \label{eq:l_Debye}
\end{equation}
{where $v_{\text{th},e}=\sqrt{k_BT_e/m_e}$ is the electron thermal speed. $\lambda_{\text{Debye}}$ is a characteristic plasma length which corresponds to the typical distance at which electromagnetic waves are screened and damped by the surrounding plasma (here mainly electrons): RPC-MIP operates efficiently when the emitter-receiver distance is greater than $2\lambda_{\text{Debye}}$}, (2) to take into account the magnetisation of the environment, and/or (3) to include the presence of an ion sheath surrounding the spacecraft and the probe as a result of the negative spacecraft charging \citep{Gilet2017, Gilet2020, Wattieaux2019, Wattieaux2020}.
In passive mode, the emitting antennas are turned off and the plasma plays the role of a natural emitter of electromagnetic waves. The spectra measured corresponds to the Fourier Transform of the voltage in this case.

At 67P at low activity ($>2$~au), the cometary plasma is dominated by warm electrons, with a mean energy typically around 5--10~eV but varying between 2--20~eV \citep{Wattieaux2020}. Observations from RPC-LAP \citep{Edberg2015}, RPC-MIP \citep{Heritier2017}, and RPC-IES \citep{Myllys2019} showed that the (warm) electron number density $n_{e,\text{warm}}$ followed a $1/r$ dependence with cometocentric distance.  \citet{Galand2016}, \citet{Heritier2017}, and \citet{Heritier2018} demonstrated that the warm electrons were produced by electron-impact ionisation and, to a lesser extent, photoionisation, and are lost through transport: The plasma density is well captured by Eq.~\ref{eq:multi_ni} following a $1/r$ dependence (see Section~\ref{section:2:4:1}). The warm electron number density was found to be slightly decreasing with increasing heliocentric distance at Rosetta's position, a result of the increase in $r$ with heliocentric distance throughout the mission. It is more pronounced when the electron number density is normalised to a given cometocentric distance $r$ \citep{Myllys2019}.
Moreover, $n_{e,\text{warm}}$ increased during solar events due to an increase in the ionising electron population \citep{Hajra2018}. Its dependence with season and hemisphere is discussed in \citet{Heritier2018}.

Near perihelion, the warm population around 67P was primarily driven by photoionisation \citep{Heritier2018}. The newborn, warm photoelectrons are quickly cooled through collisions with neutral species and the cometary electron population becomes \emph{cold} \citep{Gilet2020}. The relative importance of the warm versus cold electron populations observed in the inner coma of 67P during the Rosetta mission is reviewed in Section~\ref{section:3:1:2}. 
\subsubsection{From warm to cold cometary electrons\label{section:3:1:2}}

Since the newborn electrons have average energies ($\sim10$~eV, see Section~\ref{section:3:1:1}) well above the cometary neutrals \citep[$0.01$~eV near the surface, see][and the chapters by Marschall et al. and Biver et al. in this volume]{Gombosi1986}, if the coma is dense enough, electrons may undergo efficient energy degradation through collisions with neutrals. 
The lowest electron temperature which can be reached is the temperature of the cometary neutrals.

Evidence of cold electrons was provided by electron spectrometer measurements from the ICE spacecraft in the tail of 21P \citep{MeyerVernet1986, Zwickl1986}. Indirect evidence during 1P's flyby by Giotto near 1~au from the analysis of observed ion composition by the use of an ionospheric model and observed neutral densities and composition \citep{Eberhardt1995}: They derived electron temperature as low as $10^2$~K ($\sim0.1$~eV) at a cometocentric distance of 2000~km. 

At 67P, the RPC-LAP and RPC-MIP sensors detected cold electrons \citep{Eriksson2017,Gilet2017}.  Both instruments have their own limitations in the cold electron detectability  \citep[i.e. the highly negative potential of Rosetta when the cold electron density is high can completely hide cold electrons from RPC-LAP, while assumptions used in the complex model used for RPC-MIP analysis of cold electrons may not always be valid, for example when the ratio of warm and cold temperatures is too small, see][]{Gilet2020, Wattieaux2019}. Despite these limitations, general trends were inferred \citep{Engelhardt2018, Odelstad2018, Gilet2020, Wattieaux2020}. Though largely fluctuating, the contribution of the cold electrons to the total electron population was observed to mainly increase with local outgassing (hence latitude through season) and to decrease with cometocentric distance $r$ and heliocentric distance $r_h$. Cold electrons were always detected when Rosetta was within the diamagnetic cavity where cold electrons made up close to 100\% of the total electron density. Post-perihelion, the largest contribution of cold electrons was found to be over the southern, summer/autumn hemisphere, reaching values up to 70\% to 90\%  of the total electron population, with temperatures as low as 0.05~eV and number densities as high as 1000\,cm$^{-3}$. \citet{Gilet2020} ruled out the influence of the neutral composition as H$_2$O (abundant in the northern hemisphere) and CO$_2$ (more abundant in the southern hemisphere) have similar momentum-transfer cross sections. However, one may note that H$_2$O has higher total cross-sections than CO$_2$ (see Fig.~\ref{fig:electron_spectrum}): Electrons, on average, lose more momentum by colliding with CO$_2$ than with H$_2$O, although collisions with the former are scarcer.
     
\begin{figure}[ht]
    \centering
    \includegraphics[width=\linewidth,trim=20px 800px 130px 720px,clip]{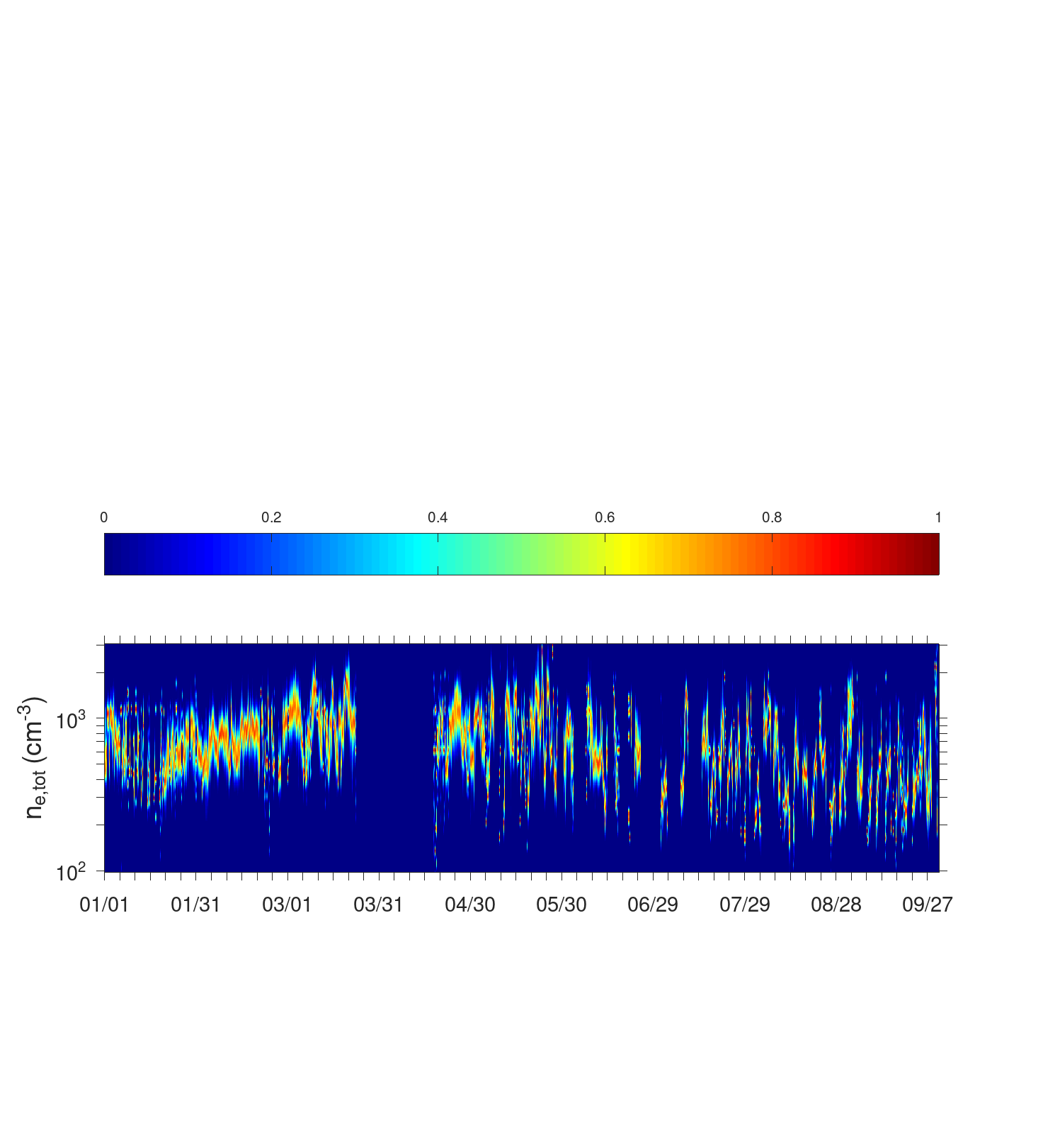}\\
    \includegraphics[width=\linewidth,trim=20px 470px 130px 800px,clip]{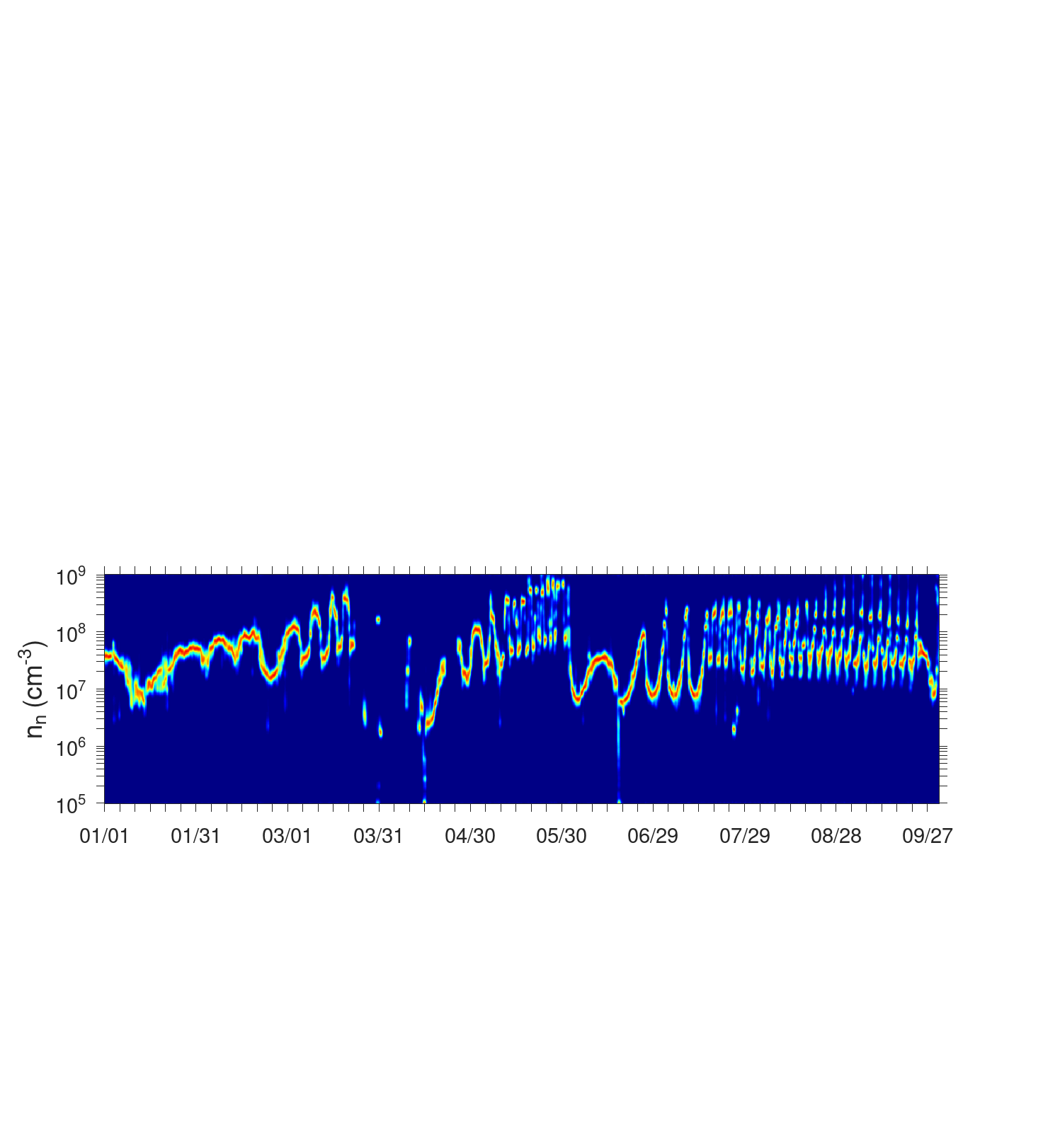}\\
    \includegraphics[width=\linewidth,trim=20px 370px 130px 900px,clip]{Chapter_latex/figures/Fig_11a.png}\\
    \includegraphics[width=\linewidth,trim=20px 470px 130px 800px,clip]{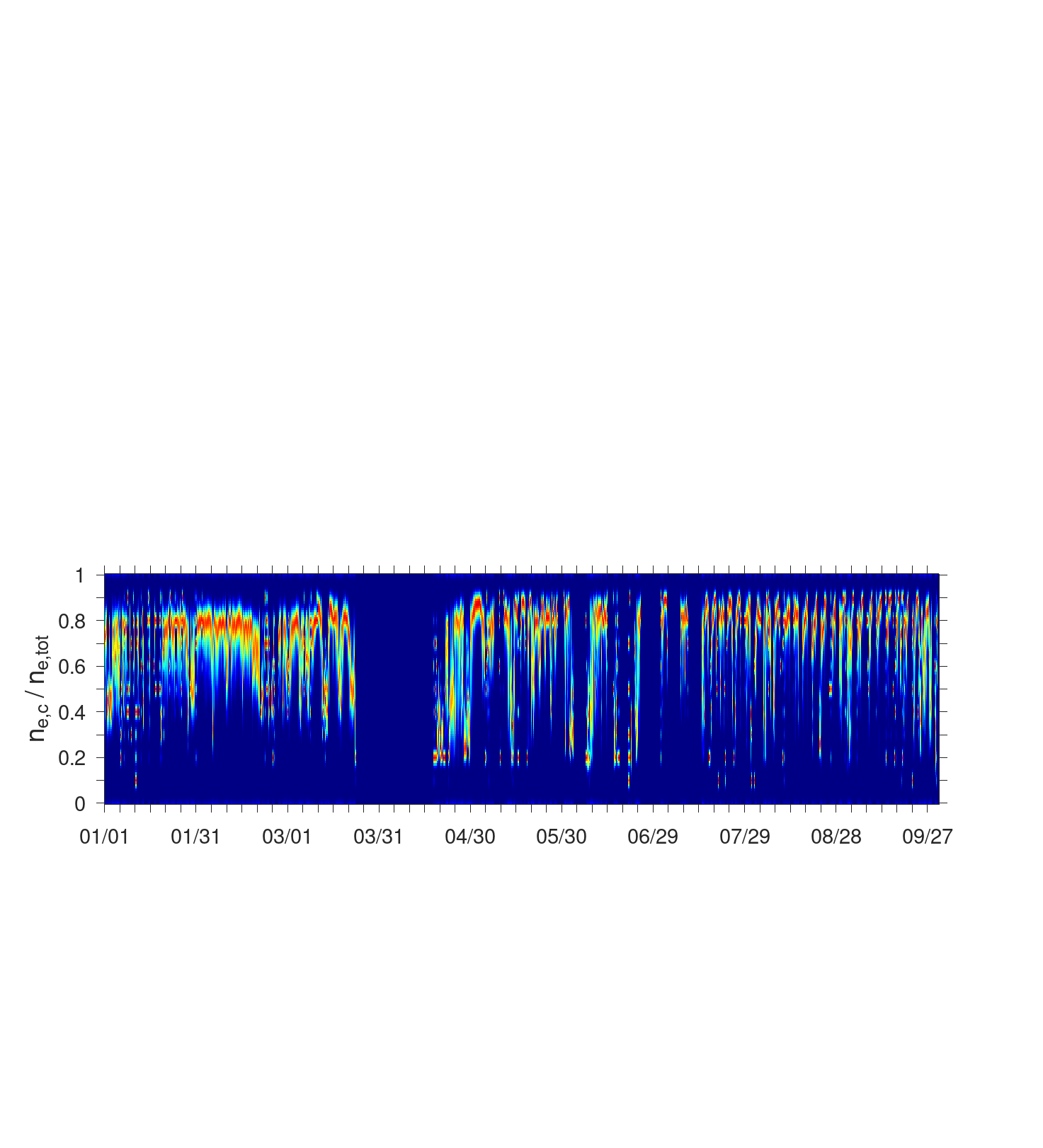}\\
    \includegraphics[width=\linewidth,trim=20px 470px 130px 800px,clip]{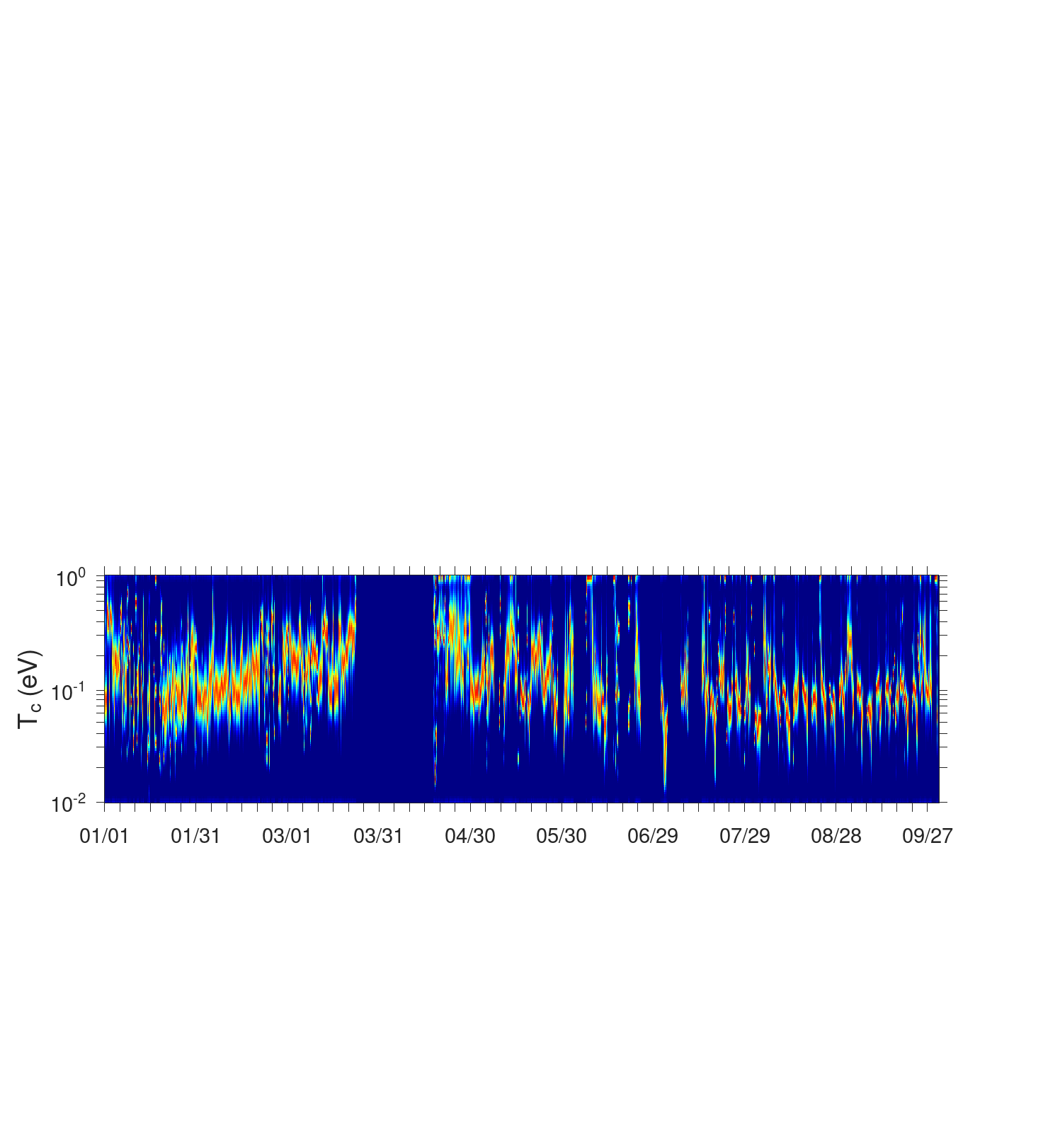}\\
    \includegraphics[width=\linewidth,trim=20px 420px 130px 800px,clip]{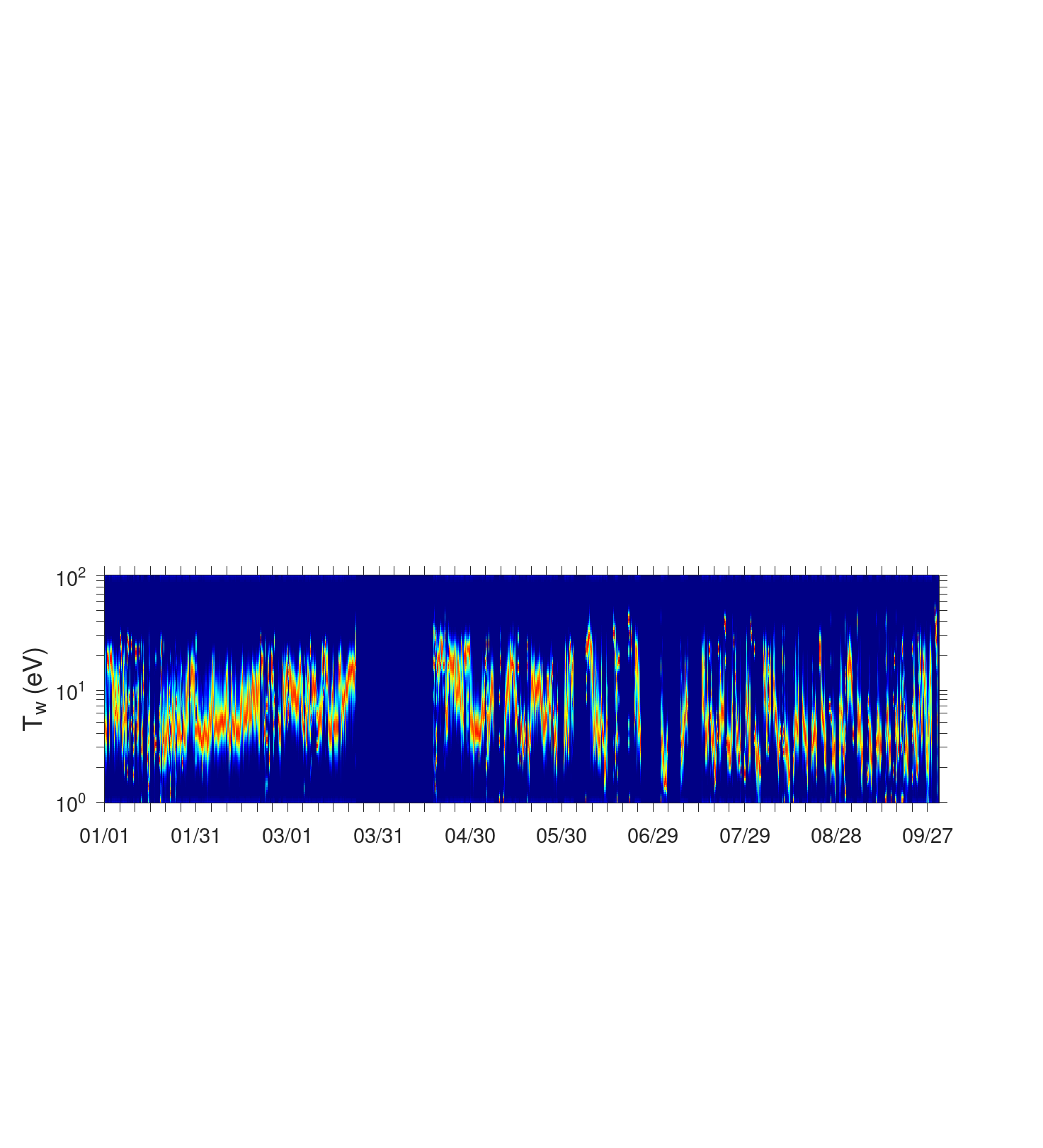}\\
    \caption{Evolution of the total neutral number density $n_n$, electron number density $n_e$, fraction of cold electrons $n_{e,\text{cold}}/n_{e,\text{total}}$, and temperatures of both populations $T_\text{cold}$ and $T_\text{warm}$ from January to September 2016 (DD/MM) at Rosetta. The dispersion of the results was computed and normalised over time intervals of 6 h. Colour bar corresponds to the occurrence. Adapted from \citet{Wattieaux2020}.}
    \label{fig:cold_e}
\end{figure}

At 1P, the presence of cold electrons was predicted by ionospheric models and showed that, for high outgassing activity,  the bulk of the cometary plasma is cold, with electron temperatures reaching values as low as the neutral temperature within the diamagnetic cavity (See Section\,\ref{section:3:3}), whereas warm electrons, constantly produced, but in minority, continue to be present \citep{Mendis1985, Korosmezey1987, Gan1990, Eberhardt1995}. At larger cometocentric distances, the electron temperature is expected to increase as: ($i$) electron and neutral densities decrease, hence their collision probability decreases and ($ii$) there is efficient heating of the cold electrons through Coulomb collisions between the cold electrons and the newborn, more energetic, electrons.

In order to assess where electrons may undergo significant collisions, a heuristic approach may be applied. Similar to planetary atmospheres, one may estimate the position of the so-called \emph{exobase} \citep[also \emph{critical level},][]{Jeans1923,Spitzer1949,Chamberlain1963}, that is, the boundary where the gas (here electrons) transitions from a continuum flow (collisional) to a free-molecular regime (collisionless). It is also defined as the critical limit where the atom/molecule undergoes one collision on average along its path outwards to infinity. For this, one may evaluate the \emph{Knudsen} number:
\[
\text{Kn}\sim \dfrac{\ell_{e^-,\text{mfp}}}{L}
\]
where $\ell_{e^-,\text{mfp}}$ stands for the mean free path (i.e. the mean distance covered by a particle between two successive collisions, here $(n_n \sigma_{e,n})^{-1}$ where $\sigma_{e,n}$ is the electron-neutral total cross section as electrons mainly collide with neutrals) and $L$ stands for a characteristic length (for instance, the electron number density scale height $|\nabla \log n_i|^{-1}$, here $\sim r$). The electron exobase is expected to be located at a comeocentric distance:
\begin{equation}
    \text{Kn}\sim 1 \longrightarrow r_{e^-, \text{exo}} = L \sim \frac{1}{n_n \sigma_{e,n}}
    \label{eq:el_exobase}
\end{equation}
Cold electrons were expectedly detected near perihelion \citep[around 1.24~au, with $Q\sim 10^{28}$~s$^{-1}$,][]{Hansen2016,Engelhardt2018, Gilet2020} when the electron exobase predicted by Eq.~\ref{eq:el_exobase} can reach a few hundred kilometres.

Surprisingly, cold electrons were also detected  away from perihelion beyond heliocentric distances of 3~au \citep[$Q<10^{26}$~s$^{-1}$,][]{Gilet2020}, when the electron exobase predicted by Eq.~\ref{eq:el_exobase} is located below the surface \citep{Engelhardt2018,Gilet2020}. 
 Hence, no efficient cooling of the electrons by collisions with neutrals was anticipated. 
The simple approach based on the Knudsen number implies a radial flow for the electrons, which is not sufficient to generate a cold electron population as observed with Rosetta \citep{Gilet2020}.
3D modelling from Particle-in-Cell (PiC) simulations, applied to low outgassing comets, have shown that electrons are trapped in the close environment of the nucleus \citep{Deca2017,Deca2019,Sishtla2019}. This region is set up by an ambipolar electric field produced by the cometary plasma. Based on a 3D kinetic test particle model, \citet{Stephenson2022}  demonstrated that the trapping increases significantly the path of electrons in the dense coma close to the nucleus: This allows electrons to undergo substantial collisions and become cold despite low neutral densities. In the light of the complex, electromagnetic environment at low outgassing, the electron exobase needs to be moved closer to the nucleus than originally predicted through the Knudsen number.

\subsubsection{Accelerated electrons\label{section:3:1:3}}

The most energetic electron population is often referred to as \emph{suprathermal}, either as their energy exceeds $k_BT_e$ where $T_e$ is the mean temperature of the whole electron population, or as their energy distribution departs from a Maxwellian distribution at high energies (that is, $f_e(E)\,\mathrm{d}E\not\propto \sqrt{E} \exp(-E/k_BT_e)\,\mathrm{d}E$). Note that the warm electron distribution in energy also departs from a Maxwellian distribution. The electron population under focus in this section corresponds to electrons with energies above the mean energy of the whole electron population dominated by cold and warm electrons (and above the upstream solar wind electrons, i.e. around 10~eV). Their energy distribution tends not to decrease exponentially (i.e. $f_e(E)\not\propto \sqrt{E} \exp(-E/k_BT_e)$) but rather follows a power law (i.e. $f_e(E)\propto E^{-\kappa}$, as in kappa distributions). The reason for such a behaviour in a plasma and its possible ubiquity in nature are still debated and beyond the scope of this chapter  \citep{Pierrard2016}.

Let us review the findings on the \emph{hot} electron population detected by Rosetta by the RPC-IES electron spectrometer \citep{Broiles2016, Broiles2016b, Myllys2019}. It was derived by fitting a suprathermal double-kappa function on the electron velocity distribution. From such a fitting, characteristics of the warm and hot populations were derived. 
While the bulk of the electron population is dominated by warm and/or cold electrons (see Section~\ref{section:3:1:1} and Section~\ref{section:3:1:2}), the hot electrons with temperatures of a few tens of eV are rarefied with typical densities below a few cm$^{-3}$  \citep{Myllys2019}.
Their density was found to decrease significantly from 3-10~cm$^{-3}$ near perihelion down to less than 0.1~cm$^{-3}$ at large heliocentric distances \citep[3.0-3.8~au,][]{Broiles2016, Myllys2019}. Their temperature does not exhibit any strong dependence with heliocentric distance or cometocentric distance. The dependence of the hot population with $r$ is significantly less than for the warm population, with a dependence more in $r^{-2}$ than in $r^{-1}$, though the uncertainty is large and the dependence breaks below 50~km.
The energy distribution measured by RPC-IES needs to be corrected to take into account the spacecraft potential which is ranging between $-20$ to $-5$~eV typically (see Fig.\,\ref{fig:electron_spectrum} and Section~\ref{section:3:1:1}). 

The source of the hot electron population has been under debate. It was first proposed by \citet{Broiles2016} to be suprathermal halo solar wind electrons \citep{Pierrard2016}. \citet{Myllys2019} however showed that the  hot electron population observed at Rosetta is denser (by almost an order of magnitude) and colder than the solar wind halo component. Furthermore, they found an increase in the electron temperature with increasing invariant kappa index, an opposite trend to the one derived for the halo solar wind electrons \citep{Pierrard2016}. Hence, the bulk of the hot electrons has a different origin. 

At large heliocentric distances, \cite{Madanian2016} suggested that electrons could be accelerated towards the comet by an ambipolar electric field and that this field was the result of the electron pressure gradient due to plasma inhomogeneity observed close to the comet. When an electron is born far from the comet or comes from the space environment, such as the solar wind electrons, it is accelerated towards the cometary nucleus when falling into the potential well set up by the ambipolar electric field. 
Using a self-consistent collisionless PiC model applicable to weakly outgassing comets, \citet{Deca2017} showed that the suprathermal electrons present close to the comet are originating from the solar wind and  accelerated by the ambipolar electric field set up by cometary plasma.
This explanation is consistent with the large-scale acceleration process for the source of the suprathermal electrons responsible for auroral emissions \citep{Galand2020,Stephenson2022}.  These  atomic emissions observed in the Far UltraViolet (FUV) by the UV spectrograph Alice onboard Rosetta were shown to result from the dissociative excitation of molecules, such as water and CO$_2$, by energetic ($>15-20$~eV) electrons. Through a multi-instrument analysis, it was found that the variation in the FUV brightnesses correlated very closely with the variation in the electron-impact emission frequency, driven by the RPC-IES electron flux. This remained true even during solar events and when the FUV emissions were observed at the limb, far from the location of Rosetta where the RPC-IES was measuring the suprathermal electrons. These accelerated solar wind electrons play a critical role in the cometary environment, as they are responsible not only for the auroral emissions, but also for the bulk of the ionisation, and hence of the cometary plasma, at large heliocentric distances  \citep[see][and Section~\ref{section:2:2:2}]{Galand2016, Heritier2018}. 

Near perihelion, the source of the suprathermal electrons is still under debate. Indeed, with the formation of a dense ionosphere under solar illumination, a diamagnetic cavity (see Section\,\ref{section:3:3}) is formed around the cometary nucleus and solar wind electrons do not have (easy) access to the inner ionosphere any more. 
\cite{Madanian2017} observed a modest but consistent drop in the
RPC-IES electron fluxes over the 40~eV to 100~eV range between outside and inside the diamagnetic cavity. They showed that the lower fluxes measured inside the cavity are however too high to be explained by photoelectrons alone. They proposed two mechanisms to explain the observations: a trapping mechanism of the photoelectrons inside the cavity and/or the penetration of part of the solar wind electrons into it. 


\subsection{Effect on the ion dynamics\label{section:3:2}}

     Electrons play a critical role in the plasma and ion dynamics. Indeed, as their mass is much lower than that of positive charges, such as protons, they have a greater mobility and react quicker to changes in the electromagnetic fields. 
     
     Ion dynamics is largely driven by macroscopic electric fields. In cometary and classical (i.e. not relativistic) plasmas, quasi-neutrality holds ($n_i\approx n_e$) and hence the charge density ($q(n_i-n_e)$) is very small at large temporal and spatial scales. This means that the Maxwell-Gauss equation is reduced to $\nabla \cdot \vec{E}\approx 0$ and it cannot be used to derive the electric field (no unique solution in absence of boundary conditions). An alternative equation needs to be used; this is the Generalised Ohm's Law (GOL) which can be derived from the electron momentum equation as follows.  
     For the electron momentum density $n_em_e\vec{V}_{e}$, its temporal evolution in conservation form is given by:
     \begin{align}
         m_e\dfrac{\partial (n_e \vec{V}_e)}{\partial t}=\:-&\overbrace{\nabla\cdot \mathsf{P}_e\vphantom{)}}^{\mathsf{P}_e  \text{ gradient}}-\overbrace{\nabla\cdot (n_e m_e \vec{V}_e\otimes\vec{V_e})}^{\text{Electron dynamic pressure gradient}}\nonumber\\
         &\underbrace{-qn_e(\vec{E}+\vec{V}_e\times\vec{B})}_{\text{Lorentz force}}+\underbrace{n_em_e\,\vec{a}_{ext}}_\text{External forces}\nonumber\\
         &+\dfrac{\delta \mathcal{P}_e}{\delta t}
         \label{eq:continuityElectrons}
     \end{align} 
    where $\vec{a}_{ext}=\vec{F}_\text{ext}/m_e$ with $\vec{F}_\text{ext}$ the set of forces acting upon electrons which are not electromagnetic and independent of the environment, for instance gravity,
    $\mathsf{P}_e$ and $\vec{V}_e\otimes\vec{V}_e$ are tensors of rank 2, which are $3\times3$ matrices in 3D, and ${\delta \mathcal{P}_e}/{\delta t}$ encompasses all microscopic source and loss of electron momentum, for instance frictional interactions between electrons and neutrals/ions \citep{Cravens1997book,Szego2000}. 
    
    The electron number density $n_e$ (a scalar), the electron mean (bulk) velocity $\vec{V}_e$ (a vector), and the electron thermal pressure $\mathsf{P}_e$ (a matrix) are mathematically defined respectively as the zeroth, first, and second central moments of the electron distribution function over velocity space and are given by:
     \begin{align}
        n_{e}(\vec{r})=& \int_{\mathbb{R}^3}f_e(\vec{r},\vec{v})\,\mathrm{d}^3\vec{v} \\
        V_{e,j}(\vec{r})=& \frac{1}{n_e(\vec{r})}\int_{\mathbb{R}^3} \! v_jf_e(\vec{r},\vec{v})\,\mathrm{d}^3\vec{v}\\
        \mathsf{P}_{e,jk}(\vec{r})=& m_e\int_{\mathbb{R}^3}\! [v_j-V_{e,j}(\vec{r})][v_k-V_{e,k}(\vec{r})]f_e(\vec{r},\vec{v})\,\mathrm{d}^3\vec{v} \\
        [\vec{V}_e\otimes\vec{V}_e]_{jk}=&V_{e,j}(\vec{r})V_{e,k}(\vec{r})
    \end{align}
    where $\vec{v}$ is the particle's velocity vector, 
    a microscopic quantity to distinguish from the particle's bulk velocity $\vec{V}(\vec{r})$, a macroscropic (averaged) quantity at the location $\vec{r}$.
    $j$ and $k$ are indices for the different components (e.g. $j,k=x,y,z$ and $\mathrm{d}^3\vec{v}=\mathrm{d}v_x\mathrm{d}v_y\mathrm{d}v_z$ in Cartesian coordinates, 9 possible combinations in 3D, but there are different for other geometries). $f_e(\vec{r}.\vec{v})$, in electrons\,m$^{-6}$\,s$^{3}$, is a fundamental function which contains all the properties of the electrons: It is the so-called velocity distribution function and corresponds to the density of particles (here electrons) in the volume element $\mathrm{d}^3\vec{r}\,\mathrm{d}^3\vec{v}$ around the position $(\vec{r},\vec{v})$ in the space-velocity domain (6 dimensions). $f_e$ gives a continuous statistical description of the behaviour of the electrons with their number at a given velocity and at a given position. $f_e(\vec{r},\vec{v})/n_e(\vec{r})$ may be interpreted as a \emph{probability distribution function} in the mathematical sense. As such, $f_e$ can be reconstructed from moments at all orders. However, this is impossible in practice: Only a limited set of equations can be solved and moments can be known or approximated. In addition, the temporal evolution of the moment of order $n$ always involves the divergence of the moment of order $n+1$ (Eq.~\ref{eq:continuityElectrons}). This is the reason why most mathematical models (e.g. MHD approach) apply closures which approximate the expressions of high order moments in terms of low order ones \citep[e.g. see][for a mathematical description]{Grad1949,Chapman1970}.
    
    Two simplifying assumptions may be done here: steady state and massless electrons. As electrons have a very low mass, they do not have inertia nor contribute to the dynamic pressure. In addition, at comets, forces such as gravity are irrelevant. Within a cometary plasma, the thermal energy is mainly carried by the electrons, while the dynamic pressure is carried by the ions. If we ignore microscopic source and loss terms of electron momenta and with the aforementioned assumptions, Eq.\,(\ref{eq:continuityElectrons}) becomes:
         \begin{equation}
         0\approx-\nabla\cdot \mathsf{P}_e-qn_e\vec{E}-qn_e\vec{V}_e\times\vec{B}\label{eq:continuityElectronsSimple}
    \end{equation}
    Instead of writing in terms of electron velocities, it is more common to use $\vec{V}$, the mean plasma velocity, and $\vec{J}$, the current density, given by:
    \begin{eqnarray}
        \vec{V}&=&\dfrac{n_im_i\vec{V}_i+n_em_e\vec{V}_e}{n_im_i+n_em_e}\nonumber\\
         \vec{J}&=&qn_i\vec{V_i}-qn_e\vec{V}_e\nonumber
    \end{eqnarray}
    and conversely
    \begin{equation*}
        \vec{V}_i\approx \vec{V}\text{ and }   \vec{V}_e\approx\vec{V}-\dfrac{\vec{J}}{qn_e}
    \end{equation*}
    if $n_i\approx n_e$ and as $m_e\ll m_i$. Eq.\,(\ref{eq:continuityElectronsSimple}) becomes:
    \begin{equation}
         0\approx-\nabla\cdot \mathsf{P}_e-qn_e\vec{E}-qn_e\vec{V}\times\vec{B}+\vec{J}\times\vec{B}\label{eq:generalisedOhmLaw00}
    \end{equation}
    that is
    \begin{equation}
         \vec{E}\approx\overbrace{-\vec{V}\times\vec{B}\vphantom{\dfrac{1}{qn_e}}}^{\substack{\text{\vphantom{p}Motional}/\\\text{\vphantom{p}convective}}}\overbrace{-\frac{1}{q n_e}\nabla\cdot \mathsf{P}_e}^{\substack{\text{Ambipolar}/\\\text{polarisation}}}\overbrace{+\frac{1}{q n_e}\vec{J}\times\vec{B}}^\text{Hall term\vphantom{p}}\label{eq:generalisedOhmLaw}
    \end{equation}
    known as the GOL. It is not the most rigorous demonstration and, for more details, the reader is referred to \citet{Somov2007} or \citet{Valentini2007}, for example. This equation is fundamental as it provides a relatively good, unique, and physical estimate of the electric field without solving the Maxwell-Gauss equation at large temporal and spatial scales. The contributions of the different terms have been assessed at low outgassing activity by \citet{Deca2019} and are detailed in the chapter by Götz et al. in this volume.
    
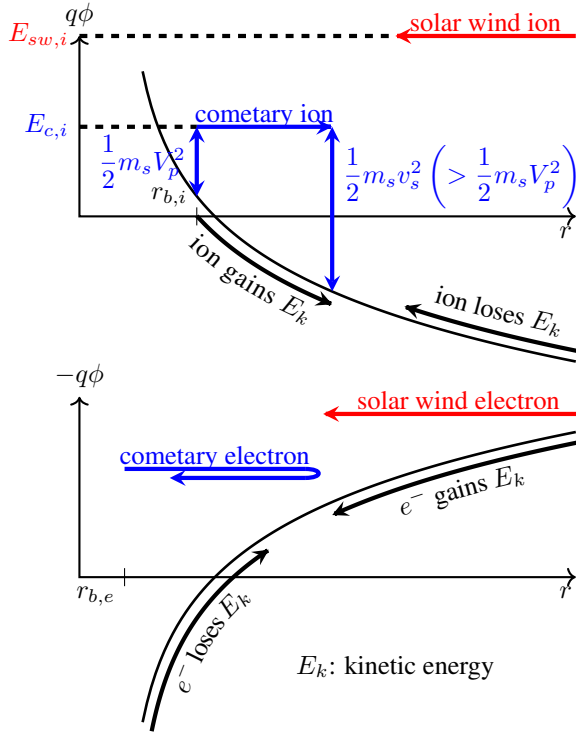
\begin{figure}[ht]
\begin{tikzpicture}[scale=1.2]
    \clip (-1.5,-6) rectangle (5,3);
    \draw [<->,thick] (-0.5,2) node (yaxis) [above] {$q\phi$}
        |- (5,0) node (xaxis) [pos=0.99,below] {$r$};
    \draw [-,thick] (0,-2) node {};
    \draw[domain=0.2:5,samples=101,line width=1] plot (\x,{-ln(\x)});
    \draw[-stealth,domain=0.5:2,samples=101,ultra thick,blue,postaction={decorate, 
                decoration={text along path,text color=blue, raise=0.2em, text={cometary ion},
                text align={align=center}}}] plot (\x+0.3,{-ln(0.5)+0.3});
    \draw[stealth-,domain=3:5,samples=101,ultra thick,red,postaction={decorate, 
                decoration={text along path,text color=red, raise=0.2em, text={solar wind ion},
                text align={align=center}}}] plot (\x,{2});
    \draw[dashed,ultra thick] (-0.5,{-ln(0.5)+0.3}) -- (0.8,{-ln(0.5)+0.3})node[pos=0,anchor=east] {\color{blue}$E_{c,i}$}; 
    \draw[dashed,ultra thick] (-0.5,2) -- (3,2)node[pos=0,anchor=east] {\color{red}$E_{sw,i}$}; 
    \draw[stealth-stealth,ultra thick,blue] (0.8,{-ln(0.8)}) -- (0.8,{-ln(0.5)+0.3}) node[pos=0.5,anchor=east]{$\dfrac{1}{2} m_s V_p^2$};
    \draw[stealth-stealth,ultra thick,blue] (2.3,{-ln(0.5)+0.3}) -- (2.3,{-ln(2.3)}) node[pos=0.3,anchor=west]{$\dfrac{1}{2} m_s v_s^2\left(>\dfrac{1}{2} m_s V_p^2\right)$};
    \node[anchor=south east] (C) at (0.8,0){$r_{b,i}$};
    \draw[-] (0.8,0.1) -- (0.8,-0.1);
    \draw[stealth-,domain=3:5,samples=101,ultra thick,black,postaction={decorate, 
                decoration={text along path, text color=black,raise=0.5em,text={ion loses {$E_k$}},
                text align={align=center}}}] plot (\x+0.1,{-ln(\x)+0.1});
    \draw[-stealth,domain=0.9:2.4,samples=101,ultra thick,black,postaction={decorate, 
                decoration={text along path, text color=black,raise=-1em,text={ion gains {$E_k$}},
                text align={align=center}}}] plot (\x-0.1,{-ln(\x)-0.1});
\begin{scope}[yshift=-4cm]
    \clip (-1.5,-3) rectangle (5,3);
    \draw [<->,thick] (-0.5,2) node (yaxis) [above] {$-q\phi$}
        |- (5,0) node (xaxis) [pos=0.99,below] {$r$};
    \draw [-,thick] (0,-2) node {};
    \draw[domain=0.2:5,samples=101,line width=1] plot (\x,{ln(\x)});
    \draw[stealth-,domain=2:5,samples=101,ultra thick,red,postaction={decorate, 
                decoration={text along path, text color=red,raise=0.2em,text={solar wind electron},
                text align={align=center}}}] plot (\x+0.2,{ln(5)+0.2});
    \draw[stealth-,domain=2.2:5,samples=101,ultra thick,black,postaction={decorate, 
                decoration={text along path, text color=black,raise=-1em,text={{$e^-$} gains {$E_k$}},
                text align={align=center}}}] plot (\x+0.1,{ln(\x)-0.1});
    \draw[-stealth,domain=0.2:1.5,samples=101,ultra thick,black,postaction={decorate, 
                decoration={text along path, text color=black,raise=-1em,text={{$e^-$} loses {$E_k$}},
                text align={align=center}}}] plot (\x+0.1,{ln(\x)-0.1});
    \draw[-,domain=0:2,samples=101,ultra thick,blue,postaction={decorate, 
                decoration={text along path, text color=blue,raise=0.2em,text={cometary electron},
                text align={align=center}}}] plot (\x,{1.2});
    \draw[stealth-,blue,ultra thick] (0.5,1.1)--(2,1.1);
    \draw[-,blue,ultra thick] (2,1.2) .. controls (2.2,1.2) and (2.2,1.1) .. (2,1.1);
    \node[anchor=north east] (C) at (0,0){$r_{b,e}$};
    \draw[-] (0,0.1) -- (0,-0.1);
    \node at (3,-1){$E_k$: kinetic energy};
\end{scope}
\end{tikzpicture}
\caption{Schematic of how ions and electrons may gain/lose energy by means of the electric potential $\phi$ (black line). This potential is driven by the electron pressure and is plotted as a function of the cometocentric distance $r$. Solar wind electrons gain energy as they fall and dive towards the nucleus, while cometary electrons born close to the nucleus at $r_{b,e}$ ($<10-100~r_c$) are trapped by the potential well. In contrast, solar wind ions are barely decelerated and cometary ions are accelerated, though the gain in kinetic energy  will depend on their birthplace $r_{b,i}$ and the potential profile. Magnetic field and collisions have been ignored here. The red and blue arrows show the direction of motion of the particles.
\label{3_2_fig3}}
\end{figure}

    In Eq.~\ref{eq:generalisedOhmLaw}, one term is dominating the ion acceleration in the inner coma, within the diamagnetic cavity, and along the magnetic field: the ambipolar/polarisation electric field, defined as $\vec{E}_{\mathsf{P}_e}\approx-\nabla\cdot \mathsf{P}_e/(q n_e)$. In the inner coma, electron and ion densities are dominated by the population of cometary origin which only depends on the cometocentric distance (assuming spherical symmetry). Let us extend this assumption to the electron pressure and  temperature. In such a case, $\vec{E}_{\mathsf{P}_e}$ is only radial and its amplitude is given by:
    \begin{equation}
        {E}_{\mathsf{P}_e}\approx -\dfrac{1}{qn_e}\dfrac{\partial \mathsf{P}_{e,rr}}{\partial r}-\dfrac{2\mathsf{P}_{e,rr}-\mathsf{P}_{e,\theta\theta}-\mathsf{P}_{e,\phi\phi}}{r} \label{eq:ambipolarField}
    \end{equation}
    where $\mathsf{P}_{e,rr}$ stands for the electron pressure component along $r$ and $\mathsf{P}_{e,\theta\theta}$ and $\mathsf{P}_{e,\phi\phi}$ stand for both perpendicular components. In spherical symmetry, $\mathsf{P}_{e,\theta\theta}=\mathsf{P}_{e,\phi\phi}$ with $\theta$ and $\phi$ the polar and azimuthal angles in spherical, polar coordinates. If the electron temperature is isotropic with a temperature $T_e(r)$ (i.e. $\mathsf{P}_e=P_e(r)\,\mathsf{I}_3=n_e(r)\,k_B\,T_e(r)\,\mathsf{I}_3$ where $\mathsf{I}_3$ is the identity matrix and $P_e(r)=n_e(r)\,k_B\,T_e(r)$ is a scalar), Eq.~\ref{eq:ambipolarField} is reduced to \citep{Cravens1984}:
    \begin{equation}
        {E}_{\mathsf{P}_e}\approx-\dfrac{1}{qn_e}\dfrac{\mathrm{d} P_e}{\mathrm{d} r}=-\dfrac{k_BT_e}{q}\dfrac{\mathrm{d} \log n_e}{\mathrm{d} r}-\dfrac{k_B}{q}\dfrac{\mathrm{d} T_e}{\mathrm{d} r}
    \end{equation}
Neglecting the electron temperature gradient and as $n_e(r)\propto 1/r$ from theory and observations \citep[e.g.  see][]{Balsiger1986,Edberg2015,Beth2019}, one gets \citep{Vigren2015}:
    \begin{equation}
    {E}_{\mathsf{P}_e}\approx \dfrac{k_BT_e}{qr}
    \label{eq:Eambi}
    \end{equation}
    As the electrons have often more energy in the form of thermal agitation (though it may not be valid in the inner coma of 1P), electrons can leave the coma faster than the ions. However, quasi-neutrality should be maintained and $\vec{E}_{\mathsf{P}_e}$ ensures to keep it that way, in particular along the magnetic field lines as other components act perpendicular to $\vec{B}$. $\vec{E}_{\mathsf{P}_e}$ appears mainly radial, oriented outwards near the nucleus: it decelerates escaping cometary electrons inwards and accelerates cometary ions outwards. Conversely, ions coming from outside would be deflected and repelled from the nucleus, while electrons would be attracted and accelerated as they approach the nucleus (see Fig.~\ref{3_2_fig3}).
    
    Eq.~\ref{eq:Eambi} should be considered with great caution as it is based upon some assumptions (e.g. constant and isotropic $T_e$ through the coma)\ difficult to verify with a single spacecraft and, above all, a negatively charged one. It provides insight on how the electric field and associated potential might behave in the inner ionosphere but should not be applied over large scales (e.g. from the surface to the infinity in terms of cometocentric distances) or at low outgassing activity. Indeed, the associated potential would be in $\log(r)$ trapping all cometary electrons, preventing them to reach infinity while cometary ions could escape, leading to a charge imbalance. In addition, PiC simulations performed by \citet{Deca2019} at low outgassing activity show that $\vec{E}_{\mathsf{P}_e}$ is not spherically symmetric near the nucleus (despite having assumed a spherically symmetric outgassing) and hence it cannot be described by Eq.~\ref{eq:Eambi}. 
    
From this approach, one would expect ions to be accelerated and exceed the local neutral speed at which they are born in the first tens or hundreds kilometres above the surface. However, this was not the case at 1P for example. \citet{Schwenn1987} showed that the radial component of the ion velocity at 1P remained steady, close to the neutral speed between the closest approach and 20,000~km. This means either that ions and neutrals are collisionally coupled and ions might not experience any radial acceleration by means of electromagnetic forces. As aforementioned, $\vec{E}_{\mathsf{P}_e}$ depends on the electron/plasma temperature. \citet{Eberhardt1995} showed that the plasma temperature in the same region was cold, increasing gradually from 100~K (0.01~eV) at 2000~km to 900~K (0.09~eV) at 8500~km. This must be compared with the kinetic energy of newborn water ions which is around 0.1~eV. Therefore, at 1P, near the nucleus, newborn ions are slightly supersonic (their initial speed exceeds $\sqrt{k_B(T_i+T_e)/m_i}$). This translates into the presence of an inner shock ahead of the contact surface \citep{Cravens1989,Goldstein1989}. As ions decelerate when approaching the contact surface, supersonic ions become subsonic and a shock forms before being deflected. The existence of this shock was anticipated before Giotto \citep[see][]{Houpis1980}. However, it was difficult and hard to observe as Giotto was too fast.

The situation is different at 67P. Although cold electrons were observed during Rosetta's escort phase, the mean electron temperature was mostly around $5$--$15$~eV. Nevertheless, \cite{Gilet2020} showed that within the diamagnetic cavity the electron population was dominated by the cold component ($<0.1$~eV), consistent with previous observations by \citet{Odelstad2018}. It is then not clear whether or not ions were subsonic/supersonic at birth, if that depends on the cometary activity and location (e.g. within and outside the diamagnetic cavity) or on the ability of solar wind ions and electrons to access the inner ionosphere. This aspect should be further investigated in the future.

\subsection{Diamagnetic cavity\label{section:3:3}}

Amongst the different regions met at comets and the boundaries separating the different environments, covered in the chapter by Götz et al. in this volume, one in particular draws our attention regarding the inner cometary ionosphere: the \emph{diamagnetic cavity}. The adjective `diamagnetic' comes from an analogy to materials which, if subject to an external magnetic field, generate in response an induced magnetic field in the opposite direction in order to reduce or nullify the total magnetic field strength. The diamagnetic cavity is characterised by an almost null magnetic strength (below instrumental detection) and a steep magnetic field gradient at its edge, the latter having different nicknames in the literature: \emph{ionopause}, \emph{contact surface}, or the \emph{diamagnetic cavity boundary layer} (DCBL). This boundary is a tangential discontinuity whose existence was already anticipated by \citet{Biermann1967}.  The diamagnetic cavity was observed at two comets. The first time was during Giotto's flyby on 13/14 March 1986 \citep{Neubauer1986} when the spacecraft entered into the cavity for two minutes crossing the boundary at around 4500--5000~km \citep{Neubauer1987}. The second time was during the Rosetta mission on 26 July 2015 \citep{Goetz2016a}, followed later by the identification of more than 600 crossings \citep{Goetz2016b}. The detections were performed a few months on either side of the perihelion passage when 67P was most active; however, no detection at the location of Rosetta could be achieved at large heliocentric distances, missing the anticipated shrinking and/or disappearance of the diamagnetic cavity. 

The origin of the boundary, the physics at play at the boundary, and how to maintain it were ardently debated in the 1990s following the detection of the diamagnetic cavity at 1P during the Giotto flyby. Debates converged in one idea: As the coma of 1P was extremely dense, the force induced by the magnetic field gradient (pressure and tension) may be balanced by the force exerted by the neutrals on the ions by means of collisions, the so-called \emph{ion-neutral drag}, of which the strength depends on $|\vec{V}_i-\vec{V}_{n}|$ \citep{Cravens1986,Ip1987}. Indeed, at the microscopic level, as ions and neutrals may have very different velocities and neutrals are much more abundant, collisions tend to reduce the gap between both speeds and force the velocity of ions to be closer to that of neutrals. However, to balance the magnetic pressure in the right direction, ions are required to be slower than neutrals at the boundary and beyond, with for instance a null or even inward ion velocity.  This strong velocity requirement was in general not fulfilled at 67P during Rosetta's many hundred crossings of the boundary, questioning the validity of the simple ion-neutral drag model at low outgassing conditions. The question of whether or not this region is associated with collisions is open. Nevertheless, statistical studies have provided valuable insight. For instance, at 67P, \citet{Henri2017} showed that the DCBL crossings occurred close to the electron-neutral exobase (see Section~\ref{section:3:1:2}), meaning that collisions between electrons and neutrals might play a critical role in the formation of this cavity. Further evidence that collisions and electron cooling might have a significant role in its formation was the presence of a shock on the inner edge of the diamagnetic cavity at 1P \citep{Gombosi1996,Rubin2009}. The cause is quite well understood: Ions are born supersonic at 1P (i.e. $V_n \gtrsim \sqrt{k_BT_e/m_i}$) and as they are slowed down by the magnetic field gradient, they undergo deceleration from supersonic to subsonic speeds and a shock forms at the transition. Although $V_n$ is not expected to vary between comets, $T_e$ does and was extremely low at 1P compared with 67P. Such a low value is associated with high outgassing and cooling through collisions with the neutrals.

For a modelling perspective, it is critical to predict whether a diamagnetic cavity is formed and which shape and size it has, as the plasma dynamics in a magnetised and an unmagnetised {medium} are very different. On the one hand, in an unmagnetised medium, the plasma dynamics is driven by $\vec{E}_{\mathsf{P}_e}$ (see Section~\ref{section:3:2}) and symmetries may exist. On the other hand, in a magnetised medium, the symmetry is broken (in a spherically symmetric problem, there is no magnetic field) and the dynamics of ions and electrons becomes more complex. For instance, ions may travel radially in the former case but not in the latter. In addition, the ability to accurately model the transition between collisionless and collisional media with the same approach is challenging, computationally and mathematically. In the collisional case, kinetic physics including elastic and inelastic collisions is required.
3D modelling of the DCBL remains a key challenge to tackle in the future.

\section{Ion population\label{section:4}}

As for electrons, several ion populations, discriminated in energy and origin, were identified during past cometary flybys and during the course of the Rosetta mission. They consist of pickup ions, solar wind ions, and cold cometary ions; each may predominate depending on cometocentric distance and/or cometary activity. Instruments of choice to characterise these populations are ion mass spectrometers. Aboard Rosetta, these were: (1) ROSINA-DFMS, which provided detailed ion composition of the cometary ionosphere thanks to its high mass resolution mode (see Section \ref{section:2:4:2} for details), (2) RPC-ICA, able to determine the ion distribution function and distinguish in energy, mass, and direction between different populations (e.g. solar wind protons, solar wind alpha particles, pick up ions, cometary ions). In addition, there were the Langmuir probe, RPC-LAP, with the ability to measure the ion current and the electron-ion analyser, RPC-IES, which, although lacking the mass determination, measured the total ion fluxes throughout the mission. This section is dedicated to the different ion populations present within the cometary ionosphere in terms of energy and the processes which might affect the latter.
\subsection{Collisions versus acceleration\label{section:4:1}}

In a cometary ionosphere, energy is stored under different forms, either in the electromagnetic fields ($B^2/\mu_0$ and $\varepsilon_0E^2$) or within the particles, kinetic or thermal. For electrons ($T_e \sim10$~eV at most, see Section\,\ref{section:3:1:1}) which are light species, the energy is mainly thermal, whereas for ions, it is mainly kinetic.

Cometary ions are produced from neutrals (see Section~\ref{section:3:1:1}). During an ionisation, most of the energy goes to the newborn electron and hence the ion keeps the original momentum of the incident neutral. Thus, newborn ions have an initial speed of $\sim$0.5--1~km\,s$^{-1}$, namely a kinetic energy of 0.1--0.2~eV for ion masses spanning from 18 to 44~u. If most of the available energy is stored by the electrons in absence of electron and ion collisions,  ions should undergo acceleration and `drain' energy from the electrons by means of electromagnetic fields. However, when ion collisions are frequent and ion-neutral chemistry is dominant (see Section~\ref{section:2:4:1}), ions struggle to accelerate: An ion may collide with a neutral (e.g. charge exchange or proton transfer), the ion, which may have accelerated, becomes a neutral, the neutral a cold ion (or one of its fragments), reducing the energy of the  ion present albeit not affecting the total plasma charge. Although the produced ion may have a different mass from the original one, the total plasma momentum is overall reduced. In a collisional coma (high outgasssing rate), another reason why ions have difficulty to be accelerated by electromagnetic fields is because the latter is weak: The region usually corresponds to the diamagnetic cavity (i.e. low or null magnetic field), dominated by cold electrons, yielding a weak electric field from the weak electron pressure gradient (see Section \ref{section:3:3}). 

At comet~1P during the Giotto flyby at 0.89~au, ions were not accelerated in the inner ionosphere \citep{Schwenn1987}, either because the coma was dense and/or cometary electrons were too cold for causing a large enough electron pressure-gradient-driven field. Acceleration by ions was observed beyond the diamagnetic cavity, especially from 20,000~km \citep{Altwegg1993, Rubin2009}. What is the situation at 67P? During low outgassing at large heliocentric distances (beyond 2~au), a multi-instrument analysis linking datasets from the Rosetta Plasma Consortium (RPC-IES, RPC-MIP, RPC-LAP) and ROSINA (COPS, DFMS) by a physics-based model (see Section~\ref{section:2:4:1}) gave excellent agreement in terms of ionospheric density within the instrument uncertainties \citep{Galand2016,Heritier2017,Heritier2018, Vigren2019}. The model assumes that the ions are moving radially at the speed of neutrals, that is, they have not suffered from any acceleration between their birth place and the location of Rosetta.

Why the models work so well at large heliocentric distances/low outgassing activity \citep[e.g. ][]{Heritier2018} is not entirely clear, as there are theoretical as well as observational indications for the acceleration of newborn ions, flowing at speeds around an order of magnitude above the neutral gas speed. 
Based on theory and models, $\vec{E}_{\mathsf{P}_e}$ must accelerate the ions away from the coma as they are born at subsonic speeds in the case of 67P. Therefore, are the ion-neutral collisions present at 67P, as discussed in Section \ref{section:2:3}, sufficient to limit/prevent this acceleration? Simulations by \citet{VigrenEriksson2019}  indicate that this was not the case at the location of Rosetta even at perihelion, the most active outgassing phase. From RPC-ICA observations, \citet{Bercic2018} estimated the outward flow of low-energy cometary ion population during a one-month period when Rosetta was at 28~km from the nucleus at 2.5--2.7~au (low outgassing rate), around 6~km\,s$^{-1}$. Similar results were also reported by \citet{Nilsson2020}. Ion velocities from combined analysis of RPC-LAP and RPC-MIP data (and by two different approaches) have also showed similar ion speeds \citep{Johansson2021}.

While these observations appear to be consistent with the expected effect of the ambipolar electric field, one has to be aware of the technical problems with observing low energy ions (a few eVs) by detectors on a spacecraft charging down to negative potentials, typically $-5$ to $-20$~V (see Section~\ref{section:3:1:1}), accelerating the ions to similar energies. The negative spacecraft potential is believed to be somehow beneficial for RPC-ICA measurements: all the ions must reach the spacecraft and be detectable as the minimum energy threshold of RPC-ICA was below $-qV_{SC}$. Simulations of the ion measurements based on Spacecraft Plasma Interaction System (SPIS) did not clearly identify the reason(s) why low energy ions should not reach the RPC-ICA detector \citep{Bergman2020b,Bergman2021a}. Some possibilities include the obstruction of its $360^\circ\times90^\circ$ field of view, the limited angular coverage in some operational modes, and/or large uncertainties regarding the geometric factor at low energies, that is, the factor which allows to convert counts from the instrument to physical quantities such as the ion differential flux in this case. However, these simulations have also showed that ions, particularly low-energy ones, had very distorted trajectories around Rosetta prior their detection. Regardless, it is clear that the plasma density (the zeroth order moment) from the ion distribution measured by RPC-ICA is one to two orders of magnitude lower than those measured by RPC-MIP and RPC-LAP, considered as baselines, such that the direct measurements of the low energy ions on and at Rosetta is still an open question. 

The ion speed derived from combining both measurements by RPC-LAP and RPC-MIP was more indirect. The ion current measured by RPC-LAP is combined with the plasma density from RPC-MIP, assuming only one ion species at 18 or 19~u \citep[an assumption far from being true especially near perihelion, when many ions coexist and might have different speeds,][]{Beth2020}, to derive an effective speed, regardless of the direction \citep{Johansson2021}. The only indication of flow direction by RPC-LAP was the observation of a dropout in the ion current observed when the spacecraft itself obstructed the ion flow, preventing the ions to reach the probe. It occurred a few times when 67P, the spacecraft, and one of the RPC-LAP probes were aligned \citep{Odelstad2018}, as expected for a radial outflow. While completely excluding effects of the spacecraft potential on RPC-LAP is difficult, errors sufficient to explain the large difference to the speed of the neutral gas are equally difficult to conceive. Some might be still inherent to the analysis of Langmuir probes' measurements. Indeed, deriving plasma properties from $I-V$ curves has never been a straightforward and easy task and relies on assumptions. For instance,  it is common to derive the {effective} ion speed (thermal and bulk speeds are indistinguishable) based on strong assumptions \citep[see the seminal paper by][]{MottSmith1926}: a \emph{single} ion population and specific distribution functions (e.g. ring or Maxwellian distribution), drifted or not, in the Orbital Motion Limited (OML) regime (the Debye length is infinite compared with the probe's size, which is not always the case), and for an unmagnetised plasma (while at low outgassing, the environment is magnetised). Depending on the ratio between the probe's radius and the ambient Debye length, the estimates of the ion effective speed may be very different \citep{Laframboise1966} and can differ significantly from the bulk ion speed.

The multi-instrumental approach proposed at large heliocentric distances is not applicable near perihelion due to higher outgassing rates. Additional processes have to be taken into account. For instance, chemical loss through electron-ion dissociative recombination needs to be included (besides loss through transport) as well as attenuation of the EUV solar flux. Although the effect on the electron density of photoabsorption by the neutral gas is ruled out at the location of Rosetta \citep{Heritier2018}, \citet{Johansson2017} observed attenuation of the EUV solar flux near perihelion. This was attributed to dust, lessening the ionisation rate and causing in part this overestimation as models did not take into account dust absorption. Finally, large ion speeds were also reported \citep{Vigren2017,Odelstad2018} implying that acceleration takes place. If confirmed, this would reduce the electron density locally \citep{Vigren2019} as the models are based on flux conservation.

The apparent discrepancy at low activity between the success of {the} multi-instrument analysis, assuming $V_i\approx V_n$ to reproduce the measured ion density, and the observations, indicating much higher ion speeds, needs to be solved. The physics of the interplay of ion-neutral collisions in an inhomogeneous gas with an inhomogeneous and dynamic plasma, bathed in electromagnetic fields varying in space and time, is a very rich and complex matter, with much still to be explored. On the one hand, hybrid simulations may include ion-neutral collisions, but as they lack the electron kinetics, much of the detailed physics causing acceleration is partially lost \citep[e.g.][]{Koenders2015}. On the other hand, particle-in-cell (PIC) simulations may include collisions either self-consistently or by application of them as an extra layer somewhat along the lines of \citet{Stephenson2022}. This approach is expected to be more and more pertinent in the coming years.

\subsection{Pick-up ions\label{section:4:2}}

In a region where the solar wind flows, the newborn, cometary ions, initially at rest with respect to the ambient plasma, may feel the solar wind convective electric field $\mathbf{E_\text{sw}}$ and start being accelerated by it. This process, called `pickup', results, in a kinetic sense, in these accelerated ions of cometary origin having cycloid trajectories in the $\mathbf{E}\times\mathbf{B}$-drift direction, where $\mathbf{B}$ is the local magnetic field. The implicit assumption is that solar wind flow and magnetic field directions are not aligned, so that the cross product is non-zero, otherwise the pickup process would be rendered negligible. In velocity space, the accelerated pickup ions form so-called `ring'-shaped velocity distribution functions (VDFs), with a maximum velocity of the picked-up particles of the order of $2V_\text{sw}\sin \xi$, where $\xi$ is the angle between the local solar wind magnetic field $\mathbf{B}$ and solar wind velocity $\mathbf{V_\text{sw}}$. Pitch-angle scattering further distributes the energy in phase space transforming the initial ideal pickup ring or partial ring VDFs into bispherical shell VDFs \citep{Coates2004}. Such VDFs, from ring-beam to bispherical shells, have been observed at comets including 1P and 67P \citep[e.g. ][]{Coates2015,Nilsson2015a,Goldstein2015}, denoting several stages of evolution in the pickup ion process. In turn, these distributions are unstable and generate waves heating the ambient plasma \citep{Wu1972}.
A more detailed outlook of pickup ions, their effect on large plasma structures, and wave-particle interactions, is given in the chapter by Götz et al. in this volume. 

The pickup process has a drastic effect on the solar wind flow \citep{Glassmeier2017}. As the energy and momentum are directly transferred from \emph{light} solar wind to \emph{heavy} cometary ions, two effects simultaneously take place: To conserve energy, the solar wind is slowed down usptream of the cometary nucleus, and, to conserve momentum, is deflected in the direction opposite to the pickup ions' motion. Moreover, due to the instability of the VDFs created in the pickup process, part of the transferred energy is diverted towards the excitation of low-frequency waves (ion cyclotron waves, Alfvén waves, etc.), which in turn heat the plasma and produce local turbulence. Momentum transfer between solar wind and accelerated cometary ions can be particularly encouraged when VDFs are non-gyrotropic, further feeding local instabilities. Because of the implantation of heavy ions into the solar wind flow, the pickup process thus contributes to the mass-loading of the solar wind \citep{Szego2000}, an idea first put forward by \cite{Biermann1967}. 

\subsection{Solar wind versus ionospheric plasma\label{section:4:3}}

\begin{figure*}[!ht]
    \centering
    \includegraphics[width=\linewidth]{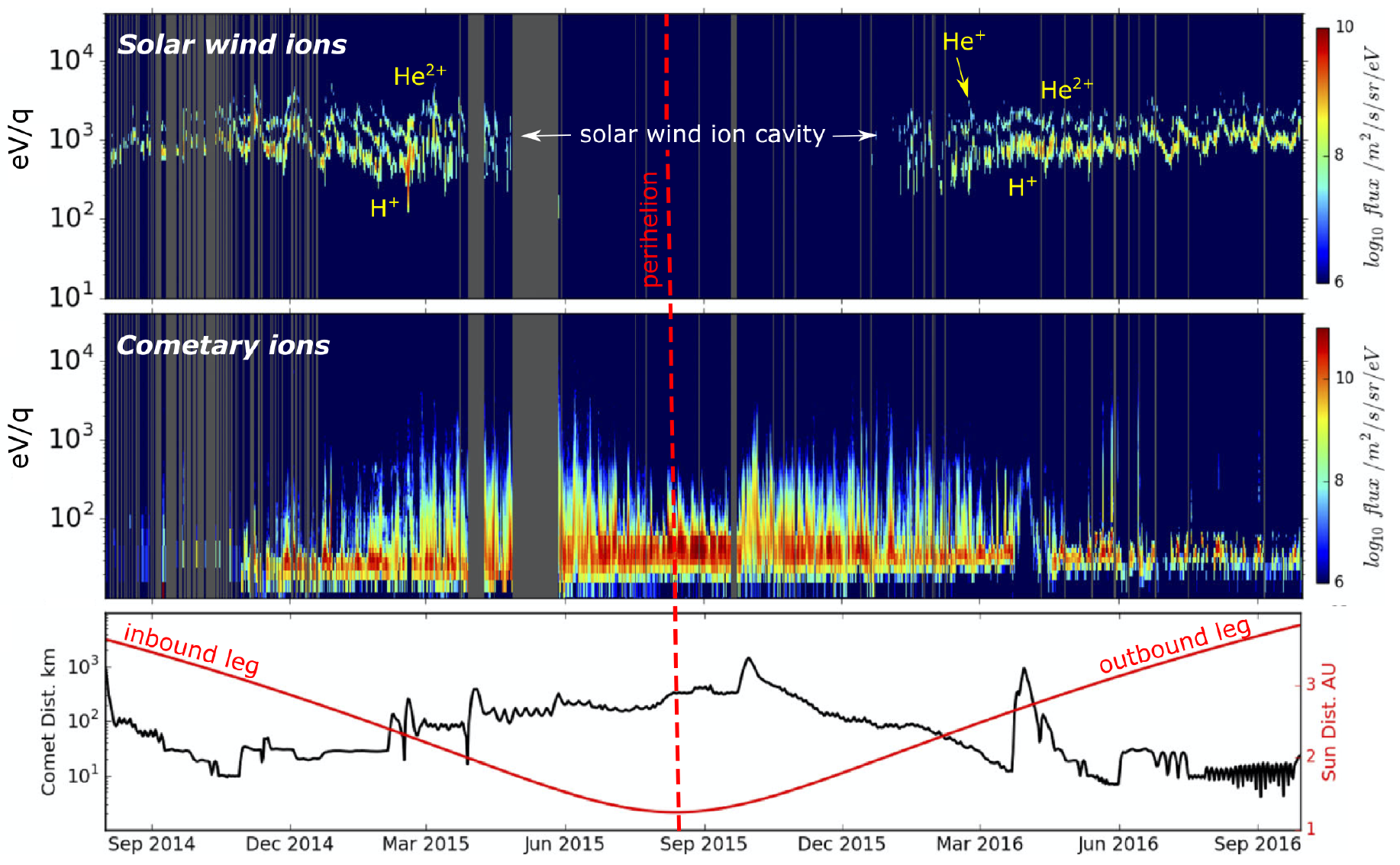}
    \caption{Overview of the solar wind (top) and cometary ion (middle) energy spectrograms during the Rosetta mission, as measured by RPC-ICA and angle-integrated over 1~hour. Heliocentric and cometocentric distances are shown for reference on the bottom panel. At this low temporal resolution, solar wind protons (around $E_p/q\sim 1$~keV\,q$^{-1}$) and alphas (around $2E_p/q$ energy, fourfold heavier but doubly-charged) can be clearly identified, whereas charge-exchange products He$^+$ (around $4E_p/q$) are more seldom seen. The \emph{solar wind ion cavity} spans April-December 2015, with no detection of solar wind ions during that period. Data gaps are highlighted as grey zones. Note that the colour bars on the first two panels do not have the same upper bounds. \citep[Adapted from Fig. 1 of][]{Nilsson2017}. \label{fig:ionfluxes_sw_cometary_ions}
    }
\end{figure*}

The interaction between the solar wind and the cometary neutral environment gives rise to several coexisting populations of ions, which may take precedence over one another in the inner coma depending on heliocentric distance (and hence, nucleus activity) and on cometocentric distance. What are the conditions for cometary ions (at high or low energy) to prevail over solar wind ions? Historical missions composing the Halley armada or more recently Rosetta give us clues to answer this question.

The transition region where cometary ions gradually become dominant is usually called the \emph{cometopause} \citep{Cravens2004}, and sometimes \emph{protonopause} \citep{Sauer1995}. It has historically been defined in two ways (see also the chapter by Götz et al. in this volume): 
\begin{enumerate}[label={[\arabic*]}]
    \item the \emph{charge-exchange collisionopause},
    \item the inner region delimited by $n_{sw} \approx n_{ci}$ (or $F_{sw} \approx F_{ci}$), where $n$ (resp. $F$) are the total number densities (energy and/or momentum fluxes) of solar wind ions ($sw$) and of cometary ions ($ci$, referred to as solely $i$ in Section~\ref{section:2}). 
\end{enumerate}
In definition [2], the cometopause may not be a sharp boundary but rather a smooth cross-over transition between dominating species in the plasma; in the case when the solar wind plasma is shocked and thermal speeds of cometary and solar wind ions are similar, density and flux ratio definitions are equivalent. In fact, the presence or not in the data of a cometopause at 1P and its physical relevance were highly debated at the time of its first discovery \citep{Gringauz1986,Reme1994}. \citet{Gombosi1987} introduced a typical ion-neutral collision scale linked to the charge-exchange collisionopause (definition [1]):
\begin{equation}
    \ell_\text{CX} = \frac{\sigma_{p,s}^{sw,\text{CX}}(E_{sw})\,Q}{4\pi\,V_p}\label{eq:cxCollisionopause}
\end{equation}
resulting, from the point of view of the ion composition, in the conversion of fast (but already decelerated) light solar wind ions $sw$ to slow heavy cometary ions. For a solar wind speed of $300$~km\,s$^{-1}$, a charge-exchange cross section of protons in water of $2\times 10^{-19}$~m$^2$ \citep{SimonWedlund2019a}, $Q\sim10^{26}$--$10^{30}$~s$^{-1}$ (representative of activity levels of 67P and 1P), and $V_p\sim600$~m\,s$^{-1}$, $r_c\lesssim\ell_\text{CX}\lesssim 2.5\times 10^4$~km. In turn, the subsolar position of the cometopause was defined by \citet{Gombosi1987} as:
\begin{equation}
    R_\text{CX} \sim \sqrt{\ell_\text{CX}\ r_0}\label{eq:cometopausePosition}
\end{equation}
where $r_0$ is the distance at which the flow speed of the shocked solar wind ions decreases down to the mean solar wind proton thermal speed. This is due to solar wind mass-loading behind the bow shock once the latter has formed (and which is situated thousands of kilometres upstream of the nucleus). In the case of 1P, $r_0\sim 1.6\times 10^{5}$\,km, so that $R_\text{CX} \sim 6\times 10^4$\,km, possibly consistent with off-Sun-comet-axis Giotto and Vega-2 observations around $1.5\times 10^5$\,km along their respective path \citep{Gombosi1987}. At 67P around perihelion ($Q\sim 5\times 10^{27}$~s$^{-1}$), numerical hybrid simulations predicted $R_\text{CX}\geq 800$~km using definition [2] to image the density cross-over \citep{SimonWedlund2017,Alho2021} with $\ell_\text{CX}\sim 130$~km (from Eq.~\ref{eq:cxCollisionopause}), which implies $r_0\sim 5000$~km (calculated by inverting Eq.~\ref{eq:cometopausePosition}).

Two intimately intertwined aspects drive the evolution of the total plasma ion composition: cometocentric distance on the one hand, and outgassing rate on the other (for a description of the cometary ion composition, see Section~\ref{section:2:4:2}). The more the coma shrinks as the result of a smaller outgassing rate, the deeper into the coma must a given spacecraft venture to observe a relatively comparable ion composition. As a rule, H$_2$O$^+$ and CO$_2^+$ are the most prominent cometary ion species in the inner ionosphere whatever the outgassing rate, provided that we are at a sufficient distance from the nucleus, typically above $R_{s,n}\sim R_\text{CX}$ (see Eq.\,\ref{eq:Rsn} and Section~\ref{section:2:4:2}). Let us illustrate this with two examples from the Rosetta mission: one at large heliocentric distance (low outgassing), one near perihelion (high outgassing). What would a mass spectrometer see? As a guide, typical ion spectrograms, captured during the Rosetta mission, of the solar wind ions and of the cometary ions are displayed in Fig.~\ref{fig:ionfluxes_sw_cometary_ions}.

From the start of the activity to medium activities, the neutral outgassing rate of a typical Jupiter-family comet \citep{Lowry2008}, such as 67P, is of the order of $10^{25}$--$10^{26}$~s$^{-1}$ \citep[see Fig.~\ref{fig:schematic} or][]{Coates2009,Hansen2016}. 
Under such conditions, $R_\text{CX} \sim r_c$ and solar wind ions are expected to have energy and momentum fluxes (although not necessarily densities) at least equal to those of cometary ions, as they pass almost undisturbed through the comet environment, with minimal mass-loading effects such as deflection and slowing down. However, the onset of magnetospheric-like behaviours, with the production of cometary pickup ions and sporadic solar wind charge-exchanged products, such as He$^+$, can already be seen from $Q \geq 10^{26}$\,s$^{-1}$, as demonstrated by Rosetta's ion energy spectrograms (from RPC-ICA and RPC-IES) during both inbound and outbound conditions \citep{Nilsson2015a,Nilsson2017,Goldtsein2017,SimonWedlund2019c}. Of note, negatively charged H$^-$ ions, thought to be arising from double charge exchange of solar wind protons, were also observed in the electron channel of RPC-IES \citep{Burch2015}. Thus, a lot of the total ion energy and momentum flux is still contained in the solar wind ions at relatively small cometocentric distances \citep{Williamson2020}, such as those probed by Rosetta ($r \leq 150$~km). From Rosetta's point of view, the cometopause transition, seen as a gradual flux cross-over between cometary and solar wind ions, occurred for $Q\approx2$--$5\times 10^{26}$~s$^{-1}$, that is, in and around January 2015 (inbound leg of 67P's orbit around the Sun) and February 2016 (outbound leg) (see chapter of Götz et al. in this volume).  

At medium-to-high outgassing activity, typically for $Q \gtrsim 10^{27}$\,s$^{-1}$, corresponding to closing in on perihelion conditions for 67P, $R_\text{CX}\gtrsim 500$~km, and the angular deflection of the solar wind ahead of the obstacle may become so large that the solar wind is effectively kept from entering the inner coma. This region, where no solar wind ions can be detected outside of occasional solar wind-driven events, such as CMEs, was dubbed by the Rosetta team the \emph{solar wind ion cavity}(SWIC). Rosetta stayed in this region from around April to December 2015 \citep[][and see Fig.\,\ref{fig:ionfluxes_sw_cometary_ions}]{Behar2017}. In it, the heavier, slower plasma of cometary origin, produced locally by photoionisation and electron impact ionisation, dominated the total ion composition and fluxes. Hence, for these relatively high outgassing rates, Rosetta was always inward of the cometopause, with the SWIC possibly being its inner edge \citep{Mandt2019}. This is mainly because: (i) Rosetta orbited close to the comet at any given time ($<500$~km outside of temporary excursions); (ii) RPC-MIP and RPC-LAP, instruments measuring $n_{ci}$, could not probe plasma densities below $\sim 50$~cm$^{-3}$. As the solar wind density is typically $n_{sw}\approx 1$--$10$~cm$^{-3}$, it was not possible to locate where $n_{sw}\approx n_{ci}$.
Thus, in a way similar to onions, the cometopause successively encompasses the ion-neutral and electron-neutral collisionopauses, where plasma-neutral collisions start to play a major role in the coma, but also the diamagnetic cavity once it forms. For more discussion on the formation of large-scale boundaries including the cometopause, the diamagnetic cavity, the collisionopauses, and the solar wind ion cavity, the reader is directed to the chapter by Götz et al. in this volume.

\section{Dusty (dirty) complex plasma\label{section:5}}

When we look at a comet in the night sky, our eyes mostly detect sunlight reflected by dust grains emitted from the comet nucleus, forming the coma and the dust tail (usually the more visible of the two tails, see the chapter by Agarwal et al. in this volume). 
This obvious presence of dust in the same volume of space as the cometary plasma immediately suggests that the coma and dust tail may be regions where there are lots of interaction between the dust and the plasma. The plasma may charge the dust, which in turn can influence the motion of the plasma, coupling them together in what is known as a \emph{dusty} or \emph{complex plasma}.

In this section, we will only provide a very limited introduction to the range of possible effects of dust-plasma effects at comets (Section~\ref{section:5:1}) and to the mechanics of dust charging (Section~\ref{section:5:2}). The main reason for spending so little space on this is very simple: In contrast to the dust-free physics of a cometary ionosphere, presented in this chapter until now, and against many expectations before the mission arrival at 67P, the analysis of the Rosetta data has as yet not provided any entirely new picture of dust-plasma interactions at comets. As we will see in Section~\ref{section:5:3}, some dust-plasma interaction signatures have been identified in the Rosetta data. But the same Section will show that the rich Rosetta data sets on plasma and cometary dust actually suggest such interactions probably have only a small impact on the plasma, at least for most of the mission and for the regions visited around this particular comet. For a more extensive treatment of dust charging and its possible effects on the comet plasma environment, we refer to previous literature on the subject \citep[see for example][]{Mendis2013a}.

\subsection{Why care about dust charging?\label{section:5:1}}

Consider free electrons moving around in the coma. If there are lots of dust grains, an electron might now and then hit one of the grains, perhaps knocking out one electron (in the process known as \emph{secondary emission}) and thereby increasing the number of \emph{free} electrons and charging the dust grain positively. The dust grain would then effectively become an enormously heavy positive ion. Alternatively, the impacting electron might stick to a dust grain, decreasing the number of free electrons and charging the dust negatively. The charge acquired on the dust grain suddenly couples it to the electromagnetic fields around it, and it will itself influence these fields and thereby other charges. Both plasma and dust interact as seen in Fig.~\ref{fig:dustcharge}.

Two main effects may ensue from dust charging and variations in the number density of free electrons. The first one affects the collective behaviour of the plasma and its wave modes, such as the frequency at which it naturally pulsates, $\omega_{pe}\propto \sqrt{n_e}$ (see Eq.~\ref{eq:omega_pe}). One can show that the decrease in the free-electron number density due to attachment to dust grains changes the dispersion relation of \emph{ion acoustic waves} in such a way as to push them to higher frequency, also decreasing their Landau damping \citep{Barkan1996}. It also allows a completely new wave mode, known as \emph{dust acoustic waves}, which, because of the huge dust grain inertia, propagates much slower than usual plasma waves \citep{Merlino2014}. Moreover, electromagnetic modes can be influenced as the charged dust provides the plasma with a higher inertia, and with a continuous distribution of dust masses: This may be interpreted as a continuum of cyclotron resonances. However, due to the large dust grain mass (see the chapter by Engrand et al. in this volume for a discussion of dust size distribution) and weak magnetic field (of the order of 10~nT) at a comet outside 1~au, the effects on electromagnetic waves are expected only at very low frequency such as to not be of practical relevance. Consequently  acoustic modes are  the most relevant wave modes at comets. For a discussion of the physics related to cometary plasma waves (albeit not dust-related), see the chapter by Götz et al. in this volume.

But dust charging can also affect the dynamics of the dust. From the point of view of cometary evolution, the most important of these effects is the so-called \emph{levitation}, in which dust grains can be expelled from the cometary nucleus by electrostatic repulsion if they and their immediate neighbourhood are charged negatively when bombarded by electrons from the surrounding plasma. This effect could potentially be important as it may increase the nucleus dust production from the surface. We refer to the chapter by Agarwal et al. in this volume for further discussion on processes that lift off dust grains from the nucleus. In contrast, farther out from the nucleus' surface, a charged dust grain is subject not only to the forces of gravity and radiation pressure (i.e. a momentum transfer between the impacting solar photons and atoms, molecules, and dust grains in the coma) acting on any grain, but also to the electromagnetic fields in the plasma. Due to the large dust grain inertia, fields acting over large scales in space and time are most significant: We will see below that this effect actually was observed in situ by Rosetta. In addition, the attachment of charges (electrons) to a dust grain causes a repelling electrostatic force between various parts of the grain itself. This force may contribute to erosion of the grain, either by more or less continuous ablation with small fractions of the grain splitting off or fragmentation, when the grain breaks in two (or more) pieces of approximately the same size. Note, however, that other processes may also lead to the same result (see the chapter by Engrand et al. in this volume).

\subsection{Dust charging\label{section:5:2}}

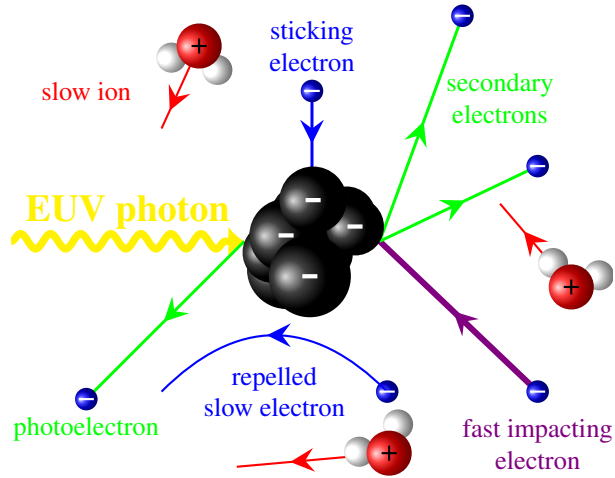
\begin{figure}[ht]
	\begin{tikzpicture}[thick,scale=1, every node/.style={scale=1}]
    \clip (-4,-3.5) rectangle (4,3.2);

    \draw [thick,-{Latex[width=4mm, length=6]},decorate,decoration={snake,amplitude=1mm,segment length=4mm,post length=2mm},yellow,line width=1mm] (-4,0) -- (-0.9,0)node[pos=0.5,anchor=south]{\Large\textbf{EUV photon}};
    \draw [-,thick,green,line width=0.4mm] (-0.9,0) -- (-2.9,-2)node[
    sloped,
    pos=0.5,
    allow upside down]{\arrowIn};
    \shade [ball color=blue,shading angle=45] (-3,-2.1) circle (0.15);
    \draw [thick,white] (-3.1,-2.1) -- (-2.9,-2.1);
    \draw [-,thick,color=red!50!blue,line width=0.8mm] (3,-2) -- (0.9,0) node[
    sloped,
    pos=0.5,
    allow upside down]{\arrowIn};
    
    \draw [-,thick,green,line width=0.4mm] (0.9,0) -- (3,1)node[
    sloped,
    pos=0.5,
    allow upside down]{\arrowIn};
    \draw [-,thick,green,line width=0.4mm] (0.9,0) -- (2,3)node[
    sloped,
    pos=0.5,
    allow upside down]{\arrowIn};
   
    \shade [ball color=blue,shading angle=45] (3,-2) circle (0.15);
    \draw [thick,white] (3.1,-2) -- (2.9,-2);
    \shade [ball color=blue,shading angle=45] (2,3) circle (0.15);
    \shade [ball color=blue,shading angle=45] (3,1) circle (0.15);
    \draw [thick,white] (3.1,1) -- (2.9,1);
    \draw [thick,white] (2.1,3) -- (1.9,3);
        \draw [-,thick,blue,line width=0.4mm] (0,2) -- (0,0)node[
    sloped,
    pos=0.3,
    allow upside down]{\arrowIn};
    \shade [ball color=blue,shading angle=45] (0,2) circle (0.15);
    \draw [thick,white] (0.1,2) -- (-0.1,2);

    \def\r{0.4}

        \pgfmathsetseed{5};
    \foreach \i in {1}{
        \pgfmathsetmacro{\radius}{3}
        \pgfmathsetmacro{\theta}{150}
        \pgfmathsetmacro{\phi}{120}
        \pgfmathsetmacro{\xcoor}{\radius*cos(\phi)}
        \pgfmathsetmacro{\ycoor}{\radius*sin(\phi)}
        \pgfmathsetmacro{\xcoora}{\radius*cos(\phi)+0.1}
        \pgfmathsetmacro{\ycoora}{\radius*sin(\phi)+0.1}
        \pgfmathsetmacro{\xcoorb}{\radius*cos(\phi)-0.1}
        \pgfmathsetmacro{\ycoorb}{\radius*sin(\phi)-0.1}
        \draw [-,thick,red] (\xcoor,\ycoor) -- (-2,1.5)node[
    sloped,
    pos=0.7,
    allow upside down]{\arrowIn};
        \begin{scope}[shift={(\xcoor,\ycoor)},rotate=45]
        
        \shade [ball color=white,shading angle=45] ({\r*cos(\theta-52.225)},{-\r*sin(\theta-52.225)}) circle (0.2);
        \shade [ball color=white,shading angle=45] ({\r*cos(\theta+52.225)},{-\r*sin(\theta+52.225)}) circle (0.2);
        \shade [ball color=red,shading angle=45] (0,0) circle (0.3);
        \end{scope}
        \draw [thick,black] (\xcoor,\ycoora) -- (\xcoor,\ycoorb);
        \draw [thick,black] (\xcoora,\ycoor) -- (\xcoorb,\ycoor);
    }
    
        \foreach \i in {1}{
        \pgfmathsetmacro{\radius}{3}
        \pgfmathsetmacro{\theta}{-80}
        \pgfmathsetmacro{\phi}{-70}
        \pgfmathsetmacro{\xcoor}{\radius*cos(\phi)}
        \pgfmathsetmacro{\ycoor}{\radius*sin(\phi)}
        \pgfmathsetmacro{\xcoora}{\radius*cos(\phi)+0.1}
        \pgfmathsetmacro{\ycoora}{\radius*sin(\phi)+0.1}
        \pgfmathsetmacro{\xcoorb}{\radius*cos(\phi)-0.1}
        \pgfmathsetmacro{\ycoorb}{\radius*sin(\phi)-0.1}
        \draw [-,thick,red] (\xcoor,\ycoor) -- (-1,-3)node[
    sloped,
    pos=0.6,
    allow upside down]{\arrowIn};
        \begin{scope}[shift={(\xcoor,\ycoor)},rotate=45]
        
        \shade [ball color=white,shading angle=45] ({\r*cos(\theta-52.225)},{-\r*sin(\theta-52.225)}) circle (0.2);
        \shade [ball color=white,shading angle=45] ({\r*cos(\theta+52.225)},{-\r*sin(\theta+52.225)}) circle (0.2);
        \shade [ball color=red,shading angle=45] (0,0) circle (0.3);
        \end{scope}
        \draw [thick,black] (\xcoor,\ycoora) -- (\xcoor,\ycoorb);
        \draw [thick,black] (\xcoora,\ycoor) -- (\xcoorb,\ycoor);
    }
    
            \foreach \i in {1}{
        \pgfmathsetmacro{\radius}{3.5}
        \pgfmathsetmacro{\theta}{-30}
        \pgfmathsetmacro{\phi}{-10}
        \pgfmathsetmacro{\xcoor}{\radius*cos(\phi)}
        \pgfmathsetmacro{\ycoor}{\radius*sin(\phi)}
        \pgfmathsetmacro{\xcoora}{\radius*cos(\phi)+0.1}
        \pgfmathsetmacro{\ycoora}{\radius*sin(\phi)+0.1}
        \pgfmathsetmacro{\xcoorb}{\radius*cos(\phi)-0.1}
        \pgfmathsetmacro{\ycoorb}{\radius*sin(\phi)-0.1}
        \draw [-,thick,red] (\xcoor,\ycoor) -- (2.5,0.5)node[
    sloped,
    pos=0.6,
    allow upside down]{\arrowIn};
        \begin{scope}[shift={(\xcoor,\ycoor)},rotate=45]
        
        \shade [ball color=white,shading angle=45] ({\r*cos(\theta-52.225)},{-\r*sin(\theta-52.225)}) circle (0.2);
        \shade [ball color=white,shading angle=45] ({\r*cos(\theta+52.225)},{-\r*sin(\theta+52.225)}) circle (0.2);
        \shade [ball color=red,shading angle=45] (0,0) circle (0.3);
        \end{scope}
        \draw [thick,black] (\xcoor,\ycoora) -- (\xcoor,\ycoorb);
        \draw [thick,black] (\xcoora,\ycoor) -- (\xcoorb,\ycoor);
    }
    
    \pgfmathsetseed{5};
    \foreach \i in {1,2,3,4,5,6,7,8,9,10}{
        \pgfmathsetmacro{\radius}{0.2+((rand+1)*0.5)*0.5}
        \pgfmathsetmacro{\ro}{0.2+(1+rand)*0.2}
        \pgfmathsetmacro{\theta}{rand*360}
        \pgfmathsetmacro{\phi}{rand*360}
        \pgfmathsetmacro{\xcoor}{\radius*cos(\phi)}
        \pgfmathsetmacro{\ycoor}{\radius*sin(\phi)}
        \pgfmathsetmacro{\xcoora}{\radius*cos(\phi)+0.1}
        \pgfmathsetmacro{\ycoora}{\radius*sin(\phi)+0.1}
        \pgfmathsetmacro{\xcoorb}{\radius*cos(\phi)-0.1}
        \pgfmathsetmacro{\ycoorb}{\radius*sin(\phi)-0.1}
        \begin{scope}[shift={(\xcoor,\ycoor)},rotate=45]
            \shade [ball color=black,shading angle=45] (0,0) circle ({\ro});
            \draw [ultra thick,white,rotate=-45] (-0.1,0) -- (0.1,0);
        \end{scope}
        
    }
    \node[anchor=center,green,align=center] at (2.5,1.9)   (a) {secondary\\electrons};
    \node[anchor=center,color=red!50!blue,align=center] at (3,-2.7)   (a) {fast impacting\\electron};
    \node[anchor=center,green,align=center] at (-3,-2.5)   (a) {photoelectron};
    \node[anchor=center,red] (b) at (-3,2) {slow ion};
    \node[anchor=center,blue,align=center] (c) at (0,2.6) {sticking\\electron};
    \shade [ball color=blue,shading angle=45] (1,-2) circle (0.15);
    \draw [thick,white] (0.1,2) -- (-0.1,2);
    \draw[blue] (1,-2) .. controls (0,-1) and (-1,-1) .. (-2,-2)node[
    sloped,
    pos=0.5,
    allow upside down]{\arrowIn};
    \draw [thick,white] (1.1,-2) -- (0.9,-2);
    \node[anchor=center,blue,align=center] at (-0.5,-2)   (d) {repelled\\slow electron};
	\end{tikzpicture}
    \caption{Schematic of various effects influencing the charging of a dust grain (conglomerated black spheres). $n_i$ and $n_e$ are assumed equal within the figure (not every ion or electron is represented) but the higher thermal speed of the electrons leads to more hits on a dust grain thereby charging the grain negatively when impacting electrons stick to the grain. Sufficiently energetic free electrons (magenta line) and solar EUV photons (yellow) can kick out electrons from the grain (green lines), gradually shifting the total charge towards positive values and leading to a non-uniform charging of the grain.
    \label{fig:dustcharge}
    }
\end{figure}

What are the effects influencing the charge of a dust grain? As discussed earlier, a plasma consists of free charges, positive and negative, moving around at random velocities distributed in accordance with their temperatures ($T_s$ for ion species $s$ or $T_e$ for free electrons) and masses ($m_s$ and $m_e$). One of their typical speeds is the thermal speed $v_{\text{th}} = \sqrt{k_BT/m}$ corresponding to the \emph{standard deviation} around their mean bulk velocity $\vec{V}$.  Ions and electrons have masses which differ by about 4 orders of magnitude ($m_s/m_e \sim 3.3\times 10^4$ for water-group ions, typical for a cometary coma), while their temperatures are usually alike. In addition, in the inner cometary ionosphere, $|\vec{V}_s|\gg v_{th,s}$ for ions whereas $|\vec{V}_e|\ll v_{th,e}$. As a consequence, electrons typically move much faster (typically a few hundred times faster) than ions such that in a given time span, a dust grain (or any `body' in the ionised cometary gas) will be hit more frequently by electrons than by ions. The grain thus preferentially charges negatively, until reaching an equilibrium  potential $V_d$ sufficient to stop further charging by repulsion of electrons and catching of ions. This potential can be estimated as:
\begin{equation}
    V_d \sim  -\dfrac{3 k_BT_e}{q}
    \label{eq:Vd}
\end{equation}
It can be noted that these considerations for dust grains apply similarly to a spacecraft, so that the spacecraft potential (discussed in Section~\ref{section:3:1:1}) is in principle set by a similar equation. However, analysis and modelling have shown that due to the design of the Rosetta spacecraft, its electrostatic potential in the inner coma was rather determined by the plasma density \citep{Johansson2021}.

From the grain potential $V_d$, we can calculate its charge:
\begin{equation}
q_d=C_d \:V_d
\label{eq:qd}
\end{equation}
where $C_d$ is the dust's capacitance. The latter can be estimated, or at least a minorant may be found, by considering a spherical grain of radius $r_d$ in vacuum such that
\begin{equation}
C_d = 4 \pi \varepsilon_0\, r_d
\label{eq:C}
\end{equation} 
Therefore,  if all grains have the same size and are all spherical, the total charge density $\rho_d$ (in A\,s\,m$^{-3}$) carried by the dust is obtained by {combining Eqs.~\ref{eq:Vd}--\ref{eq:C}}:
\begin{equation}\label{eq:rhod}
   \rho_d= q_d\ n_d \sim -12 \pi \varepsilon_0 r_d\, n_d\, k_B\, T_e/q
\end{equation}
where $n_d$ is the number of dust grains per unit volume.

In reality, the attachment of free electrons is not the only factor influencing the grain potential (Figure~\ref{fig:dustcharge}). The emission of electrons from sunlit dust grains due to the photoelectric effect will tend to charge the grains positively whereas, if the impacting flux of free electrons is sufficiently high, the negative charging by electrons attaching to grains will dominate. In practice, the limiting density is on the order of $10^2$~cm$^{-3}$ at 1~au (scaling with $r_h^{-2}$),
depending on the dust grain properties. In the dense parts of a coma, we expect to find mainly negatively-charged dust grains. Another major effect is secondary emission, occurring when an electron of sufficiently high energy kicks out one or more electrons from a grain. It is typically important when the incoming electron has an energy of $\sim 100\text{s}$~eV and does not matter much below 10~eV, although this depends on the detailed dust properties, often  not very well known (size and mass distributions, shape, composition). 

A further complicating factor is the time it takes to charge a dust grain to its equilibrium potential $V_d$ (Eq.~\ref{eq:Vd}). 
The total electric current collected by the grain $I_e$ [A], due to the randomly moving plasma and mainly driven by the free electrons, is approximately given by $I_e\approx J_eS_d$ where $S_d = 4 \pi r_d^2$ [m$^2$] is the grain surface and $J_e$ [A\,m$^{-2}$]  is the electron thermal current density, driven by free electrons moving towards the grain. {$J_e$ is} given by \citep[e.g.][]{MottSmith1926}:
\begin{equation}
    J_e = - q n_e \sqrt{\frac{k_B T_e}{2 \pi m_e}}.
\end{equation}
Determining $q_d$ as a function of time requires solving a differential equation, exactly as for a resistor–capacitor circuit. The characteristic time, or time constant, for charging a dust grain from $0$ to $V_d$ is given by:
\begin{equation}
    \mathcal{T}_d \sim \frac{C_dV_d}{I_e} = 3 \varepsilon_0 \dfrac{\sqrt{2 \pi m_e\, k_BT_e}}{n_e\, q^2\, r_d} = 3\sqrt{2 \pi}\, \dfrac{\lambda_\text{Debye}}{r_d\,\omega_{p,e}}
    \label{eq:tau_charge}
\end{equation}
{where $\lambda_\text{Debye}$ and $\omega_{p,e}$ are defined by Eqs.~\ref{eq:omega_pe}-\ref{eq:l_Debye}. Each dust grain can be seen as a capacitor and a resistor in series with its characteristic capacitance $C_d$ and resistance $R_d=V_d/I_e$ [$\Omega$], depending not only on their size and shape but also on the local electron number density and temperature. The current is provided to this circuit by means of free electrons in the plasma falling on the dust grain.} From Eq.~\ref{eq:tau_charge}, we note that large grains charge up faster than small ones (as $C\propto r_d$ and $I_e\propto r_d^2$, bigger grains collect more electrons). Only one of the most relevant dust charging processes at comets has been described here and a more exhaustive description is given in \citet{MeyerVernet2013}.

How long does a dust grain take to reach $\sim V_d$ for typical cometary plasma parameters? For a grain of radius $r_d = 10~\mu$m in a plasma with $n_e = 10^3$~cm$^{-3}$ and $T_e = 10$~eV, we get $\mathcal{T}_d \sim 10$~s, while a 100-nm `nanograin' requires $\sim10^3$~s (more than 15 minutes). During this time, the grain may travel through plasmas of quite varying characteristics, particularly in the inner coma \citep[see the chapter by Agarwal et al. in this volume and][for discussions of dust grain velocities]{Marschall2020}. At some distances, the different $\mathcal{T}_d$ between grains associated with the outward dust flow through a plasma such that $n_e(r)\propto 1/r$ can lead to {complex} situations where small grains may conserved the negative charge acquired closer to the nucleus (long $\mathcal{T}_d$) while larger grains may quickly charge positively by means of photoemission (short $\mathcal{T}_d$). In addition, the secondary emission, which tends to be more efficient for small grains, can lead to similar results. As dust grains may travel at different speeds and take different trajectories through the plasma, they experience different charging along their path. All this leads to an expectation of a distribution of charge states even for similar dust grains -- and, as is clear from the chapter by Engrand et al. in this volume, there is a large variation among dust grain properties.

In these circumstances, is it possible to estimate the fraction of the free electrons which will attach to dust grains? This is an important question, because this fraction must be significant in order for the dust charging to have any considerable effect on the plasma. Clearly, the answer must depend on $n_e$ and $n_d$. However, the dust size distribution is also crucial in this respect. For example, let's assume that all dust grains have the time to reach their equilibrium potential as given by Eq.~\ref{eq:Vd}. The charge carried by a grain depends on its capacitance, which scales linearly with its size (as seen in the example of a sphere, Eq.~\ref{eq:C}). Contrarily, the mass of a grain scales with its volume $\mathcal{V}_d\propto r^3_d$. As a consequence, for a fixed total dust mass, the plasma is more depleted in free electrons and dust charging is more efficient if the dust is mostly made of lots of small grains than of a few large ones \citep{Vigren2015}. For a given dust size distribution, detailed calculations can be used to estimate the total amount of dust charging. 

As the size distribution of dust in the innermost coma of comet~67P is quite well constrained by the dust instruments (see the chapter by Engrand et al. in this volume), \citet{Vigren2021} carried out such calculations and constrained the possible electron depletion in this environment. Their model, combining observed plasma densities and electron energies with dust size distributions derived from the dust observations at Rosetta by means of Eq.~(\ref{eq:rhod}), shows that because of the preponderance of large grains, the fraction of electrons attaching to dust grains must have been small or even negligible during most parts of the Rosetta mission. Quite extreme assumptions on the dust distribution had to be taken even to push the electron depletion to a few percent. A similar negative conclusion was reached by \citet{Vigren2022}, using analogous arguments, for positive charging of dust grains. 

There may be exceptions at specific locations and times, as in outbursts  \citep[see, for example, the 2016 Feb 19 event studied by][]{Grun2016} which could lead to a locally high density of small grains, though no clear signatures of obvious electron depletion have yet been identified in the Rosetta data set. An apparent absorption of solar EUV radiation has also been used to argue that the dust size distribution at large distances from Rosetta (several thousand kilometres or more) may include larger amounts of grains down to the size of tens of nanometres \citep{Johansson2017}. As few such grains have been observed close to the nucleus, they have presumably resulted from one or several erosion and fragmentation processes (see discussion in Section~\ref{section:5:1} and the chapter by Engrand et al. in this volume) which possibly could lead to stronger dust-plasma coupling farther from the nucleus than the regions investigated by Rosetta. One may note that very small dust grains, with masses down to $10^{-23}$~kg, were observed by the Vega spacecraft $\sim10^5$~km from the nucleus of comet Halley \citep{Sagdeev1989}.

While there may still be much to learn from the Rosetta data set, the combination of the observed average properties of dust and plasma quite clearly indicates that the typical electron depletion in the innermost coma of 67P is small, with correspondingly weak impact of the dust on the overall plasma dynamics. However, charged dust grains of very small size were indeed observed in some events, as discussed in Section~\ref{section:5:3}.

\subsection{Charged dust dynamics at Comet~67P\label{section:5:3}}

How do charged dust grains move in the coma? Let us start by considering neutral dust grains. Close to the nucleus, the grain is dragged by the outflowing cometary gas. As the neutral gas density decreases with cometocentric distance (see Eq.~\ref{eq:n_n}), the neutral drag force decays rapidly with distance and is hence stronger close to the nucleus. The very weak gravitational force of a comet nucleus can also mostly be neglected for most dust grains farther outside \citep[however, see][for examples of gravitationally bound dust motion around comet 67P]{Davidsson2015}. In addition, the solar radiation pressure impresses a constant force on every grain in the anti-sunward direction. Grains emitted from the nucleus in the direction of the Sun can therefore be turned back by the radiation pressure and return towards the  nucleus and farther out downstream of the comet. The radiation pressure force on a grain depends on $r_d^2$, while the inertia scales with its volume, through its mass, which is $\propto r_d^3$. Therefore, the radiation pressure affects more efficiently the dynamics of small/light grains than that of large/heavy grains. Other effects, such as Mie scattering and Poynting–Robertson drag, are not discussed here: Fuller details are given in \citet{Burns1979}.

Charged grains are, in addition, subject to electromagnetic forces arising in the comet-plasma interaction, discussed in the chapter by Götz et al. in this volume. Due to the large mass of dust grains, compared with cometary ions and electrons (in the gas phase)  \citep[even a small 10~nm grain are estimated to have a mass of order $10^5$~u by][]{Gombosi2015dust}, the gyroradius of dust grains generally becomes so large that the grain's gyromotion can be neglected within the inner coma explored by Rosetta at 67P. Nevertheless, the magnetic field indirectly leads to the appearance of a convection electric field which may efficiently affect the dust dynamics. If this force is constant over the region traversed by the dust grain, the dust's trajectory will describe a parabola of which its axis of symmetry is along this force. However, electric and magnetic fields change in time, and, as aforementioned in Section \ref{section:5:2}, the charge of a grain may evolve as well, as it travels through plasma regions of different $J_e(\vec{r})$, while grains eventually erode and/or fragment which yields changes in the dust's mass, size, charge, and number. Therefore, the picture of motion in a conservative field must be applied with caution.

The RPC-IES spectrometer was designed for the measurement of (positive) ions and electrons. However, it was able to detect charged dust grains. As there was no possibility for direct mass separation, charged dust signatures could not be unambiguously separated from those of ions and electrons by IES. That said, some signatures stand out from ordinary charged particle signatures and better fit to expectations for charged dust. \citet{Burch2015dust} found a few events with negatively charged grains moving outward from the nucleus at kinetic energies of a few hundreds of eVs (if assumed to be singly-charged; higher charge states would have correspondingly higher energy) as well as much more energetic grains (1--17~keV\,q$^{-1}$), also negatively charged, arriving from the approximate Sun's direction. All these detections were from the very early phase of the Rosetta escort of 67P, from 23 August to 1 September 2014, when Rosetta was at 50--60~km sunward of the nucleus and the outgassing activity was still very low. \citet{Gombosi2015dust} analysed the observations of outward moving grains in terms of neutral gas drag acceleration. The anti-sunward moving grains were attributed to radiation pressure acting on small grains likely created from larger grains by erosion at large distances from the nucleus. For both types of grains, consistency between model and data resulted for grains' sizes within the 30--80~nm range. The convection electric field, changing in time in response to the interplanetary magnetic field, was discussed as a possible reason for the intermittency of the observations, deviating grains from the densest regions away or towards Rosetta.

\citet{Llera2020} could identify another event, for a similar period (19 Sep 2014), with positively-charged and negatively-charged grains over a broad energy range, as expected from the competing charging processes discussed in Section~\ref{section:5:1}. Both showed signatures very similar to the negatively charged antisunward moving grains previously reported. Interestingly, while grains of both signs had an antisunward velocity component, their trajectory was symmetric with respect to the antisunward component. This is what is to be expected for charged dust grains subject to a combination of the solar radiation pressure, same for both signs, and the solar wind convection electric field in the comet reference frame, which  accelerates grains of opposite charge in opposite directions.

These are, as yet, the only events with signatures which have been attributed to charged dust grains in the tens of nm-size range, all from the initial part of the Rosetta mission. Why are there so few? Regarding the outflowing grains discussed by \citet{Burch2015dust}, their signature is quite weak and may be swamped with electron fluxes in the same range, which were significant through most of the mission. The higher energy signature of the (roughly) antisunward grains should not suffer as much, and here it seems likely they are just not present in significant numbers in the inner coma. According to the interpretation by \citet{Johansson2017} of  anomalously low flux of solar radiation in the EUV observed at Rosetta around perihelion when 67P's activity peaked, nm-size dust grains were mainly present at cometocentric distances of at least a few thousand kilometres during this period. This may point to the fragmentation of dust grains occurring farther out at high activity than at low activity. Alternatively, the dust itself may already have different properties depending on $Q$ and $r_h$. 

Rosetta has as yet only showed some tantalising glimpses of dust-plasma interactions, concentrated on the early phase of very low activity, while mission wide modelling indicated that the dust impact on the plasma mostly is small or negligible. Nevertheless, the Rosetta data set are far from exhausted and may still hold more clues to the presence and importance of dust charging.

\section{Summary \& open questions\label{section:6}}
{Prior to the 1980s, comets and their plasma environment had relied only on speculation and simulation work, either in the laboratory or starting with the first numerical simulations. Thanks to the historical missions that followed, from the first flybys of comets in the 1980s to Rosetta's 2-year mission escorting comet~67P around the Sun 30 years later, many questions regarding the ionosphere of comets, its formation, structure and evolution at different outgassing activities have now been answered. Table~\ref{tab:highlights} presents highlights of our current understanding sorted by the themes structuring this chapter, from the physics presiding at the creation of a cometary ionosphere (Section\,\ref{section:2}, presented as a tutorial-style toolbox), the description of the plasma population, electrons (Section\,\ref{section:3}) and ions (Section\,\ref{section:4}), and the puzzling and still largely unexplored role of charged dust in the overall dynamics and composition of the cometary ionosphere (Section\,\ref{section:5}).}

Many open questions remain, some of the more prominent ones are summarised in Table\,\ref{tab:openQuestions}.
With the advent of new missions, several of these questions will hopefully be answered. Ideally, multi-point, multi-instrument, in situ exploration of the cometary coma, either through a flyby or an extended escort phase of the comet nucleus ``à-la-Rosetta'', are necessary to address them with more than one spacecraft if possible as already proposed by the community for the ESA Voyage 2050 programme call for white papers \citep{Goetz2021a}. This is among the goals of the highly anticipated ESA/JAXA \emph{Comet Interceptor} mission \citep{Snodgrass2019}, currently in-the-works as an international endeavour: boasting simultaneous $3$-point measurements thanks to a mother spacecraft A complemented by two smaller probes, B1 and B2, the mission is planned to be launched in 2029 and to fly by a `dynamically new comet' yet to be discovered (that is, visiting our inner Solar System for the first time). Of particular interest for cometary ionospheric physics, probe A, farthest from the cometary nucleus, hosts a Dust, Field, and Plasma (DFP) set of sensors to probe the dust, electromagnetic fields, (dusty) plasma and energetic particles (electrons, ions, and neutral atoms). This set is complemented by a neutral pressure gauge and mass spectrometer probing the major volatile species (MANIaC). On probe B2 which will get closest to the cometary nucleus, the DFP package is reduced to a dust impact sensor and counter, and a magnetometer. Probe B1 hosts a plasma suite, which includes a time-of-flight mass spectrometer and a magnetometer. All these sensors will be probing the plasma environment around the target to deepen our understanding of cometary ionospheres, including formation, loss, and interaction with the space environment.

No new dedicated cometary mission, with sample return or not, is envisaged in the decade following Comet Interceptor, despite a strong call from the community, as proposed for the ESA Voyage 2050 programme call \citep{BockeleeMorvan2021,Goetz2021a}. Because of their complexity, costs, and necessary multidisciplinary approach, future cometary missions will likely be international collaborations between space agencies (as for Comet Interceptor) and address a broad range of physics including plasma physics. By escorting comet~67P around perihelion for two years, Rosetta is a successful proof of concept showing that a spacecraft orbiting a comet for a long period of time can be designed, flown, and can harvest an amazingly rich dataset, despite the weak nucleus gravity. With our increasing capacity to detect more and more new targets of interest, interstellar objects or not, one can envision multiple flybys during an extended mission phase, with several small-scale missions, like Comet Interceptor or along the lines of the original Halley Armada. The next passage of comet 1P/Halley in our vicinity in 2061, together with other comets from different reservoirs, may provide an interesting window of opportunity.
\begin{table*}[ht]
    \centering
    \renewcommand{\arraystretch}{1}
    \begin{tabular}{p{.2\textwidth}p{.75\textwidth}}
        Topic & Highlights  \\
        \hline
        Birth of cometary ionosphere: & $\bullet$ Identification of ionospheric sources and losses, and of their relative contribution, for different outgassing activities (Section~\ref{section:2:2:4})\\
        &$\bullet$ CMEs/CIRs impact on the ionospheric sources and densities (Sections~\ref{section:2:2:4} and \ref{section:2:4:1})\\
        &$\bullet$ Plasma density profile observed all the way down to the surface (Section~\ref{section:2:4:1})\\
        &$\bullet$ First detection \& confirmation of ion species (up to 39~u/q); Evolution of the ion composition over heliocentric distance and season (Section~\ref{section:2:4:2})\\
        \hline
        Electron population: &$\bullet$ Three populations confirmed: cold, warm, and hot (Section~\ref{section:3:1})\\
        &$\bullet$ Symbiosis between the three populations: coexist and are dependent on each other (Section~\ref{section:3:1})\\
        &$\bullet$ Hot electrons: Similarities with auroral Physics at Earth (Section~\ref{section:3:1:3})\\
        &$\bullet$  Detection of cold electrons even when the coma is quasi-collisionless (Section~\ref{section:3:1:2})\\
        &$\bullet$ Evidence of the role played by electrons in the location of the diamagnetic cavity boundary (see Section~\ref{section:3:3})\\
        \hline
        Ion population: & $\bullet$ Three populations: slow cometary ions, early picked-up cometary ions, solar wind ions (Sections~\ref{section:4:2} and \ref{section:4:3})\\
        &$\bullet$ Evidence of chemistry  (ion-neutral collisions) taking place in the coma (Sections~\ref{section:2:4:2} and \ref{section:4:1})\\
        &$\bullet$ Evidence for accelerated cometary ions (Sections~\ref{section:3:2} and \ref{section:4:1})\\
        &$\bullet$ {Cometary ion density (but not flux) dominating over that of solar wind ions even at very low outgassing rate  (Section~\ref{section:2:2:4})}\\
        \hline
        Dusty plasma: & $\bullet$ No evidence yet of a dusty plasma at comet 67P with Rosetta (Section~\ref{section:5:3}) \\
        &$\bullet$ Observations of a few events of charged dust grains, positive and negative (Section~\ref{section:5:3})\\
        \hline
    \end{tabular}
    \caption{Highlights of this chapter on cometary ionospheres\label{tab:highlights}}
    \bigskip
    \bigskip
    \bigskip
    \begin{tabular}{p{.95\linewidth}}
        Open questions\\
        \hline
        \bf{Birth of cometary ionosphere:}\\
         $\bullet$ What is the plasma balance near perihelion? What are the key plasma source and loss?  \\
         $\bullet$ What is the 3D distribution of the ionospheric densities under different outgassing activities?\\
         $\bullet$ What is the composition of heavy ions ($>$40~u\,q$^{-1}$)?\\
         $\bullet$ What is the change, if any, of the ion composition in different plasma regions and boundaries? \\
         $\bullet$ What is the role played by the dust regarding EUV absorption under high activity? \\
        \hline
        \bf{Electron population}\\
         $\bullet$ What is the energy distribution of electrons at high temporal and energy resolutions? How does it vary in space? \\
         $\bullet$ What is the exact origin of cold electrons in a quasi-collisionless coma?\\
         $\bullet$ What is the role of cold electron in the presence of a diamagnetic cavity? \\
         $\bullet$ What is the source of hot electrons at high activity, when a diamagnetic cavity is formed?\\
         \hline
        \bf{Ion population:}\\
         $\bullet$ What is the energy distribution of cometary, slow ions?\\
         $\bullet$ Are ions accelerated near the surface at high activity and, if so, what is the process responsible for it?\\
         $\bullet$ What is driving the cometary ion dynamics ?\\
         $\bullet$ What is the exact role of electrons on the ion acceleration?\\
         $\bullet$ How to reconcile the observed high ion bulk speeds and the observed electron densities explained by a model excluding ion acceleration? \\ 
         $\bullet$ How to reconcile the observed high ion bulk speeds and the observed ion composition attesting of ion-neutral chemistry in the coma?\\
         \hline
        \bf{Dusty plasma:}\\
         $\bullet$ Does dust have significant impact on the cometary ionosphere for more active comets?\\
         {$\bullet$ Are dusty plasma interactions the same at all comets and, if not, what are the conditions driving those effects?}\\
        \hline
    \end{tabular}
    \caption{Non-exhaustive list of open questions regarding cometary ionospheres. \label{tab:openQuestions}}
\end{table*}

\vskip .5in
\noindent \textbf{Acknowledgments.} \\

We would like to warmly thank P.~Stephenson (Imperial College London, UK) and G.~Wattieaux (Universit\'e of Toulouse, France) for their help and support, in particular in providing un/processed dataset shown in Fig.~\ref{fig:electron_spectrum} and in generating Fig.~\ref{fig:cold_e}, respectively. We warmly thank Z.~Lewis for her careful reading of the chapter and her suggestions.
A.B. and his work at Ume\aa\ University was supported by the Swedish National Space Agency (SNSA) grant 108/18. C.S.W. thanks the Austrian Science Fund (FWF) P32035-N36. Work by M.G. at Imperial College London was supported in part by STFC of UK under grant ST/W001071/1.
\clearpage
\bibliographystyle{sss-three.bst}
\bibliography{biblio.bib}
\end{document}